\documentclass[twocolumn,longauth,traditabstract]{aa}

\usepackage{graphicx,amsmath}
\usepackage{epsf}
\usepackage{txfonts}
\usepackage[hyperindex,breaklinks, colorlinks, citecolor=blue]{hyperref}  
\usepackage{natbib}
\usepackage{upgreek}
\usepackage{url}
\usepackage{fixltx2e}
\usepackage{color}
\usepackage{txfonts}
\usepackage{aas_macros}
\usepackage[mathscr]{euscript}
\usepackage{verbatim}
\newcommand{\healpix}{{\tt HEALPix}}
\def\ge{\geqslant}

\bibpunct{(}{)}{;}{a}{}{,} 

\def\setsymbol#1#2{\expandafter\def\csname #1\endcsname{#2}}
\def\getsymbol#1{\csname #1\endcsname}

\def\Planck{\textit{Planck}}

\def\all2013resultspapers{\nocite{planck2013-p01, planck2013-p02, planck2013-p02a, planck2013-p02d, planck2013-p02b, planck2013-p03, planck2013-p03c, planck2013-p03f, planck2013-p03d, planck2013-p03e, planck2013-p01a, planck2013-p06, planck2013-p03a, planck2013-pip88, planck2013-p08, planck2013-p11, planck2013-p12, planck2013-p13, planck2013-p14, planck2013-p15, planck2013-p05b, planck2013-p17, planck2013-p09, planck2013-p09a, planck2013-p20, planck2013-p19, planck2013-pipaberration, planck2013-p05, planck2013-p05a, planck2013-pip56, planck2013-p06b}}

\newbox\tablebox    \newdimen\tablewidth
\def\leaderfil{\leaders\hbox to 5pt{\hss.\hss}\hfil}
\def\endPlancktable{\tablewidth=\columnwidth 
    $$\hss\copy\tablebox\hss$$
    \vskip-\lastskip\vskip -2pt}

\def\tablenote#1 #2\par{\begingroup \parindent=0.8em
    \abovedisplayshortskip=0pt\belowdisplayshortskip=0pt
    \noindent
    $$\hss\vbox{\hsize\tablewidth \hangindent=\parindent \hangafter=1 \noindent
    \hbox to \parindent{$^#1$\hss}\strut#2\strut\par}\hss$$
    \endgroup}
\def\doubleline{\vskip 3pt\hrule \vskip 1.5pt \hrule \vskip 5pt}

\def\L2{\ifmmode L_2\else $L_2$\fi}

\def\DeltaT{\ifmmode \Delta T\else $\Delta T$\fi}
\def\deltat{\ifmmode \Delta t\else $\Delta t$\fi}
\def\fknee{\ifmmode f_{\rm knee}\else $f_{\rm knee}$\fi}
\def\Fmax{\ifmmode F_{\rm max}\else $F_{\rm max}$\fi}
\def\solar{\ifmmode{\rm M}_{\mathord\odot}\else${\rm M}_{\mathord\odot}$\fi}
\def\Msolar{\ifmmode{\rm M}_{\mathord\odot}\else${\rm M}_{\mathord\odot}$\fi}
\def\Lsolar{\ifmmode{\rm L}_{\mathord\odot}\else${\rm L}_{\mathord\odot}$\fi}

\def\inv{\ifmmode^{-1}\else$^{-1}$\fi}
\def\mo{\ifmmode^{-1}\else$^{-1}$\fi}
\def\sup#1{\ifmmode ^{\rm #1}\else $^{\rm #1}$\fi}
\def\expo#1{\ifmmode \times 10^{#1}\else $\times 10^{#1}$\fi}
\def\,{\thinspace}
\def\lsim{\mathrel{\raise .4ex\hbox{\rlap{$<$}\lower 1.2ex\hbox{$\sim$}}}}
\def\gsim{\mathrel{\raise .4ex\hbox{\rlap{$>$}\lower 1.2ex\hbox{$\sim$}}}}

\def\simprop{\mathrel{\raise .4ex\hbox{\rlap{$\propto$}\lower 1.2ex\hbox{$\sim$}}}}
\def\deg{\ifmmode^\circ\else$^\circ$\fi}
\def\pdeg{\ifmmode $\setbox0=\hbox{$^{\circ}$}\rlap{\hskip.11\wd0 .}$^{\circ}
          \else \setbox0=\hbox{$^{\circ}$}\rlap{\hskip.11\wd0 .}$^{\circ}$\fi}
\def\arcs{\ifmmode {^{\scriptstyle\prime\prime}}
          \else $^{\scriptstyle\prime\prime}$\fi}
\def\arcm{\ifmmode {^{\scriptstyle\prime}}
          \else $^{\scriptstyle\prime}$\fi}
\newdimen\sa  \newdimen\sb
\def\parcs{\sa=.07em \sb=.03em
     \ifmmode \hbox{\rlap{.}}^{\scriptstyle\prime\kern -\sb\prime}\hbox{\kern -\sa}
     \else \rlap{.}$^{\scriptstyle\prime\kern -\sb\prime}$\kern -\sa\fi}
\def\parcm{\sa=.08em \sb=.03em
     \ifmmode \hbox{\rlap{.}\kern\sa}^{\scriptstyle\prime}\hbox{\kern-\sb}
     \else \rlap{.}\kern\sa$^{\scriptstyle\prime}$\kern-\sb\fi}
\def\ra[#1 #2 #3.#4]{#1\sup{h}#2\sup{m}#3\sup{s}\llap.#4}
\def\dec[#1 #2 #3.#4]{#1\deg#2\arcm#3\arcs\llap.#4}
\def\deco[#1 #2 #3]{#1\deg#2\arcm#3\arcs}
\def\rra[#1 #2]{#1\sup{h}#2\sup{m}}

\def\dots{\relax\ifmmode \ldots\else $\ldots$\fi}
\def\WHzsr{\ifmmode $W\,Hz\mo\,sr\mo$\else W\,Hz\mo\,sr\mo\fi}
\def\mHz{\ifmmode $\,mHz$\else \,mHz\fi}
\def\GHz{\ifmmode $\,GHz$\else \,GHz\fi}
\def\mKs{\ifmmode $\,mK\,s$^{1/2}\else \,mK\,s$^{1/2}$\fi}
\def\muKs{\ifmmode \,\mu$K\,s$^{1/2}\else \,$\mu$K\,s$^{1/2}$\fi}
\def\muKRJs{\ifmmode \,\mu$K$_{\rm RJ}$\,s$^{1/2}\else \,$\mu$K$_{\rm RJ}$\,s$^{1/2}$\fi}
\def\muKHz{\ifmmode \,\mu$K\,Hz$^{-1/2}\else \,$\mu$K\,Hz$^{-1/2}$\fi}
\def\MJysr{\ifmmode \,$MJy\,sr\mo$\else \,MJy\,sr\mo\fi}
\def\MJysrmK{\ifmmode \,$MJy\,sr\mo$\,mK$_{\rm CMB}\mo\else \,MJy\,sr\mo\,mK$_{\rm CMB}\mo$\fi}
\def\microns{\ifmmode \,\mu$m$\else \,$\mu$m\fi}

\def\muK{\ifmmode \,\mu$K$\else \,$\mu$\hbox{K}\fi}
\def\microK{\ifmmode \,\mu$K$\else \,$\mu$\hbox{K}\fi}
\def\muW{\ifmmode \,\mu$W$\else \,$\mu$\hbox{W}\fi}
\def\kms{\ifmmode $\,km\,s$^{-1}\else \,km\,s$^{-1}$\fi}
\def\kmsMpc{\ifmmode $\,\kms\,Mpc\mo$\else \,\kms\,Mpc\mo\fi}
\setsymbol{LFI:center:frequency:70GHz:units}{70.3\,GHz}
\setsymbol{LFI:center:frequency:44GHz:units}{44.1\,GHz}
\setsymbol{LFI:center:frequency:30GHz:units}{28.5\,GHz}

\setsymbol{LFI:center:frequency:70GHz}{70.3}
\setsymbol{LFI:center:frequency:44GHz}{44.1}
\setsymbol{LFI:center:frequency:30GHz}{28.5}

\setsymbol{LFI:center:frequency:LFI18:Rad:M:units}{71.7\GHz}
\setsymbol{LFI:center:frequency:LFI19:Rad:M:units}{67.5\GHz}
\setsymbol{LFI:center:frequency:LFI20:Rad:M:units}{69.2\GHz}
\setsymbol{LFI:center:frequency:LFI21:Rad:M:units}{70.4\GHz}
\setsymbol{LFI:center:frequency:LFI22:Rad:M:units}{71.5\GHz}
\setsymbol{LFI:center:frequency:LFI23:Rad:M:units}{70.8\GHz}
\setsymbol{LFI:center:frequency:LFI24:Rad:M:units}{44.4\GHz}
\setsymbol{LFI:center:frequency:LFI25:Rad:M:units}{44.0\GHz}
\setsymbol{LFI:center:frequency:LFI26:Rad:M:units}{43.9\GHz}
\setsymbol{LFI:center:frequency:LFI27:Rad:M:units}{28.3\GHz}
\setsymbol{LFI:center:frequency:LFI28:Rad:M:units}{28.8\GHz}
\setsymbol{LFI:center:frequency:LFI18:Rad:S:units}{70.1\GHz}
\setsymbol{LFI:center:frequency:LFI19:Rad:S:units}{69.6\GHz}
\setsymbol{LFI:center:frequency:LFI20:Rad:S:units}{69.5\GHz}
\setsymbol{LFI:center:frequency:LFI21:Rad:S:units}{69.5\GHz}
\setsymbol{LFI:center:frequency:LFI22:Rad:S:units}{72.8\GHz}
\setsymbol{LFI:center:frequency:LFI23:Rad:S:units}{71.3\GHz}
\setsymbol{LFI:center:frequency:LFI24:Rad:S:units}{44.1\GHz}
\setsymbol{LFI:center:frequency:LFI25:Rad:S:units}{44.1\GHz}
\setsymbol{LFI:center:frequency:LFI26:Rad:S:units}{44.1\GHz}
\setsymbol{LFI:center:frequency:LFI27:Rad:S:units}{28.5\GHz}
\setsymbol{LFI:center:frequency:LFI28:Rad:S:units}{28.2\GHz}

\setsymbol{LFI:center:frequency:LFI18:Rad:M}{71.7}
\setsymbol{LFI:center:frequency:LFI19:Rad:M}{67.5}
\setsymbol{LFI:center:frequency:LFI20:Rad:M}{69.2}
\setsymbol{LFI:center:frequency:LFI21:Rad:M}{70.4}
\setsymbol{LFI:center:frequency:LFI22:Rad:M}{71.5}
\setsymbol{LFI:center:frequency:LFI23:Rad:M}{70.8}
\setsymbol{LFI:center:frequency:LFI24:Rad:M}{44.4}
\setsymbol{LFI:center:frequency:LFI25:Rad:M}{44.0}
\setsymbol{LFI:center:frequency:LFI26:Rad:M}{43.9}
\setsymbol{LFI:center:frequency:LFI27:Rad:M}{28.3}
\setsymbol{LFI:center:frequency:LFI28:Rad:M}{28.8}
\setsymbol{LFI:center:frequency:LFI18:Rad:S}{70.1}
\setsymbol{LFI:center:frequency:LFI19:Rad:S}{69.6}
\setsymbol{LFI:center:frequency:LFI20:Rad:S}{69.5}
\setsymbol{LFI:center:frequency:LFI21:Rad:S}{69.5}
\setsymbol{LFI:center:frequency:LFI22:Rad:S}{72.8}
\setsymbol{LFI:center:frequency:LFI23:Rad:S}{71.3}
\setsymbol{LFI:center:frequency:LFI24:Rad:S}{44.1}
\setsymbol{LFI:center:frequency:LFI25:Rad:S}{44.1}
\setsymbol{LFI:center:frequency:LFI26:Rad:S}{44.1}
\setsymbol{LFI:center:frequency:LFI27:Rad:S}{28.5}
\setsymbol{LFI:center:frequency:LFI28:Rad:S}{28.2}

\setsymbol{LFI:white:noise:sensitivity:70GHz:units}{134.7\muKs}
\setsymbol{LFI:white:noise:sensitivity:44GHz:units}{164.7\muKs}
\setsymbol{LFI:white:noise:sensitivity:30GHz:units}{143.4\muKs}

\setsymbol{LFI:white:noise:sensitivity:70GHz}{134.7}
\setsymbol{LFI:white:noise:sensitivity:44GHz}{164.7}
\setsymbol{LFI:white:noise:sensitivity:30GHz}{143.4}

\setsymbol{LFI:white:noise:sensitivity:LFI18:Rad:M:units}{512.0\muKs}
\setsymbol{LFI:white:noise:sensitivity:LFI19:Rad:M:units}{581.4\muKs}
\setsymbol{LFI:white:noise:sensitivity:LFI20:Rad:M:units}{590.8\muKs}
\setsymbol{LFI:white:noise:sensitivity:LFI21:Rad:M:units}{455.2\muKs}
\setsymbol{LFI:white:noise:sensitivity:LFI22:Rad:M:units}{492.0\muKs}
\setsymbol{LFI:white:noise:sensitivity:LFI23:Rad:M:units}{507.7\muKs}
\setsymbol{LFI:white:noise:sensitivity:LFI24:Rad:M:units}{462.2\muKs}
\setsymbol{LFI:white:noise:sensitivity:LFI25:Rad:M:units}{413.6\muKs}
\setsymbol{LFI:white:noise:sensitivity:LFI26:Rad:M:units}{478.6\muKs}
\setsymbol{LFI:white:noise:sensitivity:LFI27:Rad:M:units}{277.7\muKs}
\setsymbol{LFI:white:noise:sensitivity:LFI28:Rad:M:units}{312.3\muKs}
\setsymbol{LFI:white:noise:sensitivity:LFI18:Rad:S:units}{465.7\muKs}
\setsymbol{LFI:white:noise:sensitivity:LFI19:Rad:S:units}{555.6\muKs}
\setsymbol{LFI:white:noise:sensitivity:LFI20:Rad:S:units}{623.2\muKs}
\setsymbol{LFI:white:noise:sensitivity:LFI21:Rad:S:units}{564.1\muKs}
\setsymbol{LFI:white:noise:sensitivity:LFI22:Rad:S:units}{534.4\muKs}
\setsymbol{LFI:white:noise:sensitivity:LFI23:Rad:S:units}{542.4\muKs}
\setsymbol{LFI:white:noise:sensitivity:LFI24:Rad:S:units}{399.2\muKs}
\setsymbol{LFI:white:noise:sensitivity:LFI25:Rad:S:units}{392.6\muKs}
\setsymbol{LFI:white:noise:sensitivity:LFI26:Rad:S:units}{418.6\muKs}
\setsymbol{LFI:white:noise:sensitivity:LFI27:Rad:S:units}{302.9\muKs}
\setsymbol{LFI:white:noise:sensitivity:LFI28:Rad:S:units}{285.3\muKs}

\setsymbol{LFI:white:noise:sensitivity:LFI18:Rad:M}{512.0}
\setsymbol{LFI:white:noise:sensitivity:LFI19:Rad:M}{581.4}
\setsymbol{LFI:white:noise:sensitivity:LFI20:Rad:M}{590.8}
\setsymbol{LFI:white:noise:sensitivity:LFI21:Rad:M}{455.2}
\setsymbol{LFI:white:noise:sensitivity:LFI22:Rad:M}{492.0}
\setsymbol{LFI:white:noise:sensitivity:LFI23:Rad:M}{507.7}
\setsymbol{LFI:white:noise:sensitivity:LFI24:Rad:M}{462.2}
\setsymbol{LFI:white:noise:sensitivity:LFI25:Rad:M}{413.6}
\setsymbol{LFI:white:noise:sensitivity:LFI26:Rad:M}{478.6}
\setsymbol{LFI:white:noise:sensitivity:LFI27:Rad:M}{277.7}
\setsymbol{LFI:white:noise:sensitivity:LFI28:Rad:M}{312.3}
\setsymbol{LFI:white:noise:sensitivity:LFI18:Rad:S}{465.7}
\setsymbol{LFI:white:noise:sensitivity:LFI19:Rad:S}{555.6}
\setsymbol{LFI:white:noise:sensitivity:LFI20:Rad:S}{623.2}
\setsymbol{LFI:white:noise:sensitivity:LFI21:Rad:S}{564.1}
\setsymbol{LFI:white:noise:sensitivity:LFI22:Rad:S}{534.4}
\setsymbol{LFI:white:noise:sensitivity:LFI23:Rad:S}{542.4}
\setsymbol{LFI:white:noise:sensitivity:LFI24:Rad:S}{399.2}
\setsymbol{LFI:white:noise:sensitivity:LFI25:Rad:S}{392.6}
\setsymbol{LFI:white:noise:sensitivity:LFI26:Rad:S}{418.6}
\setsymbol{LFI:white:noise:sensitivity:LFI27:Rad:S}{302.9}
\setsymbol{LFI:white:noise:sensitivity:LFI28:Rad:S}{285.3}

\setsymbol{LFI:knee:frequency:70GHz:units}{29.5\mHz}
\setsymbol{LFI:knee:frequency:44GHz:units}{56.2\mHz}
\setsymbol{LFI:knee:frequency:30GHz:units}{113.7\mHz}

\setsymbol{LFI:knee:frequency:70GHz}{29.5}
\setsymbol{LFI:knee:frequency:44GHz}{56.2}
\setsymbol{LFI:knee:frequency:30GHz}{113.7}

\setsymbol{LFI:knee:frequency:LFI18:Rad:M:units}{16.3\mHz}
\setsymbol{LFI:knee:frequency:LFI19:Rad:M:units}{15.1\mHz}
\setsymbol{LFI:knee:frequency:LFI20:Rad:M:units}{18.7\mHz}
\setsymbol{LFI:knee:frequency:LFI21:Rad:M:units}{37.2\mHz}
\setsymbol{LFI:knee:frequency:LFI22:Rad:M:units}{12.7\mHz}
\setsymbol{LFI:knee:frequency:LFI23:Rad:M:units}{34.6\mHz}
\setsymbol{LFI:knee:frequency:LFI24:Rad:M:units}{46.2\mHz}
\setsymbol{LFI:knee:frequency:LFI25:Rad:M:units}{24.9\mHz}
\setsymbol{LFI:knee:frequency:LFI26:Rad:M:units}{67.6\mHz}
\setsymbol{LFI:knee:frequency:LFI27:Rad:M:units}{187.4\mHz}
\setsymbol{LFI:knee:frequency:LFI28:Rad:M:units}{122.2\mHz}
\setsymbol{LFI:knee:frequency:LFI18:Rad:S:units}{17.7\mHz}
\setsymbol{LFI:knee:frequency:LFI19:Rad:S:units}{22.0\mHz}
\setsymbol{LFI:knee:frequency:LFI20:Rad:S:units}{8.7\mHz}
\setsymbol{LFI:knee:frequency:LFI21:Rad:S:units}{25.9\mHz}
\setsymbol{LFI:knee:frequency:LFI22:Rad:S:units}{15.8\mHz}
\setsymbol{LFI:knee:frequency:LFI23:Rad:S:units}{129.8\mHz}
\setsymbol{LFI:knee:frequency:LFI24:Rad:S:units}{100.9\mHz}
\setsymbol{LFI:knee:frequency:LFI25:Rad:S:units}{38.9\mHz}
\setsymbol{LFI:knee:frequency:LFI26:Rad:S:units}{58.9\mHz}
\setsymbol{LFI:knee:frequency:LFI27:Rad:S:units}{104.4\mHz}
\setsymbol{LFI:knee:frequency:LFI28:Rad:S:units}{40.7\mHz}

\setsymbol{LFI:knee:frequency:LFI18:Rad:M}{16.3}
\setsymbol{LFI:knee:frequency:LFI19:Rad:M}{15.1}
\setsymbol{LFI:knee:frequency:LFI20:Rad:M}{18.7}
\setsymbol{LFI:knee:frequency:LFI21:Rad:M}{37.2}
\setsymbol{LFI:knee:frequency:LFI22:Rad:M}{12.7}
\setsymbol{LFI:knee:frequency:LFI23:Rad:M}{34.6}
\setsymbol{LFI:knee:frequency:LFI24:Rad:M}{46.2}
\setsymbol{LFI:knee:frequency:LFI25:Rad:M}{24.9}
\setsymbol{LFI:knee:frequency:LFI26:Rad:M}{67.6}
\setsymbol{LFI:knee:frequency:LFI27:Rad:M}{187.4}
\setsymbol{LFI:knee:frequency:LFI28:Rad:M}{122.2}
\setsymbol{LFI:knee:frequency:LFI18:Rad:S}{17.7}
\setsymbol{LFI:knee:frequency:LFI19:Rad:S}{22.0}
\setsymbol{LFI:knee:frequency:LFI20:Rad:S}{8.7}
\setsymbol{LFI:knee:frequency:LFI21:Rad:S}{25.9}
\setsymbol{LFI:knee:frequency:LFI22:Rad:S}{15.8}
\setsymbol{LFI:knee:frequency:LFI23:Rad:S}{129.8}
\setsymbol{LFI:knee:frequency:LFI24:Rad:S}{100.9}
\setsymbol{LFI:knee:frequency:LFI25:Rad:S}{38.9}
\setsymbol{LFI:knee:frequency:LFI26:Rad:S}{58.9}
\setsymbol{LFI:knee:frequency:LFI27:Rad:S}{104.4}
\setsymbol{LFI:knee:frequency:LFI28:Rad:S}{40.7}

\setsymbol{LFI:slope:70GHz:units}{$-1.03$\mHz}
\setsymbol{LFI:slope:44GHz:units}{$-0.89$\mHz}
\setsymbol{LFI:slope:30GHz:units}{$-0.87$\mHz}

\setsymbol{LFI:slope:70GHz}{$-1.03$}
\setsymbol{LFI:slope:44GHz}{$-0.89$}
\setsymbol{LFI:slope:30GHz}{$-0.87$}

\setsymbol{LFI:slope:LFI18:Rad:M:units}{$-1.04$\mHz}
\setsymbol{LFI:slope:LFI19:Rad:M:units}{$-1.09$\mHz}
\setsymbol{LFI:slope:LFI20:Rad:M:units}{$-0.69$\mHz}
\setsymbol{LFI:slope:LFI21:Rad:M:units}{$-1.56$\mHz}
\setsymbol{LFI:slope:LFI22:Rad:M:units}{$-1.01$\mHz}
\setsymbol{LFI:slope:LFI23:Rad:M:units}{$-0.96$\mHz}
\setsymbol{LFI:slope:LFI24:Rad:M:units}{$-0.83$\mHz}
\setsymbol{LFI:slope:LFI25:Rad:M:units}{$-0.91$\mHz}
\setsymbol{LFI:slope:LFI26:Rad:M:units}{$-0.95$\mHz}
\setsymbol{LFI:slope:LFI27:Rad:M:units}{$-0.87$\mHz}
\setsymbol{LFI:slope:LFI28:Rad:M:units}{$-0.88$\mHz}
\setsymbol{LFI:slope:LFI18:Rad:S:units}{$-1.15$\mHz}
\setsymbol{LFI:slope:LFI19:Rad:S:units}{$-1.00$\mHz}
\setsymbol{LFI:slope:LFI20:Rad:S:units}{$-0.95$\mHz}
\setsymbol{LFI:slope:LFI21:Rad:S:units}{$-0.92$\mHz}
\setsymbol{LFI:slope:LFI22:Rad:S:units}{$-1.01$\mHz}
\setsymbol{LFI:slope:LFI23:Rad:S:units}{$-0.95$\mHz}
\setsymbol{LFI:slope:LFI24:Rad:S:units}{$-0.73$\mHz}
\setsymbol{LFI:slope:LFI25:Rad:S:units}{$-1.16$\mHz}
\setsymbol{LFI:slope:LFI26:Rad:S:units}{$-0.79$\mHz}
\setsymbol{LFI:slope:LFI27:Rad:S:units}{$-0.82$\mHz}
\setsymbol{LFI:slope:LFI28:Rad:S:units}{$-0.91$\mHz}

\setsymbol{LFI:slope:LFI18:Rad:M}{$-1.04$}
\setsymbol{LFI:slope:LFI19:Rad:M}{$-1.09$}
\setsymbol{LFI:slope:LFI20:Rad:M}{$-0.69$}
\setsymbol{LFI:slope:LFI21:Rad:M}{$-1.56$}
\setsymbol{LFI:slope:LFI22:Rad:M}{$-1.01$}
\setsymbol{LFI:slope:LFI23:Rad:M}{$-0.96$}
\setsymbol{LFI:slope:LFI24:Rad:M}{$-0.83$}
\setsymbol{LFI:slope:LFI25:Rad:M}{$-0.91$}
\setsymbol{LFI:slope:LFI26:Rad:M}{$-0.95$}
\setsymbol{LFI:slope:LFI27:Rad:M}{$-0.87$}
\setsymbol{LFI:slope:LFI28:Rad:M}{$-0.88$}
\setsymbol{LFI:slope:LFI18:Rad:S}{$-1.15$}
\setsymbol{LFI:slope:LFI19:Rad:S}{$-1.00$}
\setsymbol{LFI:slope:LFI20:Rad:S}{$-0.95$}
\setsymbol{LFI:slope:LFI21:Rad:S}{$-0.92$}
\setsymbol{LFI:slope:LFI22:Rad:S}{$-1.01$}
\setsymbol{LFI:slope:LFI23:Rad:S}{$-0.95$}
\setsymbol{LFI:slope:LFI24:Rad:S}{$-0.73$}
\setsymbol{LFI:slope:LFI25:Rad:S}{$-1.16$}
\setsymbol{LFI:slope:LFI26:Rad:S}{$-0.79$}
\setsymbol{LFI:slope:LFI27:Rad:S}{$-0.82$}
\setsymbol{LFI:slope:LFI28:Rad:S}{$-0.91$}

\setsymbol{LFI:FWHM:70GHz:units}{13\parcm01}
\setsymbol{LFI:FWHM:44GHz:units}{27\parcm92}
\setsymbol{LFI:FWHM:30GHz:units}{32\parcm65}

\setsymbol{LFI:FWHM:70GHz}{13.01}
\setsymbol{LFI:FWHM:44GHz}{27.92}
\setsymbol{LFI:FWHM:30GHz}{32.65}

\setsymbol{LFI:FWHM:LFI18:units}{13\parcm39}
\setsymbol{LFI:FWHM:LFI19:units}{13\parcm01}
\setsymbol{LFI:FWHM:LFI20:units}{12\parcm75}
\setsymbol{LFI:FWHM:LFI21:units}{12\parcm74}
\setsymbol{LFI:FWHM:LFI22:units}{12\parcm87}
\setsymbol{LFI:FWHM:LFI23:units}{13\parcm27}
\setsymbol{LFI:FWHM:LFI24:units}{22\parcm98}
\setsymbol{LFI:FWHM:LFI25:units}{30\parcm46}
\setsymbol{LFI:FWHM:LFI26:units}{30\parcm31}
\setsymbol{LFI:FWHM:LFI27:units}{32\parcm65}
\setsymbol{LFI:FWHM:LFI28:units}{32\parcm66}

\setsymbol{LFI:FWHM:LFI18}{13.39}
\setsymbol{LFI:FWHM:LFI19}{13.01}
\setsymbol{LFI:FWHM:LFI20}{12.75}
\setsymbol{LFI:FWHM:LFI21}{12.74}
\setsymbol{LFI:FWHM:LFI22}{12.87}
\setsymbol{LFI:FWHM:LFI23}{13.27}
\setsymbol{LFI:FWHM:LFI24}{22.98}
\setsymbol{LFI:FWHM:LFI25}{30.46}
\setsymbol{LFI:FWHM:LFI26}{30.31}
\setsymbol{LFI:FWHM:LFI27}{32.65}
\setsymbol{LFI:FWHM:LFI28}{32.66}

\setsymbol{LFI:FWHM:uncertainty:LFI18:units}{0.170\arcm}
\setsymbol{LFI:FWHM:uncertainty:LFI19:units}{0.174\arcm}
\setsymbol{LFI:FWHM:uncertainty:LFI20:units}{0.170\arcm}
\setsymbol{LFI:FWHM:uncertainty:LFI21:units}{0.156\arcm}
\setsymbol{LFI:FWHM:uncertainty:LFI22:units}{0.164\arcm}
\setsymbol{LFI:FWHM:uncertainty:LFI23:units}{0.171\arcm}
\setsymbol{LFI:FWHM:uncertainty:LFI24:units}{0.652\arcm}
\setsymbol{LFI:FWHM:uncertainty:LFI25:units}{1.075\arcm}
\setsymbol{LFI:FWHM:uncertainty:LFI26:units}{1.131\arcm}
\setsymbol{LFI:FWHM:uncertainty:LFI27:units}{1.266\arcm}
\setsymbol{LFI:FWHM:uncertainty:LFI28:units}{1.287\arcm}

\setsymbol{LFI:FWHM:uncertainty:LFI18}{0.170}
\setsymbol{LFI:FWHM:uncertainty:LFI19}{0.174}
\setsymbol{LFI:FWHM:uncertainty:LFI20}{0.170}
\setsymbol{LFI:FWHM:uncertainty:LFI21}{0.156}
\setsymbol{LFI:FWHM:uncertainty:LFI22}{0.164}
\setsymbol{LFI:FWHM:uncertainty:LFI23}{0.171}
\setsymbol{LFI:FWHM:uncertainty:LFI24}{0.652}
\setsymbol{LFI:FWHM:uncertainty:LFI25}{1.075}
\setsymbol{LFI:FWHM:uncertainty:LFI26}{1.131}
\setsymbol{LFI:FWHM:uncertainty:LFI27}{1.266}
\setsymbol{LFI:FWHM:uncertainty:LFI28}{1.287}

\setsymbol{HFI:center:frequency:100GHz:units}{100\,GHz}
\setsymbol{HFI:center:frequency:143GHz:units}{143\,GHz}
\setsymbol{HFI:center:frequency:217GHz:units}{217\,GHz}
\setsymbol{HFI:center:frequency:353GHz:units}{353\,GHz}
\setsymbol{HFI:center:frequency:545GHz:units}{545\,GHz}
\setsymbol{HFI:center:frequency:857GHz:units}{857\,GHz}

\setsymbol{HFI:center:frequency:100GHz}{100}
\setsymbol{HFI:center:frequency:143GHz}{143}
\setsymbol{HFI:center:frequency:217GHz}{217}
\setsymbol{HFI:center:frequency:353GHz}{353}
\setsymbol{HFI:center:frequency:545GHz}{545}
\setsymbol{HFI:center:frequency:857GHz}{857}

\setsymbol{HFI:Ndetectors:100GHz}{8}
\setsymbol{HFI:Ndetectors:143GHz}{11}
\setsymbol{HFI:Ndetectors:217GHz}{12}
\setsymbol{HFI:Ndetectors:353GHz}{12}
\setsymbol{HFI:Ndetectors:545GHz}{3}
\setsymbol{HFI:Ndetectors:857GHz}{4}

\setsymbol{HFI:FWHM:Maps:100GHz:units}{9\parcm88}
\setsymbol{HFI:FWHM:Maps:143GHz:units}{7\parcm18}
\setsymbol{HFI:FWHM:Maps:217GHz:units}{4\parcm87}
\setsymbol{HFI:FWHM:Maps:353GHz:units}{4\parcm65}
\setsymbol{HFI:FWHM:Maps:545GHz:units}{4\parcm72}
\setsymbol{HFI:FWHM:Maps:857GHz:units}{4\parcm39}
\setsymbol{HFI:FWHM:Maps:100GHz}{9.88}
\setsymbol{HFI:FWHM:Maps:143GHz}{7.18}
\setsymbol{HFI:FWHM:Maps:217GHz}{4.87}
\setsymbol{HFI:FWHM:Maps:353GHz}{4.65}
\setsymbol{HFI:FWHM:Maps:545GHz}{4.72}
\setsymbol{HFI:FWHM:Maps:857GHz}{4.39}

\setsymbol{HFI:beam:ellipticity:Maps:100GHz}{1.15}
\setsymbol{HFI:beam:ellipticity:Maps:143GHz}{1.01}
\setsymbol{HFI:beam:ellipticity:Maps:217GHz}{1.06}
\setsymbol{HFI:beam:ellipticity:Maps:353GHz}{1.05}
\setsymbol{HFI:beam:ellipticity:Maps:545GHz}{1.14}
\setsymbol{HFI:beam:ellipticity:Maps:857GHz}{1.19}

\setsymbol{HFI:FWHM:Mars:100GHz:units}{9\parcm37}
\setsymbol{HFI:FWHM:Mars:143GHz:units}{7\parcm04}
\setsymbol{HFI:FWHM:Mars:217GHz:units}{4\parcm68}
\setsymbol{HFI:FWHM:Mars:353GHz:units}{4\parcm43}
\setsymbol{HFI:FWHM:Mars:545GHz:units}{3\parcm80}
\setsymbol{HFI:FWHM:Mars:857GHz:units}{3\parcm67}

\setsymbol{HFI:FWHM:Mars:100GHz}{9.37}
\setsymbol{HFI:FWHM:Mars:143GHz}{7.04}
\setsymbol{HFI:FWHM:Mars:217GHz}{4.68}
\setsymbol{HFI:FWHM:Mars:353GHz}{4.43}
\setsymbol{HFI:FWHM:Mars:545GHz}{3.80}
\setsymbol{HFI:FWHM:Mars:857GHz}{3.67}

\setsymbol{HFI:beam:ellipticity:Mars:100GHz}{1.18}
\setsymbol{HFI:beam:ellipticity:Mars:143GHz}{1.03}
\setsymbol{HFI:beam:ellipticity:Mars:217GHz}{1.14}
\setsymbol{HFI:beam:ellipticity:Mars:353GHz}{1.09}
\setsymbol{HFI:beam:ellipticity:Mars:545GHz}{1.25}
\setsymbol{HFI:beam:ellipticity:Mars:857GHz}{1.03}

\setsymbol{HFI:CMB:relative:calibration:100GHz}{$\lsim 1\%$}
\setsymbol{HFI:CMB:relative:calibration:143GHz}{$\lsim 1\%$}
\setsymbol{HFI:CMB:relative:calibration:217GHz}{$\lsim 1\%$}
\setsymbol{HFI:CMB:relative:calibration:353GHz}{$\lsim 1\%$}
\setsymbol{HFI:CMB:relative:calibration:545GHz}{}
\setsymbol{HFI:CMB:relative:calibration:857GHz}{}

\setsymbol{HFI:CMB:absolute:calibration:100GHz}{$\lsim 2\%$}
\setsymbol{HFI:CMB:absolute:calibration:143GHz}{$\lsim 2\%$}
\setsymbol{HFI:CMB:absolute:calibration:217GHz}{$\lsim 2\%$}
\setsymbol{HFI:CMB:absolute:calibration:353GHz}{$\lsim 2\%$}
\setsymbol{HFI:CMB:absolute:calibration:545GHz}{}
\setsymbol{HFI:CMB:absolute:calibration:857GHz}{}

\setsymbol{HFI:FIRAS:gain:calibration:accuracy:statistical:100GHz}{}
\setsymbol{HFI:FIRAS:gain:calibration:accuracy:statistical:143GHz}{}
\setsymbol{HFI:FIRAS:gain:calibration:accuracy:statistical:217GHz}{}
\setsymbol{HFI:FIRAS:gain:calibration:accuracy:statistical:353GHz}{2.5\%}
\setsymbol{HFI:FIRAS:gain:calibration:accuracy:statistical:545GHz}{1\%}
\setsymbol{HFI:FIRAS:gain:calibration:accuracy:statistical:857GHz}{0.5\%}

\setsymbol{HFI:FIRAS:gain:calibration:accuracy:systematic:100GHz}{}
\setsymbol{HFI:FIRAS:gain:calibration:accuracy:systematic:143GHz}{}
\setsymbol{HFI:FIRAS:gain:calibration:accuracy:systematic:217GHz}{}
\setsymbol{HFI:FIRAS:gain:calibration:accuracy:systematic:353GHz}{}
\setsymbol{HFI:FIRAS:gain:calibration:accuracy:systematic:545GHz}{7\%}
\setsymbol{HFI:FIRAS:gain:calibration:accuracy:systematic:857GHz}{7\%}

\setsymbol{HFI:FIRAS:zero:point:accuracy:100GHz:units}{0.8\MJysr}
\setsymbol{HFI:FIRAS:zero:point:accuracy:143GHz:units}{}
\setsymbol{HFI:FIRAS:zero:point:accuracy:217GHz:units}{}
\setsymbol{HFI:FIRAS:zero:point:accuracy:353GHz:units}{1.4\MJysr}
\setsymbol{HFI:FIRAS:zero:point:accuracy:545GHz:units}{2.2\MJysr}
\setsymbol{HFI:FIRAS:zero:point:accuracy:857GHz:units}{1.7\MJysr}

\setsymbol{HFI:FIRAS:zero:point:accuracy:100GHz}{0.8}
\setsymbol{HFI:FIRAS:zero:point:accuracy:143GHz}{}
\setsymbol{HFI:FIRAS:zero:point:accuracy:217GHz}{}
\setsymbol{HFI:FIRAS:zero:point:accuracy:353GHz}{1.4}
\setsymbol{HFI:FIRAS:zero:point:accuracy:545GHz}{2.2}
\setsymbol{HFI:FIRAS:zero:point:accuracy:857GHz}{1.7}

\setsymbol{HFI:unit:conversion:100GHz:units}{0.2415\MJysrmK}
\setsymbol{HFI:unit:conversion:143GHz:units}{0.3694\MJysrmK}
\setsymbol{HFI:unit:conversion:217GHz:units}{0.4811\MJysrmK}
\setsymbol{HFI:unit:conversion:353GHz:units}{0.2883\MJysrmK}
\setsymbol{HFI:unit:conversion:545GHz:units}{0.05826\MJysrmK}
\setsymbol{HFI:unit:conversion:857GHz:units}{0.002238\MJysrmK}

\setsymbol{HFI:unit:conversion:100GHz}{0.2415}
\setsymbol{HFI:unit:conversion:143GHz}{0.3694}
\setsymbol{HFI:unit:conversion:217GHz}{0.4811}
\setsymbol{HFI:unit:conversion:353GHz}{0.2883}
\setsymbol{HFI:unit:conversion:545GHz}{0.05826}
\setsymbol{HFI:unit:conversion:857GHz}{0.002238}

\setsymbol{HFI:colour:correction:alpha=-2:V1.01:100GHz}{0.9893}
\setsymbol{HFI:colour:correction:alpha=-2:V1.01:143GHz}{0.9759}
\setsymbol{HFI:colour:correction:alpha=-2:V1.01:217GHz}{1.0007}
\setsymbol{HFI:colour:correction:alpha=-2:V1.01:353GHz}{1.0028}
\setsymbol{HFI:colour:correction:alpha=-2:V1.01:545GHz}{1.0019}
\setsymbol{HFI:colour:correction:alpha=-2:V1.01:857GHz}{0.9889}

\setsymbol{HFI:colour:correction:alpha=0:V1.01:100GHz}{1.0008}
\setsymbol{HFI:colour:correction:alpha=0:V1.01:143GHz}{1.0148}
\setsymbol{HFI:colour:correction:alpha=0:V1.01:217GHz}{0.9909}
\setsymbol{HFI:colour:correction:alpha=0:V1.01:353GHz}{0.9888}
\setsymbol{HFI:colour:correction:alpha=0:V1.01:545GHz}{0.9878}
\setsymbol{HFI:colour:correction:alpha=0:V1.01:857GHz}{1.0014}

\providecommand{\sorthelp}[1]{}

\def\wi{w_i}
\def\star{j}
\def\psiJ{\psi_\star}
\def\psiK{\psi_k}
\def\thetaJ{\theta_\star}
\def\thetaK{\theta_k}
\def\psiSK{\psi^\star_k}
\def\phiJ{\varphi_\star}
\def\phiK{\varphi_k}
\def\phiSK{\varphi^\star_k}
\def\betaSK{\beta}
\def\betaJI{\beta}
\def\Trans{{\rm T}}
\def\MatL{[\tens{L}]}
\def\VectG{(\tens{G})}
\def\VectN{(\tens{N})}

\def\PointJ{\mathrm{J}}
\def\PointK{\mathrm{K}}

\def\sigFWHM{\sigma_{\rm 1/2}}

\def\eg{e.g.,}
\def\exp{e}

\def\w{w_k}
\def\MatRpsi{[\tens{R}]_k}
\def\MatRpsiji{[\tens{R}]_i}
\def\co{a}
\def\coco{a^2}
\def\si{b}
\def\sisi{b^2}

\newcommand{\archeops}{{\it Archeops }}  %
\newcommand{\wmap}{\textit{WMAP}}  %
\newcommand{\hfi}{{HFI }}  %
\newcommand{\lfi}{{LFI }}  %
\newcommand{\Td}{T_{\rm obs}} 
\newcommand{\HI}{\ion {H}{i}} 
\newcommand{\NH}{N_{\rm H}} 
\newcommand{\nH}{n_{\rm H}} 
\newcommand{\Av}{A_{\rm V}}
\newcommand{\Rv}{R_{\rm V}}

\newcommand{\Nside}{N_\mathrm{side}} 
\newcommand{\dusttau}{\tau_{353}}        
\newcommand{\pv}{\polfrac_{\rm v}}   

\def\MJysr{\ifmmode {\rm \,MJy\,sr^{-1}} \else $\rm \,MJy\,sr^{-1}$ \fi}      
\def\Kkms{\ifmmode {\rm \,K\,km\,s^{-1}} \else $\rm \,K\,km\,s^{-1}$ \fi}
\def\nHUNIT{\ifmmode {\rm \,cm^{-3}} \else $\rm \,K\,km\,s^{-1}$ \fi} 
\def\NHUNIT{\ifmmode {\rm \,cm^{-2}} \else $\rm \,cm^{-2}$ \fi} 
\def\GHz{\ifmmode {\rm \,GHz} \else $\rm \,GHz$\fi}      
\def\mic{\ifmmode {\rm \,\upmu m} \else $\rm \,\upmu m$ \fi}      
\def\leaderfil{\leaders\hbox to 2pt{\hss.\hss}\hfil}

\newcommand{\plotresol}{1\deg} 
\newcommand{\nperc}{1\,\%} 

\newcommand{\sigpsyste}{\sigpolfrac^{\rm sys}}
\newcommand{\sigpstat}{\sigpolfrac^{\rm stat}}
\newcommand{\sigpsisyste}{\sigpolang^{\rm sys}}
\newcommand{\sigpsistat}{\sigpolang^{\rm stat}}

\newcommand{\db}{{\rm db}}
\newcommand{\obs}{{}}    

\newcommand{\StokesI}{I}                    
\newcommand{\StokesQ}{Q}                    
\newcommand{\StokesU}{U}                    
\newcommand{\polI}{I}                       
\newcommand{\polint}{P}                     
\newcommand{\polfrac}{p}                    
\newcommand{\polang}{\psi}                  
\newcommand{\sigpolfrac}{\sigma_{\polfrac}} 
\newcommand{\sigpolang}{\sigma_{\polang}}   
\newcommand{\DeltaAng}{\Delta \polang}          
\newcommand{\polangiau}{\psi_{\rm IAU}}         
\def\covar{\tens{C}}
\def\sigII{\covar_{II}}  
\def\sigQQ{\covar_{QQ}} 
\def\sigUU{\covar_{UU}} 
\def\sigIQ{\covar_{IQ}} 
\def\sigIU{\covar_{IU}} 
\def\sigQU{\covar_{QU}}  

\def\Scovar{\mathscr{C}}
\def\MatS{[\Scovar]_\star}
\def\MatC{[\covar]}
\def\sII{\Scovar^\star_{II}} 
\def\sIQ{\Scovar^\star_{IQ}}
\def\sIU{\Scovar^\star_{IU}} 
\def\sQQ{\Scovar^\star_{QQ}}
\def\sQU{\Scovar^\star_{QU}} 
\def\sUU{\Scovar^\star_{UU}} 
\def\CII{\covar_{II_k}} 
\def\CIQ{\covar_{IQ_k}}
\def\CIU{\covar_{IU_k}} 
\def\CQQ{\covar_{QQ_k}}
\def\CQU{\covar_{QU_k}} 
\def\CUU{\covar_{UU_k}} 

\newcommand{\pmax}{\polfrac_{\rm max}}        
\newcommand{\pzero}{\polfrac_{0}}         
\newcommand{\fsky}{f_{\rm sky}}         
\newcommand{\vect}[1]{\vec{#1}} 

\newcommand{\LoopI}{Loop~I}   
\newcommand{\PolarisFlare}{Polaris Flare}
\newcommand{\Orion}{Orion}
\newcommand{\Pipe}{Pipe}
\newcommand{\Ophiuchus}{Ophiuchus}
\newcommand{\Taurus}{Taurus}
\newcommand{\RCrA}{RCrA}
\newcommand{\Chamfil}{Cham-fil}
\newcommand{\Pyxis}{Pyxis}
\newcommand{\RII}{Aquila}         
\newcommand{\Auriga}{Auriga}
\newcommand{\RCrATail}{RCrA-Tail}
\newcommand{\Hercules}{Hercules}
\newcommand{\Libra}{Libra}
\newcommand{\ChamaeleonMusca}{Chamaeleon-Musca}
\newcommand{\Ara}{Ara}
\newcommand{\Pisces}{Pisces}
\newcommand{\RI}{Microscopium}         
\newcommand{\RIII}{Triangulum}         
\newcommand{\Perseus}{Perseus}         
\newcommand{\Pavo}{Pavo}         

\newcommand{\Fan}{Fan}          
\newcommand{\AquilaRift}{Aquila Rift}         

\newcommand{\BPM}{BPM}   
\newcommand{\SNR}{S/N}   

\newcommand{\resolution}{\theta}
\newcommand{\glon}{\ell_{\rm II}}
\newcommand{\glat}{b_{\rm II}}
\newcommand{\lag}{\delta}

\newcommand{\pmaxvalue}{19.8}       
\newcommand{\pmaxuncert}{0.7}       
\newcommand{\nsigused}{4}               
\newcommand{\randomdpsivalue}{52}               

\newcommand{\alphacorrel}{\alpha}              
\newcommand{\betacorrel}{\beta}              
\newcommand{\alphacorrelvalue}{-0.834}              
\newcommand{\betacorrelvalue}{-0.504}               
\newcommand{\maskedfraction}{21}               

\newcommand{\Healpix}{\tt HEALPix}
\def\vlsr{\ifmmode{v_{\rm lsr}}\else{$v_{\rm lsr}$}\fi}
\def\testzero{\ifmmode{v_{\rm \DeltaAng=0\degr}}\else{$\DeltaAng=0\degr$}\fi}
\def\testnoise{\ifmmode{v_{\rm \DeltaAng=52\degr}}\else{$\DeltaAng=52\degr$}\fi}

\newcommand{\Bfield}{\vec{B}}
\newcommand{\appBfield}{apparent magnetic field}
\newcommand{\Bperp}{\langle\Bfield_\perp\rangle}         

\newcommand{\Spaghettis}{filamentary features}            
\newcommand{\FilamentaryStructures}{filamentary structures}          
\newcommand{\classical}{conventional}
\newcommand{\DeltaAngName}{angle dispersion function}
\newcommand{\IntrinsicpName}{intrinsic polarization fraction}

\newcommand{\epseff}{\epsilon_{\rm eff}}

\begin{document}

\author{\small
Planck Collaboration:
P.~A.~R.~Ade\inst{78}
\and
N.~Aghanim\inst{54}
\and
D.~Alina\inst{83, 10}
\and
M.~I.~R.~Alves\inst{54}
\and
C.~Armitage-Caplan\inst{81}
\and
M.~Arnaud\inst{67}
\and
D.~Arzoumanian\inst{54}
\and
M.~Ashdown\inst{64, 6}
\and
F.~Atrio-Barandela\inst{18}
\and
J.~Aumont\inst{54}
% ...
}
\institute{\small
APC, AstroParticule et Cosmologie, Universit\'{e} Paris Diderot, CNRS/IN2P3, CEA/lrfu, Observatoire de Paris, Sorbonne Paris Cit\'{e}, 10, rue Alice Domon et L\'{e}onie Duquet, 75205 Paris Cedex 13, France\\
\and
Aalto University Mets\"{a}hovi Radio Observatory and Dept of Radio Science and Engineering, P.O. Box 13000, FI-00076 AALTO, Finland\\
\and
African Institute for Mathematical Sciences, 6-8 Melrose Road, Muizenberg, Cape Town, South Africa\\
\and
Agenzia Spaziale Italiana Science Data Center, Via del Politecnico snc, 00133, Roma, Italy\\
\and
Agenzia Spaziale Italiana, Viale Liegi 26, Roma, Italy\\
\and
Astrophysics Group, Cavendish Laboratory, University of Cambridge, J J Thomson Avenue, Cambridge CB3 0HE, U.K.\\
\and
Astrophysics \& Cosmology Research Unit, School of Mathematics, Statistics \& Computer Science, University of KwaZulu-Natal, Westville Campus, Private Bag X54001, Durban 4000, South Africa\\
\and
Atacama Large Millimeter/submillimeter Array, ALMA Santiago Central Offices, Alonso de Cordova 3107, Vitacura, Casilla 763 0355, Santiago, Chile\\
\and
CITA, University of Toronto, 60 St. George St., Toronto, ON M5S 3H8, Canada\\
\and
CNRS, IRAP, 9 Av. colonel Roche, BP 44346, F-31028 Toulouse cedex 4, France\\
\and
California Institute of Technology, Pasadena, California, U.S.A.\\
\and
Centro de Estudios de F\'{i}sica del Cosmos de Arag\'{o}n (CEFCA), Plaza San Juan, 1, planta 2, E-44001, Teruel, Spain\\
\and
Computational Cosmology Center, Lawrence Berkeley National Laboratory, Berkeley, California, U.S.A.\\
\and
Consejo Superior de Investigaciones Cient\'{\i}ficas (CSIC), Madrid, Spain\\
\and
DSM/Irfu/SPP, CEA-Saclay, F-91191 Gif-sur-Yvette Cedex, France\\
\and
DTU Space, National Space Institute, Technical University of Denmark, Elektrovej 327, DK-2800 Kgs. Lyngby, Denmark\\
\and
D\'{e}partement de Physique Th\'{e}orique, Universit\'{e} de Gen\`{e}ve, 24, Quai E. Ansermet,1211 Gen\`{e}ve 4, Switzerland\\
\and
Departamento de F\'{\i}sica Fundamental, Facultad de Ciencias, Universidad de Salamanca, 37008 Salamanca, Spain\\
\and
Departamento de F\'{\i}sica, Universidad de Oviedo, Avda. Calvo Sotelo s/n, Oviedo, Spain\\
\and
Department of Astronomy and Astrophysics, University of Toronto, 50 Saint George Street, Toronto, Ontario, Canada\\
\and
Department of Astrophysics/IMAPP, Radboud University Nijmegen, P.O. Box 9010, 6500 GL Nijmegen, The Netherlands\\
\and
Department of Physics \& Astronomy, University of British Columbia, 6224 Agricultural Road, Vancouver, British Columbia, Canada\\
\and
Department of Physics and Astronomy, Dana and David Dornsife College of Letter, Arts and Sciences, University of Southern California, Los Angeles, CA 90089, U.S.A.\\
\and
Department of Physics and Astronomy, University College London, London WC1E 6BT, U.K.\\
\and
Department of Physics, Florida State University, Keen Physics Building, 77 Chieftan Way, Tallahassee, Florida, U.S.A.\\
\and
Department of Physics, Gustaf H\"{a}llstr\"{o}min katu 2a, University of Helsinki, Helsinki, Finland\\
\and
Department of Physics, Princeton University, Princeton, New Jersey, U.S.A.\\
\and
Department of Physics, University of California, Santa Barbara, California, U.S.A.\\
\and
Department of Physics, University of Illinois at Urbana-Champaign, 1110 West Green Street, Urbana, Illinois, U.S.A.\\
\and
Dipartimento di Fisica e Astronomia G. Galilei, Universit\`{a} degli Studi di Padova, via Marzolo 8, 35131 Padova, Italy\\
\and
Dipartimento di Fisica e Scienze della Terra, Universit\`{a} di Ferrara, Via Saragat 1, 44122 Ferrara, Italy\\
\and
Dipartimento di Fisica, Universit\`{a} La Sapienza, P. le A. Moro 2, Roma, Italy\\
\and
Dipartimento di Fisica, Universit\`{a} degli Studi di Milano, Via Celoria, 16, Milano, Italy\\
\and
Dipartimento di Fisica, Universit\`{a} degli Studi di Trieste, via A. Valerio 2, Trieste, Italy\\
\and
Dipartimento di Fisica, Universit\`{a} di Roma Tor Vergata, Via della Ricerca Scientifica, 1, Roma, Italy\\
\and
Discovery Center, Niels Bohr Institute, Blegdamsvej 17, Copenhagen, Denmark\\
\and
Dpto. Astrof\'{i}sica, Universidad de La Laguna (ULL), E-38206 La Laguna, Tenerife, Spain\\
\and
European Southern Observatory, ESO Vitacura, Alonso de Cordova 3107, Vitacura, Casilla 19001, Santiago, Chile\\
\and
European Space Agency, ESAC, Planck Science Office, Camino bajo del Castillo, s/n, Urbanizaci\'{o}n Villafranca del Castillo, Villanueva de la Ca\~{n}ada, Madrid, Spain\\
\and
European Space Agency, ESTEC, Keplerlaan 1, 2201 AZ Noordwijk, The Netherlands\\
\and
Helsinki Institute of Physics, Gustaf H\"{a}llstr\"{o}min katu 2, University of Helsinki, Helsinki, Finland\\
\and
INAF - Osservatorio Astrofisico di Catania, Via S. Sofia 78, Catania, Italy\\
\and
INAF - Osservatorio Astronomico di Padova, Vicolo dell'Osservatorio 5, Padova, Italy\\
\and
INAF - Osservatorio Astronomico di Roma, via di Frascati 33, Monte Porzio Catone, Italy\\
\and
INAF - Osservatorio Astronomico di Trieste, Via G.B. Tiepolo 11, Trieste, Italy\\
\and
INAF/IASF Bologna, Via Gobetti 101, Bologna, Italy\\
\and
INAF/IASF Milano, Via E. Bassini 15, Milano, Italy\\
\and
INFN, Sezione di Bologna, Via Irnerio 46, I-40126, Bologna, Italy\\
\and
INFN, Sezione di Roma 1, Universit\`{a} di Roma Sapienza, Piazzale Aldo Moro 2, 00185, Roma, Italy\\
\and
INFN/National Institute for Nuclear Physics, Via Valerio 2, I-34127 Trieste, Italy\\
\and
IPAG: Institut de Plan\'{e}tologie et d'Astrophysique de Grenoble, Universit\'{e} Joseph Fourier, Grenoble 1 / CNRS-INSU, UMR 5274, Grenoble, F-38041, France\\
\and
Imperial College London, Astrophysics group, Blackett Laboratory, Prince Consort Road, London, SW7 2AZ, U.K.\\
\and
Infrared Processing and Analysis Center, California Institute of Technology, Pasadena, CA 91125, U.S.A.\\
\and
Institut d'Astrophysique Spatiale, CNRS (UMR8617) Universit\'{e} Paris-Sud 11, B\^{a}timent 121, Orsay, France\\
\and
Institut d'Astrophysique de Paris, CNRS (UMR7095), 98 bis Boulevard Arago, F-75014, Paris, France\\
\and
Institute for Space Sciences, Bucharest-Magurale, Romania\\
\and
Institute of Astronomy, University of Cambridge, Madingley Road, Cambridge CB3 0HA, U.K.\\
\and
Institute of Theoretical Astrophysics, University of Oslo, Blindern, Oslo, Norway\\
\and
Instituto de Astrof\'{\i}sica de Canarias, C/V\'{\i}a L\'{a}ctea s/n, La Laguna, Tenerife, Spain\\
\and
Instituto de Astronomia, Geof\'{\i}sica e Ci\^{e}ncias Atmosf\'{e}ricas, Universidade de S\~{a}o Paulo, S\~{a}o Paulo, SP 05508-090, Brazil\\
\and
Instituto de F\'{\i}sica de Cantabria (CSIC-Universidad de Cantabria), Avda. de los Castros s/n, Santander, Spain\\
\and
Jet Propulsion Laboratory, California Institute of Technology, 4800 Oak Grove Drive, Pasadena, California, U.S.A.\\
\and
Jodrell Bank Centre for Astrophysics, Alan Turing Building, School of Physics and Astronomy, The University of Manchester, Oxford Road, Manchester, M13 9PL, U.K.\\
\and
Kavli Institute for Cosmology Cambridge, Madingley Road, Cambridge, CB3 0HA, U.K.\\
\and
LAL, Universit\'{e} Paris-Sud, CNRS/IN2P3, Orsay, France\\
\and
LERMA, CNRS, Observatoire de Paris, 61 Avenue de l'Observatoire, Paris, France\\
\and
Laboratoire AIM, IRFU/Service d'Astrophysique - CEA/DSM - CNRS - Universit\'{e} Paris Diderot, B\^{a}t. 709, CEA-Saclay, F-91191 Gif-sur-Yvette Cedex, France\\
\and
Laboratoire Traitement et Communication de l'Information, CNRS (UMR 5141) and T\'{e}l\'{e}com ParisTech, 46 rue Barrault F-75634 Paris Cedex 13, France\\
\and
Laboratoire de Physique Subatomique et de Cosmologie, Universit\'{e} Joseph Fourier Grenoble I, CNRS/IN2P3, Institut National Polytechnique de Grenoble, 53 rue des Martyrs, 38026 Grenoble cedex, France\\
\and
Laboratoire de Physique Th\'{e}orique, Universit\'{e} Paris-Sud 11 \& CNRS, B\^{a}timent 210, 91405 Orsay, France\\
\and
Lawrence Berkeley National Laboratory, Berkeley, California, U.S.A.\\
\and
Max-Planck-Institut f\"{u}r Astrophysik, Karl-Schwarzschild-Str. 1, 85741 Garching, Germany\\
\and
National University of Ireland, Department of Experimental Physics, Maynooth, Co. Kildare, Ireland\\
\and
Niels Bohr Institute, Blegdamsvej 17, Copenhagen, Denmark\\
\and
Observational Cosmology, Mail Stop 367-17, California Institute of Technology, Pasadena, CA, 91125, U.S.A.\\
\and
Optical Science Laboratory, University College London, Gower Street, London, U.K.\\
\and
SISSA, Astrophysics Sector, via Bonomea 265, 34136, Trieste, Italy\\
\and
School of Physics and Astronomy, Cardiff University, Queens Buildings, The Parade, Cardiff, CF24 3AA, U.K.\\
\and
Space Sciences Laboratory, University of California, Berkeley, California, U.S.A.\\
\and
Special Astrophysical Observatory, Russian Academy of Sciences, Nizhnij Arkhyz, Zelenchukskiy region, Karachai-Cherkessian Republic, 369167, Russia\\
\and
Sub-Department of Astrophysics, University of Oxford, Keble Road, Oxford OX1 3RH, U.K.\\
\and
UPMC Univ Paris 06, UMR7095, 98 bis Boulevard Arago, F-75014, Paris, France\\
\and
Universit\'{e} de Toulouse, UPS-OMP, IRAP, F-31028 Toulouse cedex 4, France\\
% ...
}

\title{
{\Planck} intermediate results. XIX. An overview of the polarized thermal emission from Galactic dust
}

\titlerunning{The {\Planck} dust polarization sky}
\authorrunning{{\Planck} collaboration}


\abstract{
This paper presents the large-scale polarized sky as seen by {\Planck}
HFI at 353\GHz, which is the most sensitive {\Planck} channel for dust
polarization. We construct and analyse large-scale maps of dust
polarization fraction and polarization direction, while taking account
of noise bias and possible systematic effects.  We find that the
maximum observed dust polarization fraction is high ($\pmax>18\,\%$),
in particular in some of the intermediate dust column density
($\Av<1$\,mag) regions.  There is a systematic decrease in the dust
polarization fraction with increasing dust column density, and we
interpret the features of this correlation in light of both radiative
grain alignment predictions and fluctuations in the magnetic field
orientation.  We also characterize the spatial structure of the
polarization angle using the {\DeltaAngName} and find that, in nearby
fields at intermediate latitudes, the polarization angle is ordered
over extended areas that are separated by {\FilamentaryStructures},
which appear as interfaces where the magnetic field sky projection
rotates abruptly without apparent variations in the dust column
density.  The polarization fraction is found to be anti-correlated
with the dispersion of the polarization angle, implying that the
variations are likely due to fluctuations in the 3D magnetic field
orientation along the line of sight sampling the diffuse interstellar
medium.
We also compare the dust emission with the polarized synchrotron
emission measured with the {\Planck} LFI, with low-frequency radio data,
and with Faraday rotation measurements of extragalactic sources.
The two polarized components are
globally similar in structure along the plane and notably in the Fan and North Polar Spur
regions.  A detailed comparison of these three
tracers shows, however, that dust and cosmic rays generally sample different
parts of the line of sight and confirms that much of the variation observed in
the {\Planck} data is due to the 3D structure of the magnetic field. 
}


\maketitle


\section{Introduction}

\begin{figure*}[ht]
\begin{center}
\includegraphics[width=0.95\textwidth]{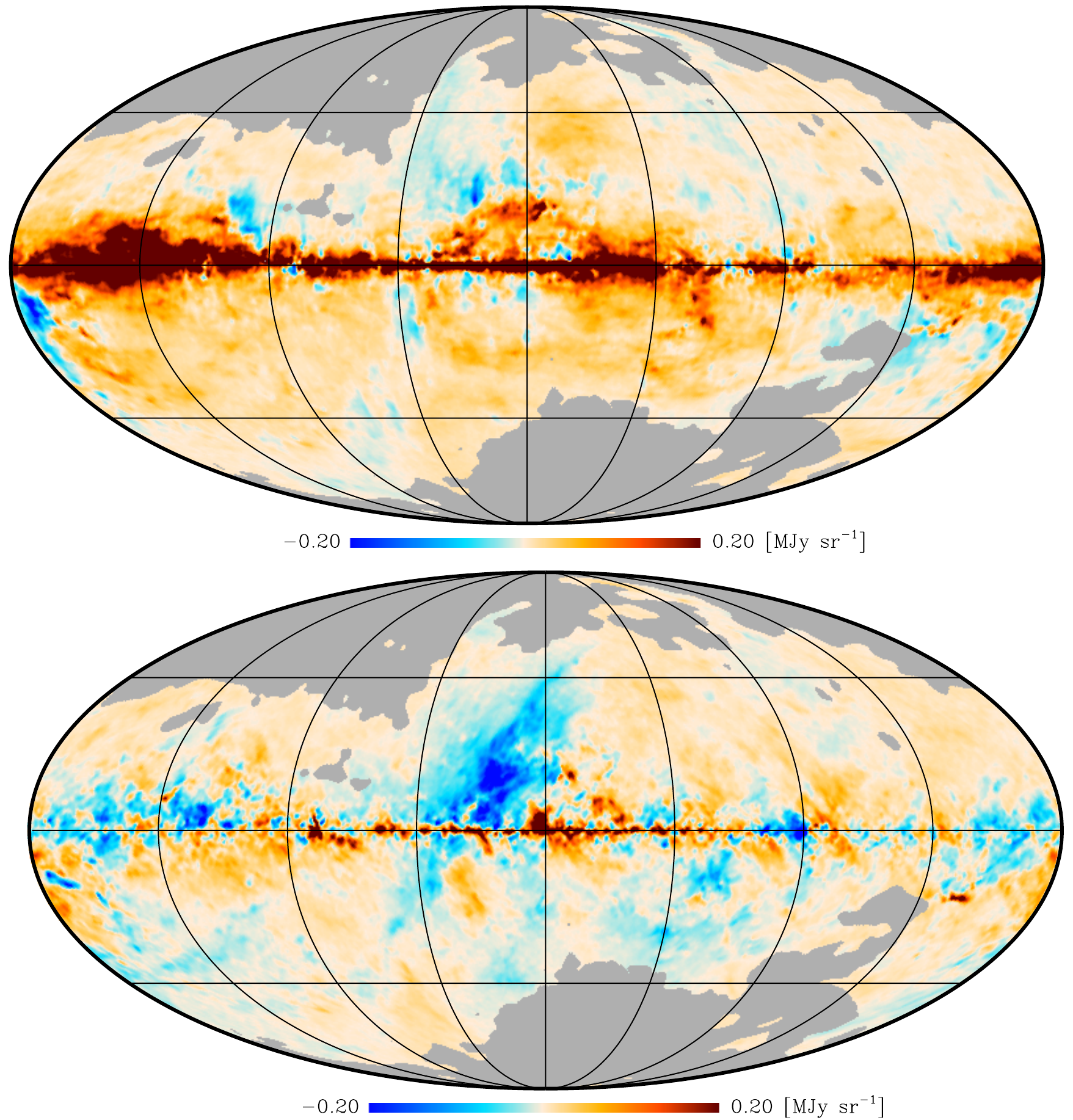}
\caption{\label{fig:rawpolarmaps}
{\Planck} 353{\GHz} polarization maps at $1^\circ$ resolution.
\emph{Upper}: $\StokesQ$ Stokes parameter map.  \emph{Lower}:
$\StokesU$ Stokes parameter map.  The maps are shown with the same
colour scale. High values are saturated to enhance mid-latitude
structures.  The values shown have been bias corrected as described in
Sect. \,\ref{sec:polarparam}. These maps, as well as those in following figures, are shown in
Galactic coordinates with the galactic center in the middle and
longitude increasing to the left.  The data is masked as described in
Sect.\,\ref{sec:othercorrections}.
}
\end{center}
\end{figure*}

\begin{figure*}[ht]
\begin{center}
\includegraphics[width=0.95\textwidth]{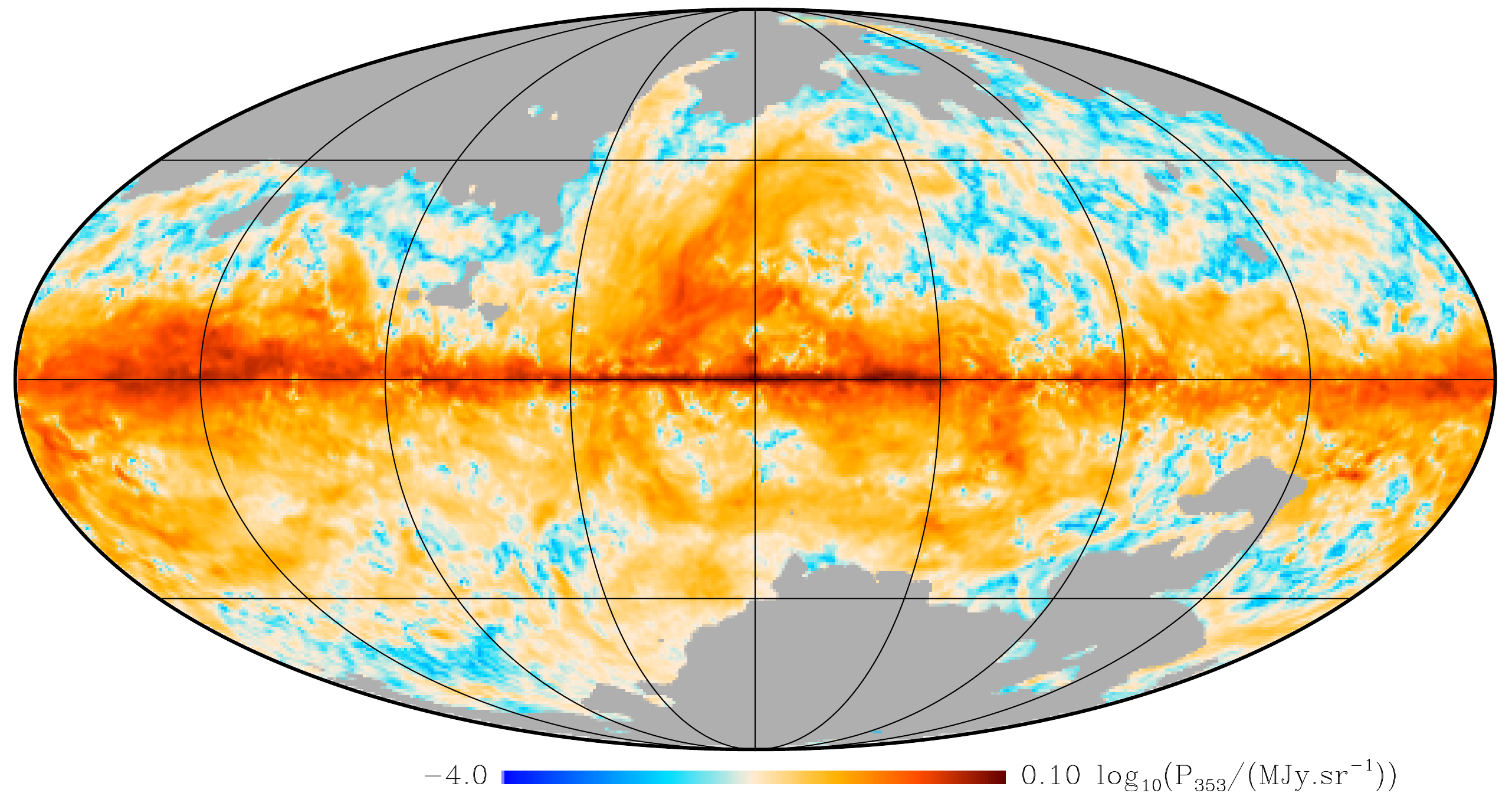}
\caption{\label{fig:rawPmap}
{\Planck} 353{\GHz} polarized intensity ($\polint$) map at $1^\circ$
resolution  in log$_{10}$ scale. The values shown have been bias corrected as described in
Sect. \,\ref{sec:polarparam}. The same mask as in
Fig.\,\ref{fig:rawpolarmaps} is applied.
The full sky map of the unpolarized intensity $\StokesI$ entering the calculation of $\polint$ is shown
in Fig. \,\ref{fig:polar_psi_and_sigpsi}.
}
\end{center}
\end{figure*}

Our Galaxy is pervaded by an interstellar magnetic field of a few microgauss,
which fills the entire disk and extends well into the halo.
This magnetic field manifests itself in a variety of ways, including
Zeeman splitting of atomic and molecular spectral lines,
Faraday rotation of polarized radio signals,
synchrotron emission from relativistic electrons,
and polarization of starlight and thermal dust emission.
With a pressure at least comparable to those of the thermal gas and
of cosmic rays, the Galactic magnetic field (GMF) plays a crucial role
in the ecosystem of our Galaxy. It governs the structure and the dynamics
of the interstellar medium (ISM), regulates the process of star formation,
accelerates cosmic rays, channels their trajectories and helps to confine them
to the Galaxy.
In addition to a large-scale regular, or coherent, component 
and an isotropic random component produced by interstellar turbulence 
\citep[with scales up to 100~pc; e.g.,][]{GaenslerJohnston1995a,Haverkorn2008}, 
the GMF also possesses an ordered random \citep[e.g.,][]{Beck2009,Jaffe2010},
or striated random \citep{JanssonFarrar2012a}, component, 
whose orientation remains nearly constant over large scales,  
but whose strength and sign vary on small scales.
Such fields are probably produced through compression or shearing
of isotropic random fields by the Galactic differential rotation,
or at large-scale spiral arm shocks, or else by rising hot 
plasma bubbles.

Our knowledge and understanding of the GMF has improved considerably
over the past few years, as a result of both progress 
in the quality (sensitivity and resolution) of radio observations
and extensive modelling efforts \citep[e.g.,][]{Sun08,SunReich2010,
Ruiz-Granados2010a, Jaffe2010, Jaffe2011,
Pshirkov2011,Fauvet2012b,JanssonFarrar2012a,JanssonFarrar2012b,Fauvet2013}.
However, the existing radio observations have inherent limitations, 
as both Faraday rotation measures (RMs) and synchrotron (total and polarized)
intensities are quantities integrated over the line of sight (LOS), which depend on the poorly constrained
density distributions of thermal and relativistic electrons, respectively.
A promising avenue to obtain a more complete and more robust picture 
of the GMF structure is to complement the radio data with {\Planck}
\footnote{\Planck~(\url{http://www.esa.int/Planck}) is a
  project of the European Space Agency (ESA) with instruments
  provided by two scientific consortia funded by ESA member states (in
  particular the lead countries France and Italy), with contributions
  from NASA (USA) and telescope reflectors provided by a collaboration
  between ESA and a scientific consortium led and funded by Denmark.}
measurements of the polarized thermal emission from interstellar dust,
which is independent of the electron densities.

A glance at the {\Planck} all-sky intensity
maps \citep{planck2013-p01} reveals that, in addition to the mottled
structure of the cosmic microwave background (CMB) at high Galactic
latitudes, the dominant pattern is that of the emission from our Galaxy. 
At the lowest frequencies, from the 30{\GHz} to 70{\GHz} bands of
the {\Planck} Low Frequency Instrument \citep[LFI,][]{bersanelli2010}, 
synchrotron emission dominates; at the highest frequencies, from the 100{\GHz} 
to 857{\GHz} bands of the High Frequency Instrument \citep[HFI,][]{lamarre2010}, 
thermal emission from interstellar dust is the dominant mechanism. These
foregrounds have to be understood and taken into account for detailed
CMB studies, but they also provide a unique opportunity to study
the Galaxy's ISM.

In particular, the thermal dust emission is linearly polarized 
\citep[e.g.,][]{Benoit2004,Vaillancourt2007}.
This polarized emission overpowers any other polarized
signal at the higher {\Planck} frequencies
\citep[e.g.,][]{Tucci2005,Dunkley2009,Fraisse2009}.
In addition to hindering the detection of the
sought-after, odd-parity, $B$-mode polarization of the CMB, the
polarized dust emission provides, in combination with the emission
spectrum itself, a powerful constraint on the physical properties 
of the dust and on the structure of the magnetic field in the Galaxy.

The linear polarization of the thermal dust emission arises from a
combination of two main factors. Firstly, a fraction of the dust grain
population is non-spherical, and this gives rise to different
emissivities for radiations with the electric vector parallel or
orthogonal to a grain's long axis. Secondly, the rotating grains 
are aligned by the interstellar magnetic field,
probably with differing efficiencies depending on grain size and
composition \citep{DraineFraisse2009}.
While the details of this process remain unclear 
\citep{Lazarian2003,Lazarian2007}, there is a consensus
that the angular momentum of a grain spun up by photon-grain interactions 
\citep{Dolginov1976,DraineWeingartner96,DraineWeingartner1997,LazarianHoang2007,HoangLazarian2008}
becomes aligned with the grain's short axis, and then 
with the magnetic field via precession \citep[e.g.,][]{Martin71}.
The end result is that, if we look across magnetic field lines,
the rotating grain will have its long axis orthogonal
to the field lines, and accordingly dust emission will be linearly polarized 
with its electric vector normal to the sky-projected magnetic field.

A related phenomenon occurs at near-UV/optical/NIR wavelengths, 
where the light from background sources becomes linearly polarized
as a result of dichroic extinction by the aligned dust grains
\citep{davis&g_51}.
Since extinction is higher for light vibrating parallel to 
the grain's long axis, i.e., perpendicular to the field lines,
the incoming light will be linearly polarized with its electric vector 
parallel to the sky-projected magnetic field.
In fact, historically, the optical polarization caused by dust extinction
led to the prediction that thermal dust emission would be polarized in
the millimetre and submillimetre domains \citep{stein_66}.

Thus, polarized thermal dust emission carries important information
on the interstellar magnetic field structure, on the grain alignment
mechanisms, and on the grain geometrical and physical properties. For example,
polarization observations between $300\mic$ and 3 mm, essentially the
domain of the {\Planck} HFI instrument,
can potentially discriminate between the polarizing grain materials, 
e.g., silicate and graphite dust versus silicate-only grains
\citep{Martin2007,DraineFraisse2009,planck2014-XXII}.

Since this far-IR emission is basically proportional to the dust mass along
the LOS, sensitivity limits explain why detailed dust polarized emission 
was observed mostly in fairly dense, complex regions of the ISM
\citep{Dotson2000,CurranChrysostomou2007,Matthews2009,Dotson2010}, 
in general close to the Galactic plane.  Measurements of the more
diffuse medium were obtained at relatively low ($\ge 2 \deg$)
angular resolution. At these large scales, the {\archeops} balloon
experiment \citep{Benoit2004,Ponthieu2005} detected the thermal dust
emission polarization at 353{\GHz}. The highest frequency channel of 
{\wmap} \citep{Page2007,Bennett2013}, 94{\GHz}, picked up the
long-wavelength Rayleigh-Jeans tail of the diffuse dust emission and its
polarization (in addition to synchrotron emission).

The {\Planck} satellite's HFI instrument leads to the first all-sky survey
of the polarized submillimetre and millimetre sky, where thermal dust emission
dominates. At 353{\GHz}, the {\Planck} data have an angular resolution
of 5\arcmin. The polarization sensitivity is expected to be such that,
at a resolution of 15\arcmin, ISM structures with $\Av=1$\,mag are
detected with a relative uncertainty on the polarization fraction of
about $40\,\%$ and an uncertainty on the polarization angle of about
$30\degr$ \citep{Pelkonen2009}. These figures improve significantly at
higher $\Av$ and/or lower resolution.
The polarized {\Planck} data bring the first all-sky
map of the polarization from a tracer of the interstellar matter. 
As such, they provide unprecedented information on the magnetic field geometry and
the dust polarization properties relevant to the disk of the Milky Way
(MW) and star forming regions, for which they provide statistical information
that is missing in stellar polarization extinction data.  It should be
emphasized, however, that the dust polarized emission provides
information mostly on the orientation of the sky-projected magnetic
field and only very indirect indication about the angle of that field
with respect to the plane of the sky, and it is expected to be almost
insensitive to the field strength.

This paper presents the {\Planck} polarization data and their
large-scale statistical properties. A companion paper
\citep{planck2014-XX}
analyses the variations of the polarization
fraction and angle described here, in comparison with the predictions
from MHD simulations.  Two other papers in this series provide a
detailed analysis of the wavelength dependence of the dust
polarization, as seen by the HFI instrument \citep{planck2014-XXII} and
a comparison between the dust polarization at visible and
submillimetre wavelengths \citep{planck2014-XXI}.

In Sect.\,\ref{sec:planckdata} we describe the data, including
discussion of systematic effects and the effects of the CMB intensity and
polarization. Maps are presented in Sect.\,\ref{sec:descplanck}, as
well as the statistics of the data. Sect.\,\ref{sec:discussion}
discusses the implications of the 353{\GHz} polarimetry for our
understanding of the GMF structure, and the conclusions are drawn in
Sect.\,\ref{sec:conclusions}.  Two appendices discuss the smoothing of
the noise covariance matrices, which is needed when the original data
are averaged, as well as the de-biasing methods for obtaining
polarization estimates.

\begin{figure*}[!h!t]
\begin{center}
\includegraphics[width=0.95\textwidth]{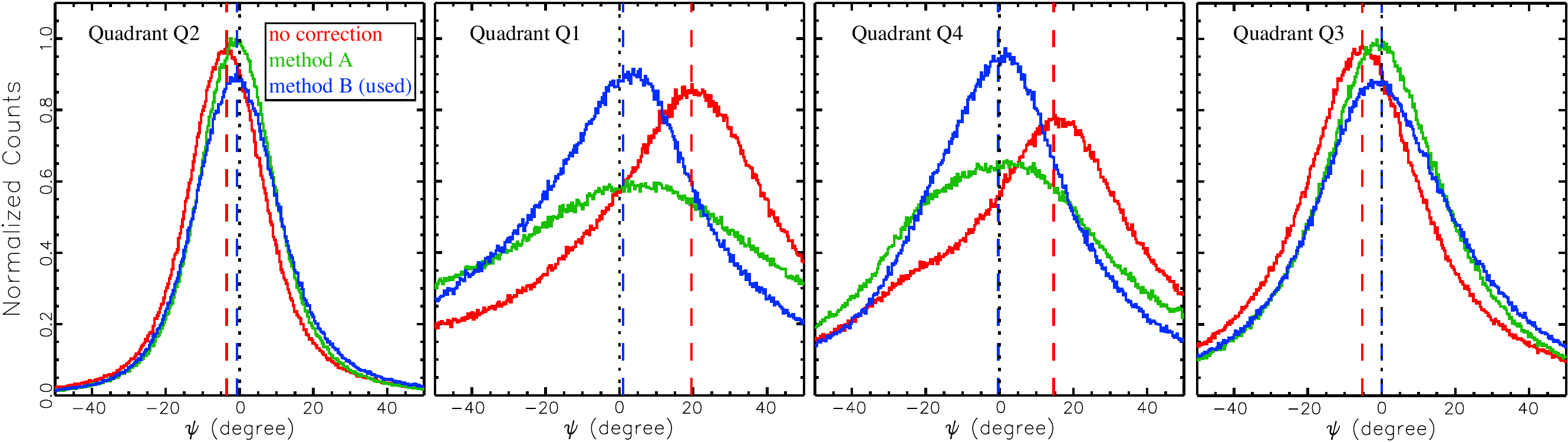}
\caption{\label{fig:phistograms_vs_bpm}
Histograms of the observed polarized angle at the full data resolution
towards the Galactic plane ($|\glat|<5\degr$) for the four Galactic
quadrants. The various curves show data uncorrected for bandpass
mismatch (red), and corrected using sky coupling coefficients derived
either from ground (method A: green) or sky measurements (method
B: dark blue).  The vertical dashed lines show the peak value
obtained from fitting the histograms with a Gaussian.
}
\end{center}
\end{figure*}

\section{Data} 
\label{sec:planckdata}

The {\Planck} mission results are presented in
\cite{planck2013-p01} and the in-flight performance of the two focal
plane instruments, the \hfi (High Frequency Instrument) and the \lfi (Low
Frequency Instrument), are given in \cite{planck2011-1.5} and
\cite{planck2011-1.4}, respectively.
The data processing and calibration of the {\hfi} data used here
are described in \cite{planck2013-p03}, \cite{planck2013-p03c},
\cite{planck2013-p03f}, \cite{planck2013-p03d} and
\cite{planck2013-p03e}.  The data processing and calibration of the
{\lfi} data are described in \cite{planck2013-p02},
\cite{planck2013-p02a}, \cite{planck2013-p02d}, and
\cite{planck2013-p02b}. 

The Planck polarization and intensity data that we use in this
analysis have been generated in exactly the same manner as the
data publicly released in March 2013 and described in
\cite{planck2013-p01} and associated papers. Note however that the
publicly available data includes only temperature maps based on the
first two surveys. \cite{planck2013-p11} shows the very good
consistency of cosmological models derived from intensity only with
polarization data at small scale (high CMB multipoles).
However, as detailed in \citet{planck2013-p03} (see their Fig. 27),
the 2013 polarization data are known to be affected by systematic
effects at low multipoles which were not yet fully corrected, and
thus, not used for cosmology.  We have been careful to check that the
Galactic science results in this paper are robust with respect to
these systematics. The error-bars we quote include uncertainties
associated with residual systematics as estimated by repeating the
analysis on different subsets of the data. We have also checked our
data analysis on the latest version of the maps available to the
collaboration, to check that the results we find are consistent within
the error-bars quoted in this paper.

The maps used include data from five
independent consecutive sky surveys (called Survey1-Survey5) for
{\hfi}, taken six months apart.  Due to the scanning strategy of the
{\Planck} mission, surveys taken one year apart (i.e. odd surveys 1
and 3 and even surveys 2 and 4) share the same observing pattern,
which is different for even and odd surveys.  Survey5 had a different
scan pattern from the other odd-numbered surveys, owing to a change in
the precession phase.  The products also include data binned into the
first and second halves of the {\Planck} stable pointing periods, or
``half-rings'' (called HR1 and HR2).  Both single-survey and half-ring
data are used for consistency checks and to assess the level of
systematic effects.  Here, we only analyse the polarization data at
353{\GHz}, which is the highest frequency {\Planck} channel with
polarization capabilities and the one with the best {\SNR} for dust
polarization.  We use the 30{\GHz} {\lfi} data in our comparison of
the dust emission at 353{\GHz} with the microwave and radio
synchrotron emission presented in Sect.\,\ref{sec:SYNCHROTRONcomp}.

In the {\Planck} map-making process \citep{planck2013-p03f},
measurements from various detectors at the same frequency are combined
to obtain the Stokes parameters ($\StokesI$, $\StokesQ$, and
$\StokesU$) at each position on the sky.  The reconstructed
polarization is a linear combination of the weighted differences
between the signal from pairs of polarization sensitive bolometers (PSBs) with
different orientations on the sky.  The resulting maps of the {\Planck} Stokes
parameters $\StokesQ$ and $\StokesU$ used in this paper are shown in
Fig.\,\ref{fig:rawpolarmaps}. The corresponding map of the observed
polarization intensity $\polint=(\StokesQ^2+\StokesU^2)^{1/2}$ is shown in
Fig.\,\ref{fig:rawPmap}. The intensity map used in this work is
shown in Fig.\,\ref{fig:polar_psi_and_sigpsi}.

\subsection{Conventions and notations}
\label{sec:conventions}

The relations between the observed Stokes parameters
($\StokesI$, $\StokesQ$, and $\StokesU$) and the polarization fraction ($\polfrac$)
and polarization angle ($\polang$) are given by
\begin{equation}
\label{equ:pem}
\polfrac=\frac{\sqrt{\StokesQ^2+\StokesU^2}}{\StokesI},
\end{equation}
and
\begin{equation}
\label{equ:thetam}
\polang=0.5 \times \arctan(\StokesU,\StokesQ),
\end{equation}
where  the two arguments function $\arctan(Y,X)$ is used to compute
$atan(Y/X)$ avoiding the $\pi$ ambiguity, such that
\begin{align}
\label{equ:quvsppsi}
\StokesQ&=\polfrac\times\StokesI\times \cos(2 \polang), \nonumber \\
\StokesU&=\polfrac\times\StokesI\times \sin(2 \polang).
\end{align}

For the Stokes parameters provided in the {\Planck} data, the angle
convention above is with respect to Galactic coordinates with
$-90\degr<\polang<+90\degr$ and $\polang=0^\circ$ towards the Galactic
north and positive towards the west (clockwise).  Note that this
convention is the one used in the {\Healpix}\footnote{\url{http://healpix.jpl.nasa.gov}} software
\citep{Gorski2005},
but is different from the IAU convention
\citep{HamakerBregman1996a}, which is $\polang=0^\circ$ towards north
and positive towards the east (counterclockwise).
The conversion between {\Planck}
Stokes parameters and the IAU convention is given by:
\begin{equation}
\polangiau=0.5 \times \arctan(-\StokesU,\StokesQ).
\end{equation}
In this paper, the tabulated angle values are given in the IAU
convention.

\begin{figure*}[!h!t]
\begin{center}
\includegraphics[width=0.95\textwidth]{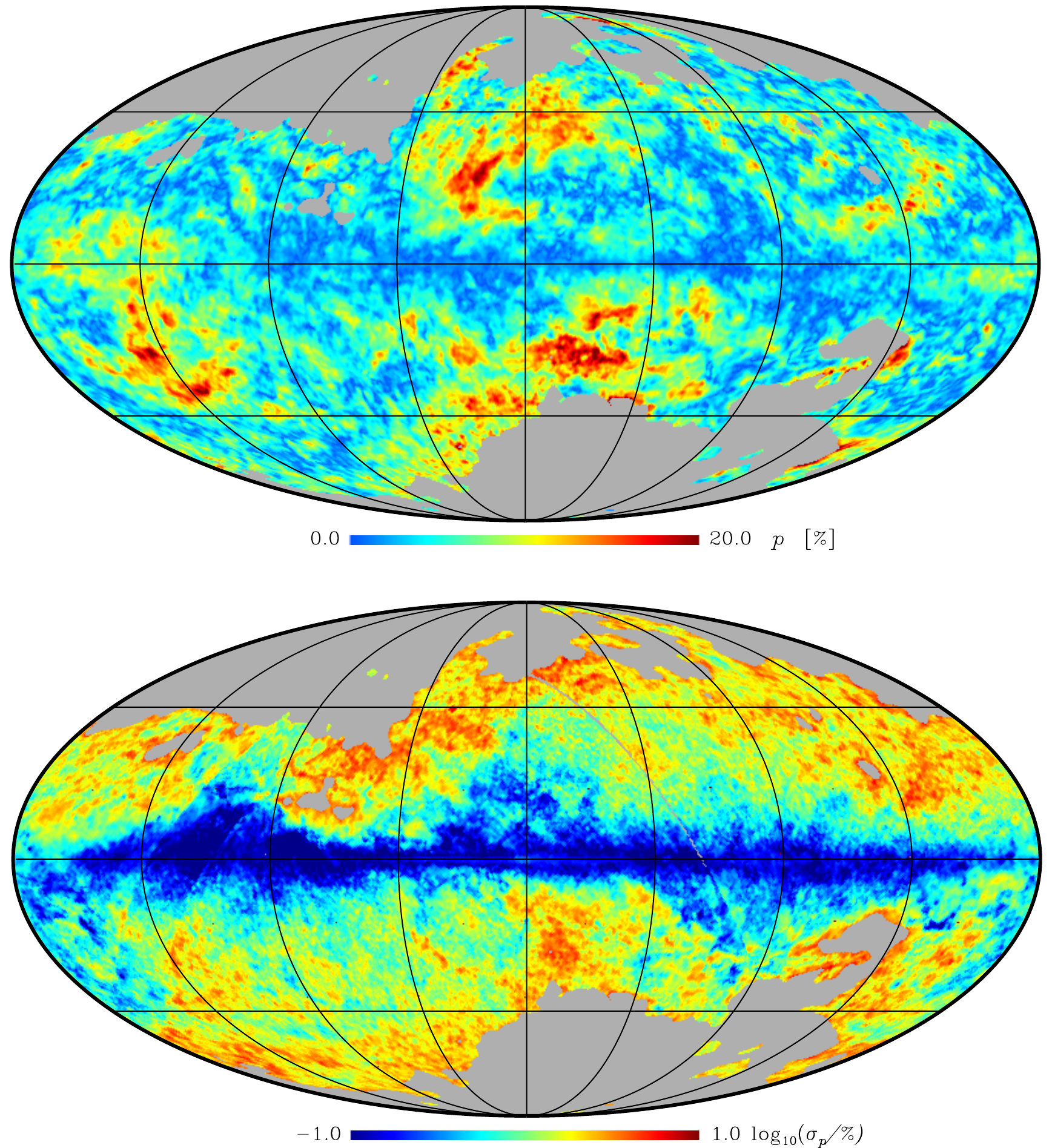}
\caption{
\emph{Upper}: Map of the 353{\GHz} polarization fraction
$\polfrac$ at $1^\circ$ resolution. The colour scale is linear and
ranges from 0\,\% to 20\,\%.  \emph{Lower}: Map of the 353{\GHz}
polarization fraction uncertainty, $\sigpolfrac$, at $1^\circ$
resolution in log$_{10}$ scale. The colour scale is from
$\sigpolfrac=0.1\,\%$ to $\sigpolfrac=10\,\%$.
The data are not shown in
the grey areas where the dust emission is not dominant or where
residuals were identified comparing individual surveys (see
Sect.\,\ref{sec:othercorrections}).
The polarization fraction is
obtained using the Bayesian method with a mean posterior estimator
(see Sect.\,\ref{sec:polarparam}). The uncertainty map includes
statistical and systematic contributions. The same mask as in
Fig.\,\ref{fig:rawpolarmaps} is applied.
\label{fig:polar_p_and_sigp}}
\end{center}
\end{figure*}

\begin{figure*}[!h!t]
\begin{center}
\includegraphics[width=0.95\textwidth]{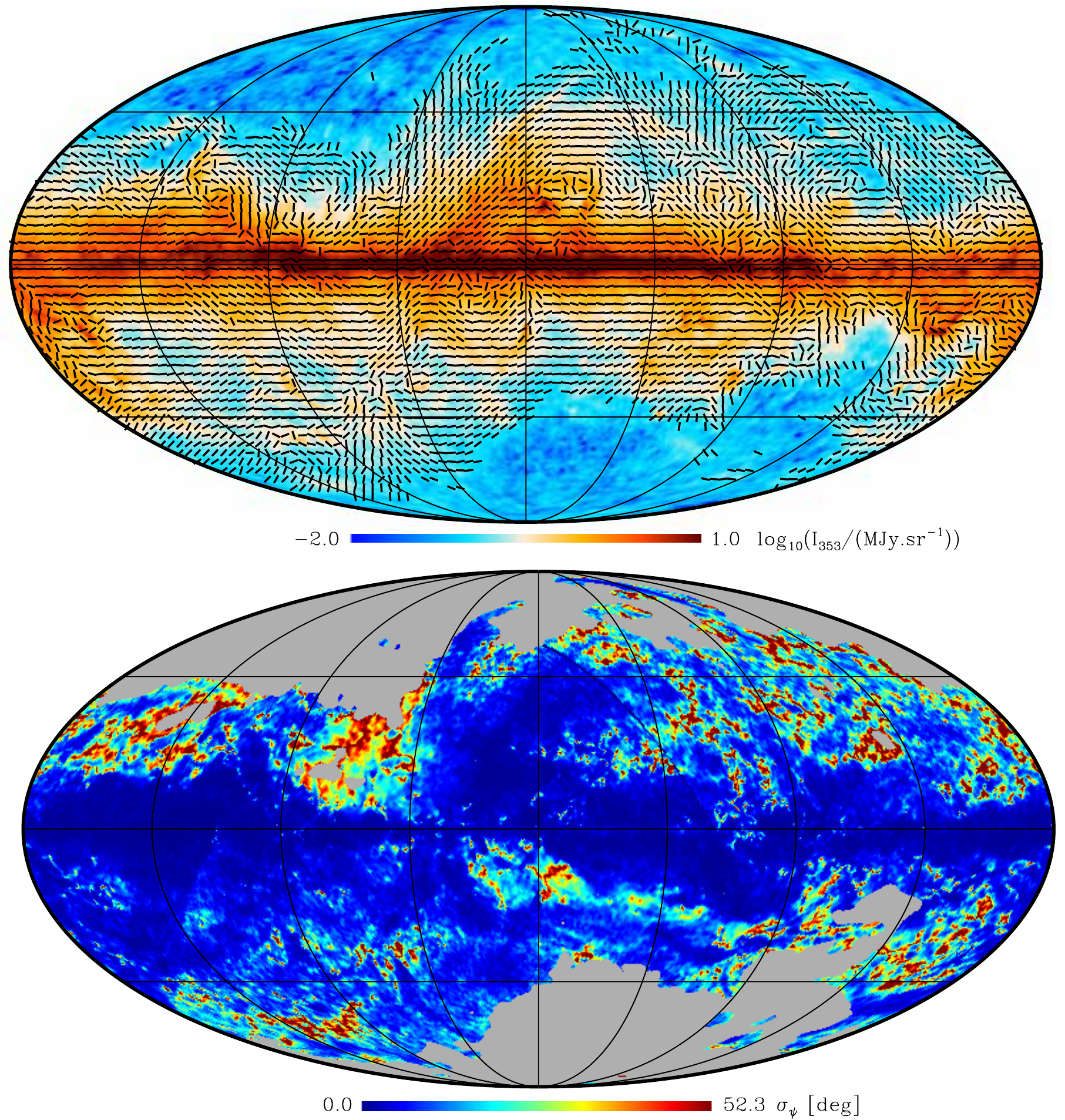}
\caption{
\emph{Upper}: Map of the {\appBfield} ($\Bperp$) orientation.  The
normalized lines were obtained by rotating the measured 353{\GHz}
polarization angles by $90^\circ$. The length of the polarization
vectors is fixed and does not reflect polarization fraction.  The
colour map shows the 353{\GHz} emission in log$_{10}$ scale and ranges
from $10^{-2}$ to $10\MJysr$.  \emph{Lower}: Map of the 353{\GHz}
polarization angle uncertainty ($\sigpolang$) at $1^\circ$
resolution. The scale is linear from
$\sigpolang=0\degr$ to $\sigpolang=52.3\degr$.  The polarization angle
is obtained using the Bayesian method with a mean posterior estimator
(see Sect.\,\ref{sec:polarparam}). The uncertainty map includes
statistical and systematic contributions. The same mask as in
Fig.\,\ref{fig:rawpolarmaps} is applied.
\label{fig:polar_psi_and_sigpsi}}
\end{center}
\end{figure*}

\subsection{Bandpass mismatch leakage correction}
\label{sec:bpm}

Owing to the way the polarization maps are constructed, any instrumental
difference between detectors of the same channel may produce a fake
polarization signal, even for unpolarized sky signal inputs.
This is the case for the bandpass mismatch ({\BPM}) between detectors
that affects  {\Planck} polarization maps.
In practice, the effect corresponds to a leakage term from intensity $\StokesI$
into polarization $\StokesQ$ and $\StokesU$. The {\BPM} polarization
leakage effect is therefore strongest in regions of high intensity,
i.e., along the Galactic plane, and affects both $\polfrac$ and
$\polang$. Note that, since the 353{\GHz} intensity data used here are
calibrated on the CMB signal, no {\BPM} leakage is produced by the CMB
anisotropies. Other astrophysical emission sources, however, produce
{\BPM} polarization leakage.

Knowing the actual {\Planck} sky scanning strategy and the
orientations of the polarization sensitive bolometers in the focal
plane, the {\BPM} polarization leakage corrections can be estimated
from the relative responses of each detector to a given sky
astrophysical emission.
The {\Planck} collaboration is exploring different methods to compute the
relative responses of detectors, as well as to produce total intensity
maps for each sky emission source. 
Two methods have been used to determine the relative responses
\citep{planck2013-p03d}.  The first one (method A) involves computing
the {\BPM} leakage between bolometers using the ground-measured bandpasses
\citep{planck2013-p03d}.  The second one (method B) deduces the
relative detector response on regions of the sky where we can obtain
$\StokesI$, $\StokesQ$, and $\StokesU$ maps for each detector
individually.  Note that this can only be performed in limited
regions of the sky, outside the Galactic plane, which have been
scanned in a large number of configurations, allowing for the full
reconstruction of $\StokesI$, $\StokesQ$, and $\StokesU$ per detector.
A comparison between the two methods is presented in
\cite{planck2013-p03d}.

When folding the above coefficients into the {\Planck} scanning
strategy, we have chosen to produce template maps $T^{X}_{b(\nu)}$ of
the {\BPM} leakage contribution for each frequency ($\nu$) channel, for each
bolometer ($b(\nu)$) and for each Stokes parameter ($X$ being
$\StokesQ$ or $\StokesU$).
The {\BPM} polarization leakage correction is
\begin{equation}
L^{X}_{\nu} = \sum_{b(\nu)}   R_{b(\nu)} \ I_{\nu} \   T^{X}_{b(\nu)} ,
\label{equ:bpm}
\end{equation}
where $R_{b(\nu)}$ represents the detector relative responses and
$I_{\nu}$ is the sky intensity.  For the purpose of the study
presented here, we only take into account {\BPM} leakage from dust thermal
emission, since this is the dominant term at 353{\GHz}.  The template maps in
Eq.\,\ref{equ:bpm} were computed using the {\Planck} thermal dust
model described in \cite{planck2013-p06b}. We used the standard
{\Planck} map-making procedure presented in \citet{planck2013-p03f}.
Note that the {\Planck} 353{\GHz} channel also includes emission from
the CO ($J=3 \to 2$) line \citep[see][]{planck2013-p03}, which should also
in principle be included in the {\BPM} leakage correction. This, however, is
relatively weak with respect to dust thermal emission and the
corresponding {\BPM} effect is expected to be small compared to that
from dust. Since we do not concentrate on regions with strong molecular
emission in this paper, no correction was applied for the CO emission
{\BPM} leakage.

Figure\,\ref{fig:phistograms_vs_bpm} shows the effect of the
correction for {\BPM} on the
observed distribution of polarization angles toward the plane of the
Milky Way ($|\glat|<5\degr$) in the four Galactic quadrants (Q1, Q2, Q3 and Q4,
defined by $0\degr<\glon<90\degr$, $90\degr<\glon<180\degr$,
$180\degr<\glon<270\degr$, and $270\degr<\glon<360\degr$, 
respectively). When no {\BPM} leakage correction is applied, angles are
observed to be distributed around $+20\degr$ and $-5\degr$ for the
inner (Q1 and Q4) and outer (Q2 and Q3) MW regions, respectively.
The difference in sign is due to the difference in average detector
orientation during Galaxy crossings, resulting from the relative
orientation of the scanning strategy and the Galactic plane.
Using the two methods discussed above for the determination of the
coupling coefficients leads to similar {\BPM} leakage estimates.  Note
also that, since the magnetic field is expected to be statistically
aligned with the Galactic plane \citep[see, e.g.,][]{Ferriere2011}, we expect
the polarization direction towards the plane to be on average around
$\polang=0\degr$. The fact that both correction methods bring the peak
of the histograms toward this value confirms the validity of the
{\BPM} correction method used here.  In the following, we adopted the
coefficients from method B.  We note, however, that although the
situation is clearly improved by the {\BPM} leakage correction, the average
observed angle distributions still peak a few degrees away from
$\polang=0\degr$, with the same sign pattern as for the uncorrected
data. This could in principle be due to incomplete correction. However,
preliminary tests have shown that the remaining correction could be
due to non-linearity in the analogue-to-digital conversion (ADC) of the
signal, which produces an additional correction with the same sign as
observed here and roughly the right amplitude.

We do not attempt here to fully assess the quality of the different
corrections, but simply use them to estimate where on the sky the
uncertainties in the corrections are small enough to be unimportant
for this study.  A plot of the {\BPM}-leakage-corrected polarization
angle versus the uncorrected polarization angle shows the magnitude of
the correction, while the correlation coefficient gives a quantitative
measure.  For the different corrections considered above, the
correlation coefficient is over 0.95 for most regions of the sky at
$|\glat|>5\degr$.  Above $|\glat| = 10\degr$, the correlation
coefficients are above 0.98, implying that the correction becomes very
small.  This is a natural result of the fact that the intensity that
is leaking into polarization is brightest towards the Galactic plane.
As measured from the difference between method A and B, the
corresponding uncertainties on the polarization angle $\polang$ and
fraction $\polfrac$ are $|\Delta \polang| < 10\degr$ and $\Delta
\polfrac<1\,\%$, respectively, towards the inner Galactic plane.
These uncertainties become less than the random errors away from the
plane.  However, {\BPM} leakage corrections are probably not the
dominant uncertainty at high galactic latitudes and very low signal
levels, where other systematic effects remaining in the data become
more important (see Sect.\,\ref {sec:othercorrections}).  For this
reason, we do not discuss specifically the polarization properties in
the lowest brightness sky area in this paper and deffer this
discussion to future papers.

The above discussion applies to the {\hfi} data, but we will also
compare the thermal dust emission at 353{\GHz} to the 30{\GHz}
emission from {\lfi}, which has a similar bandpass leakage issue.  The
{\lfi} {\BPM} correction is discussed in \cite{planck2013-p02}, where
the principle difference is the presence of multiple astrophysical
foregrounds, with different spatial and spectral distributions.  The
component separation products are therefore used in the {\lfi} {\BPM}
correction.  From a comparison of the different surveys, we estimate
that the uncertainties are of the order $10\mu$K in the polarized
intensity and dominated by the noise rather than the leakage except in
the innermost plane ($|\glon|<30\degr$ and $|\glat|<3\degr$), where the effect
is only slighly above the noise level.  For the polarization angle, we
estimate the uncertainties as roughly $15\degr$ in the plane
($|\glat|<5\degr$) and $35\degr$ away.  Again the uncertainty appears
dominated by noise, with no obvious structure related to the bandpass
leakage or scan pattern.  We have also cross-checked with {\wmap}
23\,{\GHz} data and verified that the results in
Sect.\,\ref{sec:SYNCHROTRONcomp} are very similar.

\begin{figure*}[!h!t]
\begin{center}
\includegraphics[width=0.95\textwidth]{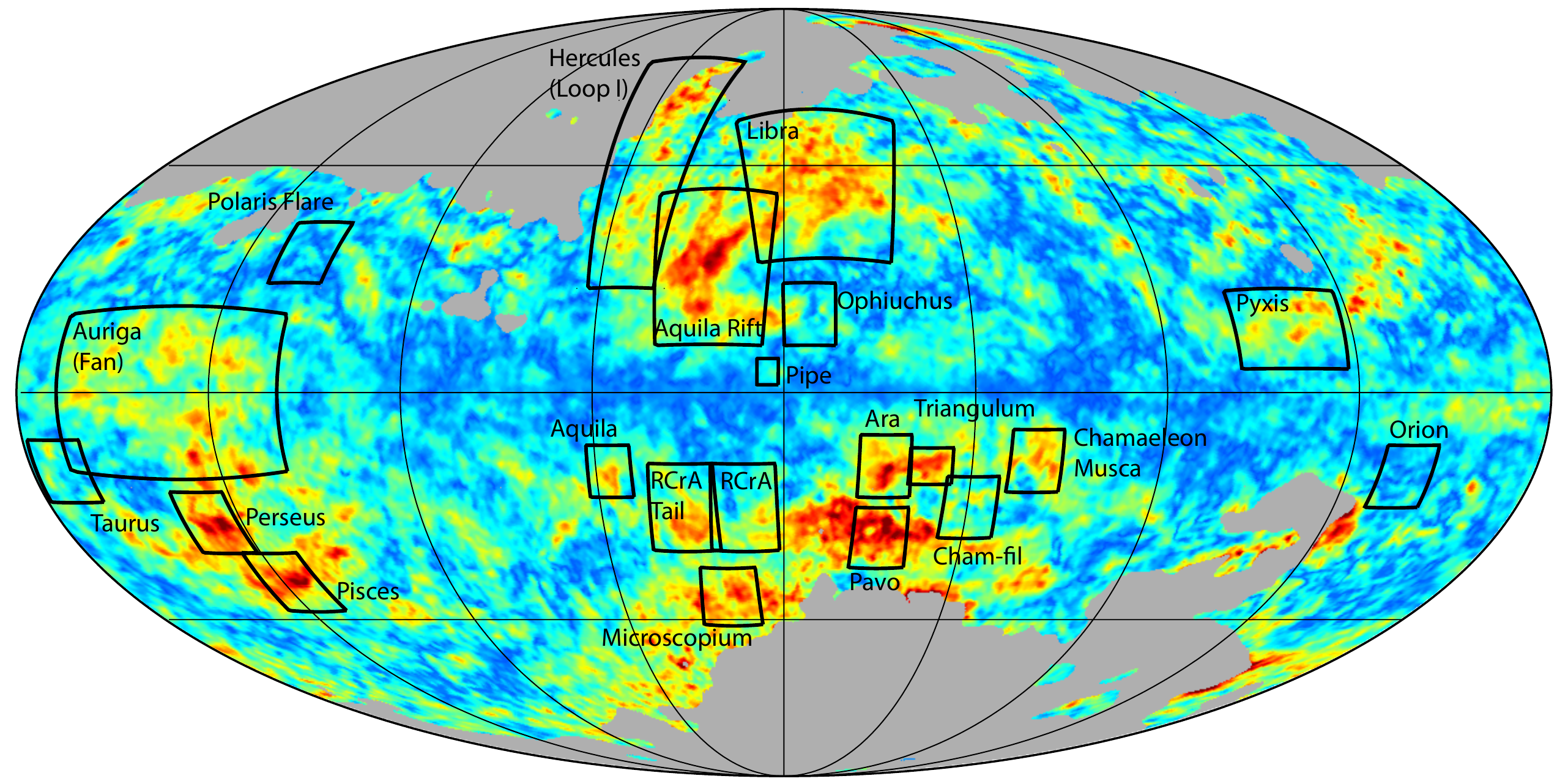}
\caption{\label{fig:overlays}
Map of polarization fraction $\polfrac$ from Fig.\,\ref{fig:polar_p_and_sigp} with selected regions
marked; statistics of these regions are given in Table\,\ref{tab:pol_regions}.
}
\end{center}
\end{figure*}

\begin{table*}[tmb]
\begingroup
\newdimen\tblskip \tblskip=5pt
\caption{
Polarization characteristics of individual regions 
shown in Fig.\,\ref{fig:overlays}, computed at $1\degr$ resolution.  The
table gives the region name (column 1), the Galactic coordinates and
extent of the region (columns 2--5), the minimum, mean, median, maximum, and
standard deviation of $\polfrac$ over the region (columns 6--10) and the
median and standard deviation of $\polang$ (columns 11--12).
Note that the values of $\polang$ are given in the IAU convention. Regions are
ordered by increasing median $\polfrac$.
}
\label{tab:pol_regions}
\nointerlineskip
\vskip -3mm
\footnotesize
\setbox\tablebox=\vbox{
   \newdimen\digitwidth 
   \setbox0=\hbox{\rm 0} 
   \digitwidth=\wd0 
   \catcode`*=\active 
   \def*{\kern\digitwidth}
   \newdimen\signwidth 
   \setbox0=\hbox{+} 
   \signwidth=\wd0 
   \catcode`!=\active 
   \def!{\kern\signwidth}
%
\tabskip=0pt
\halign{
\hbox to 1.20in{#\leaderfil}\tabskip 2.2em&    
\tabskip=1em 
\hfil#\hfil&\hfil#\hfil&\hfil#\hfil&\hfil#\hfil&\hfil#\hfil&\hfil#\hfil&\hfil#\hfil&\hfil#\hfil&\hfil#\hfil&\hfil#\hfil&\hfil#\tabskip=0pt\cr
\noalign{\doubleline}
\omit Region\hfil&    
$\glon$ & $\glat$ & $\Delta\glon$ & $\Delta\glat$ &
min($\polfrac$)&mean($\polfrac$) & med($\polfrac$) & max($\polfrac$) & stdev($\polfrac$) & med($\polang$) & stdev($\polang$) \cr
\omit\hfil  \hfil& 
[$\degr$] & [$\degr$] & [$\degr$] & [$\degr$] & [\%] & [\%] & [\%] & [\%] & [\%] & [$\degr$] & [$\degr$] \cr
\noalign{\vskip 5pt\hrule\vskip 5pt}
           \PolarisFlare  &       !120.0&           !27.0  &        12.0  &        12.0  &        0.10  &        *3.11  &        *2.94  &        *7.40  &        1.50  &      176.72  &       46.23 \cr
                    \Orion  &       !211.0&       $-16.0$  &        12.0  &        12.0  &        0.08  &        *3.22  &        *2.97  &       10.23  &        1.73  &      177.17  &       42.87 \cr
                      \Pipe  &         **!0.0&            *!4.5  &         *5.5  &         *5.5  &        0.31  &        *3.85  &        *3.53  &        *8.45  &        1.90  &      143.13  &       16.85 \cr
           \Ophiuchus  &        **$-6.0$&        !15.0  &        12.0  &        12.0  &        0.11  &        *5.11  &        *4.59  &       12.22  &        2.60  &        **0.84  &       20.69 \cr
                 \Taurus  &       !173.0&        $-15.0$  &        12.0  &        12.0  &        0.16  &        *5.08  &        *4.83  &       11.62  &        2.19  &      129.00  &       60.11 \cr
                    \RCrA  &        *!10.0&       $-22.0$  &        15.0  &        17.0  &        0.30  &        *6.80  &        *6.71  &       13.97  &        2.94  &       *11.62  &       15.42 \cr
               \Chamfil  &       !315.0  &       $-22.0$  &        12.0  &        12.0  &        1.40  &        *6.95  &        *6.78  &       15.29  &        2.22  &       *14.32  &        *8.56 \cr
                   \Pyxis  &      $-120.0$  &        !12.0  &        25.0  &        15.0  &        0.34  &        *7.09  &        *6.96  &       16.71  &        3.03  &      171.04  &       15.33 \cr
                        \RII  &        *!42.0  &       $-15.0$  &        10.0  &        10.0  &        0.88  &        *7.71  &        *7.10  &       14.63  &        3.00  &       *58.61  &       12.94 \cr
                  \Auriga  &       !145.0  &             *!0.0  &        50.0  &        30.0  &        0.12  &        *7.55  &        *7.58  &       18.64  &        2.76  &        **1.69  &       12.20 \cr
               \RCrATail  &        *!25.0  &       $-22.0$  &        15.0  &        17.0  &        1.66  &        *8.63  &        *8.40  &       15.53  &        3.16  &      170.71  &       14.65 \cr
               \Hercules  &        *!40.0  &            !45.0  &        15.0  &        50.0  &        0.37  &        *8.67  &        *8.59  &       37.49  &        3.69  &       *65.26  &       58.68 \cr
                     \Libra  &       *$-10.0$  &        !40.0  &        30.0  &        30.0  &        0.34  &        *9.35  &        *9.90  &       21.39  &        3.42  &       *20.03  &       23.72 \cr
\ChamaeleonMusca  &       !300.0  &       $-13.0$  &        12.0  &        12.0  &        0.89  &        *9.29  &        *9.98  &       15.08  &        3.15  &       *15.06  &       10.80 \cr
            \AquilaRift   &        *!18.0  &            !24.0  &        25.0  &        30.0  &        0.12  &       10.25  &       10.21  &       20.15  &        3.55  &       *50.91  &       13.09 \cr
                        \Ara  &       !336.0  &       $-14.0$  &        12.0  &        12.0  &        3.15  &       11.18  &       10.85  &       21.09  &        2.99  &      177.49  &        *8.90 \cr
                   \Pisces  &       !133.0  &       $-37.0$  &        12.0  &        12.0  &        4.32  &       12.10  &       11.72  &       20.81  &        3.22  &       *15.60  &        *4.99 \cr
                         \RI   &        *!15.0  &       $-40.0$  &        12.0  &        12.0  &        6.20  &       11.78  &       11.76  &       18.63  &        2.27  &       *24.66  &       10.80 \cr
                        \RIII  &       *$-35.0$  &   $-14.0$  &        10.0  &         7.0  &        5.21  &       12.12  &       12.12  &       17.14  &        2.82  &        **6.66  &        *4.95 \cr
                 \Perseus  &       !143.0  &       $-25.0$  &        12.0  &        12.0  &        5.66  &       12.68  &       12.68  &       21.10  &        3.20  &        **9.68  &        *5.96 \cr
                      \Pavo  &       !336.0  &       $-28.0$  &        12.0  &        12.0  &        3.60  &       14.13  &       14.33  &       21.77  &        3.61  &       *14.29  &        *7.99 \cr
\noalign{\vskip 5pt\hrule\vskip 3pt}}}
\endPlancktable                    
\endgroup
\end{table*}                        

\subsection{Deriving polarization parameters}
\label{sec:polarparam}

The polarization parameters $\polI$, $\polfrac$, and $\polang$ are
derived from the observed Stokes parameters $\StokesI$, $\StokesQ$,
and $\StokesU$ using the Bayesian method described in
\cite{planck2014-XXIII98}.  This method extends that described in
\cite{Quinn2012} by using the full $3\times3$ noise covariance matrix
of each pixel.  The effective ellipticity, as defined in
\cite{planck2014-XXIII98}, characterizes the shape of the noise
covariance matrix and couples all the terms in $\StokesQ$ and
$\StokesU$.  $\epseff = 1$ corresponds to the case described in
\cite{Quinn2012}, whereas $\epseff > 1$ means that the relation
between $\sigQQ, \sigQU, \sigUU$ is not trivial, and there are
asymmetries in the noise covariance matrix.  We calculated $\epseff$
for the {\Planck} data used here. At $1\degr$ resolution it is
normally distributed with a mean value of $1.12$ and a standard
deviation of $0.04$.  At the full {\Planck} resolution, the
distribution of $\epseff$ is a bit wider (standard deviation of
$0.05$), but the mean value does not change.  Thus, although they are
not very strong, the asymmetries of the noise covariance matrix cannot
be neglected, and the Bayesian method is well suited for the analysis
of the data.  We use a flat prior on all 3 parameters $\polfrac$,
$\polang$ and $\StokesI$ over a range centered on the {\classical}
value of each parameter, and a width corresponding to $20\sigma$,
where $\sigma$ is the {\classical} estimate for the uncertainty (see
Appendix\,\ref{sec:methodOne}). The range on $\polfrac$ and $\polang$
is further limited to $-1<\polfrac<1$ and $-90\degr<\polang<90\degr$,
respectively.  We compute the 3D posterior probability distribution
function (PDF) using $2^7$ values on each axis over the parameter
range.  The values of the polarization parameters are obtained using
the mean posterior (MP) estimator on the posterior 3D PDF.  A
comparison between the polarization parameters and uncertainties
obtained with this method and using the {\classical} approach
described in Appendix\,\ref{sec:methodOne} is shown in
Fig.\,\ref{fig:classical_vs_ml} for the {\Planck} data at $1\degr$
resolution.

When spatial smoothing is applied to the polarization data, Stokes
parameter maps are convolved with a Gaussian kernel of the appropriate
width using the dedicated smoothing software part of the {\Healpix}
library, which guarantees proper transport of $\StokesQ$ and
$\StokesU$. The maps are then resampled to larger pixel size (as
specified by the {\Healpix} $\Nside$ parameter) so as to preserve full
sampling of the data (pixel size smaller than 1/2.4 times the data
FWHM resolution). The corresponding smoothing of data covariances was
performed using the method described in Appendix\,\ref{sec:noise}.
The corresponding smoothed maps of $\polfrac$ and $\polang$ are then
computed as described above.  The statistical uncertainties in
$\polfrac$ and $\polang$ ($\sigpstat$ and $\sigpsistat$, respectively)
have been estimated as described in Appendix\,\ref{sec:Bayesian}.

\subsection{Impact of systematic effects, CIB, ZL and CMB}
\label{sec:othercorrections}

We assessed the level of contamination by systematic effects
comparing the maps of $\polfrac$ and $\polang$ obtained at $1\degr$
resolution for the full {\Planck} data with those obtained for the
various individual {\Planck} surveys (see
Sect.\,\ref{sec:planckdata}). We constructed maps of systematic
uncertainties on $\polfrac$ and $\polang$ ($\sigpsyste$ and
$\sigpsisyste$, respectively) by averaging these differences over the
{\Planck} individual surveys.  These were added to the statistical
uncertainty maps $\sigpstat$ and $\sigpsistat$, to obtain the total
uncertainty maps used in the rest of the analysis.

In this paper, we only show the {\Planck} polarization data and derived quantities,
where the systematic uncertainties are small, and where the dust
signal dominates total emission. For this purpose, we defined a mask
such that $\sigpsyste < 3\,\%$ and $I_{353} > 0.1\MJysr$. We defined the
mask at a resolution of $1\degr$ and smoothed it to $3\degr$
resolution to avoid complex edges. As a result, the maps shown exclude
$\maskedfraction\,\%$ of the sky. Note that a different mask is used
for the polarization angle dispersion function, as defined in
Sect.\,\ref {sec:polstruct}.

The cosmic infrared background (CIB) is due to emission from a large
number of distant galaxies with random orientations and is expected to
be, on average, unpolarized.  However, it can contribute
non-negligible emission at 353{\GHz} in low brightness regions of the
sky and hence reduces the apparent degree of dust polarization.  The
zero level of the 353{\GHz} intensity map has been established by
correlation with Galactic {\HI}, using the method described in
\cite{planck2013-p06b}, as was done for the publicly released 2013
maps. This offset is $0.0887\MJysr$ (uncertainty $0.0068\MJysr$) and
was subtracted from the intensity map we use, which therefore does not
contain the CIB monopole.  We added the corresponding uncertainty to
the intensity variance, so that the statistical uncertainties on
$\polfrac$ include the uncertainty on the CIB subtraction.

The zodiacal light (ZL) has a smooth distribution on the
sky. From the model constrained by its detection in the {\Planck}
bands \citep{planck2013-pip88}, its median intensity at 353{\GHz} is
$1.9\times 10^{-2}\MJysr$ over the sky area studied here, and reaches
$\simeq4.3\times 10^{-2}\MJysr$) in dust lanes near the ecliptic
plane. Its polarization in the submillimetre is currently
unconstrained observationally.  Since this intensity is subdominant
over most of the sky fraction and the polarization level of ZL is
currently unknown, we apply no correction for the possible
contribution of ZL.  We note that, if ZL was assumed unpolarized,
subtracting its intensity would raise the observed polarization levels
by about $0.5\,\%$ of the observed polarization fraction, on average
over the sky region studied here, and would not change the observd
polarization angles.  We have checked that no noticeable systematic
variation of the polarization fraction is detected in our maps along
zodiacal dust lanes.

CMB fluctuations are polarized at a level of 0.56 mK \citep{Kovac2002}
at a resolution of about $1\deg$, which
corresponds to $1.6\times 10^{-4}\MJysr$ at 353{\GHz}. In the mask we
use here, the effect of CMB polarized fluctuations is therefore
negligible and we did not attempt to correct for those fluctuations.

No additional correction was applied to the data.

\subsection{External data}
\label{sec:SYNCHROTRONdata}

In Sect.\,\ref{sec:SYNCHROTRONcomp}, we compare the {\Planck} HFI polarization maps with low-frequency radio
and microwave observations that are dominated by synchrotron emission
over most of the sky.  These include:
\begin{itemize}
\item the 408\,MHz total intensity map of \citet{haslam:1982} from the
  LAMBDA\footnote{\url{http://lambda.gsfc.nasa.gov}} site;
\item the 1.4{\GHz} total intensity map of the northern \citep{Reich82,Reich86}
  and southern \citep{Reich01} sky;
\item the 1.4{\GHz} polarized intensity maps of the northern \citep{Reich82}
  and southern \citep{Reich86} sky;    
\end{itemize}

For the analysis in Sect.\,\ref{sec:SYNCHROTRONcomp}, the {\Planck}
HFI and LFI maps are smoothed to 1$\degr$ FWHM resolution to match
these radio data and downgraded to $\Nside=256$. Most of the 1.4{\GHz}
maps are available on the Bonn survey site\footnote{\url{
http://www.mpifr-bonn.mpg.de/survey.html}. The southern part of the
1.4{\GHz} total intensity data was provided by W. Reich (private
communication).} as FITS images in Cartesian coordinates.  They are
converted into {\Healpix} using the procedure described in
\cite{Paradis2012} and are made available in this form on the CADE
site \footnote{Analysis Center for Extended Data, \textcolor{magenta} {http://cade.irap.omp}
}.
The resolution of the observations is
roughly 1$\degr$, so no additional smoothing is applied to the radio
data.  The total intensity map at 1.4{\GHz} is estimated to have an
offset of 2.8\,K \citep{Reich04} due to the combination of zero-level
calibration uncertainty, unresolved extragalactic sources, and the
CMB, so this was subtracted from the data.

The total intensity data include thermal bremsstrahlung
(free-free) emission, particularly in the plane. This is not
negligible at 408 MHz or 1.4{\GHz}. We use the {\wmap} MEM free-free
solution \citep{Gold2011} to subtract it.  We note that this
free-free template likely includes anomalous dust emission, and
there are indications that it is an overestimate by roughly
20 to 30\,\% \citep{Alves2010, Jaffe2011}.  Since synchrotron dominates over
free-free emission at low radio frequencies, even on the Galactic plane, the
uncertainties on the free-free correction are not expected to
affect the qualitative comparison with dust emission in this
paper.  But the MEM template is not sufficiently accurate to
correct for free-free when the synchrotron is subdominant at
30{\GHz}.  Furthermore, the 30{\GHz} total intensity also includes
anomalous dust emission for which we have no correction. We
therefore do not use 30{\GHz} in total intensity, but only in
polarization.

\section{Description of the {\Planck} polarization maps}
\label{sec:descplanck}

Figure\,\ref{fig:polar_p_and_sigp} shows the maps of the polarization
fraction ($\polfrac$) at a resolution of $1\degr$.
Figure\,\ref{fig:polar_psi_and_sigpsi} shows the map of the
polarization direction, also at a resolution of $1\degr$. Both figures
also show the corresponding map of the total uncertainty, which
includes the contribution from statistical and systematic uncertainty
estimates, as described in Sect.\,\ref{sec:othercorrections}.  The
maps were masked as described in Sect.\,\ref{sec:othercorrections} in
regions where large residual systematic uncertainties were evident or
where the total intensity at 353{\GHz} is not dominated by dust
emission.  Figures\,\ref{fig:polar_p_and_sigp} and
\ref{fig:polar_psi_and_sigpsi} were constructed using the mean
posterior method described in Appendix\,\ref{sec:Bayesian} and are
discussed in Sects.\,\ref{sec:polfrac} and \ref{sec:polang}.  In
Fig.\,\ref{fig:overlays} we highlight several regions of interest that
we will discuss below; parameters of these regions are given in
Table\,\ref{tab:pol_regions}.

\subsection{Polarization fraction}
\label{sec:polfrac}

\begin{figure}[!h!t]
\begin{center}
\includegraphics[width=9cm]{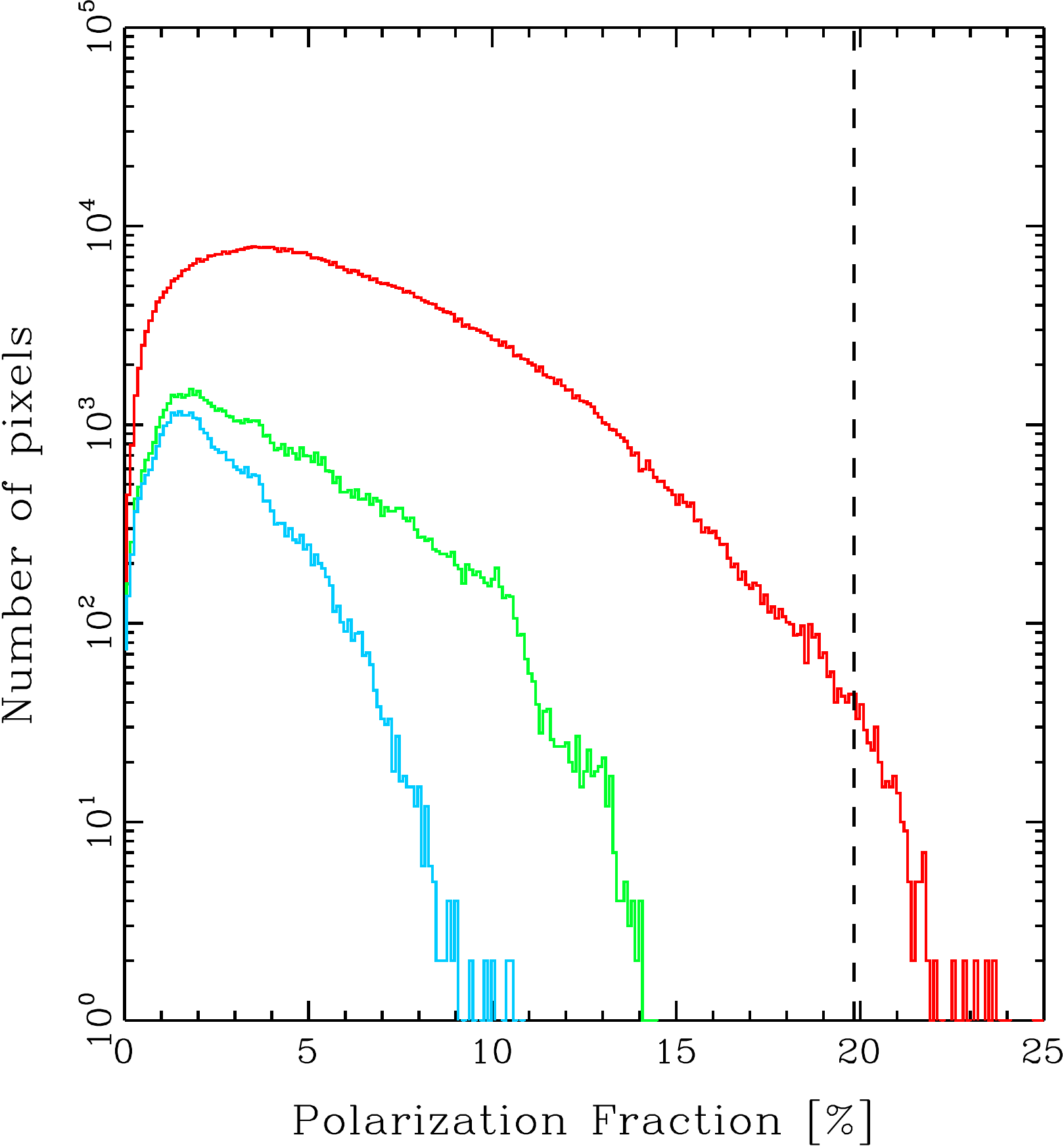}
\caption{\label{fig:phistograms}
Histograms of the observed polarization fraction at $1\degr$
resolution for the whole sky shown in Fig.\,\ref{fig:rawpolarmaps}
(\emph{red}), the Galactic plane within $|\glat|<5\degr$
(\emph{green}) and the inner Galactic plane within $|\glat|<5\degr$
and $|\glon|<90\degr$ (\emph{blue}).  The vertical \emph{dashed} line
shows the maximum value $\pmax$ discussed in
Sect.\,\ref{sec:maxpolfrac}.
}
\end{center}
\end{figure}

\begin{figure*}[!h!t]
\begin{center}
\includegraphics[width=0.95\textwidth]{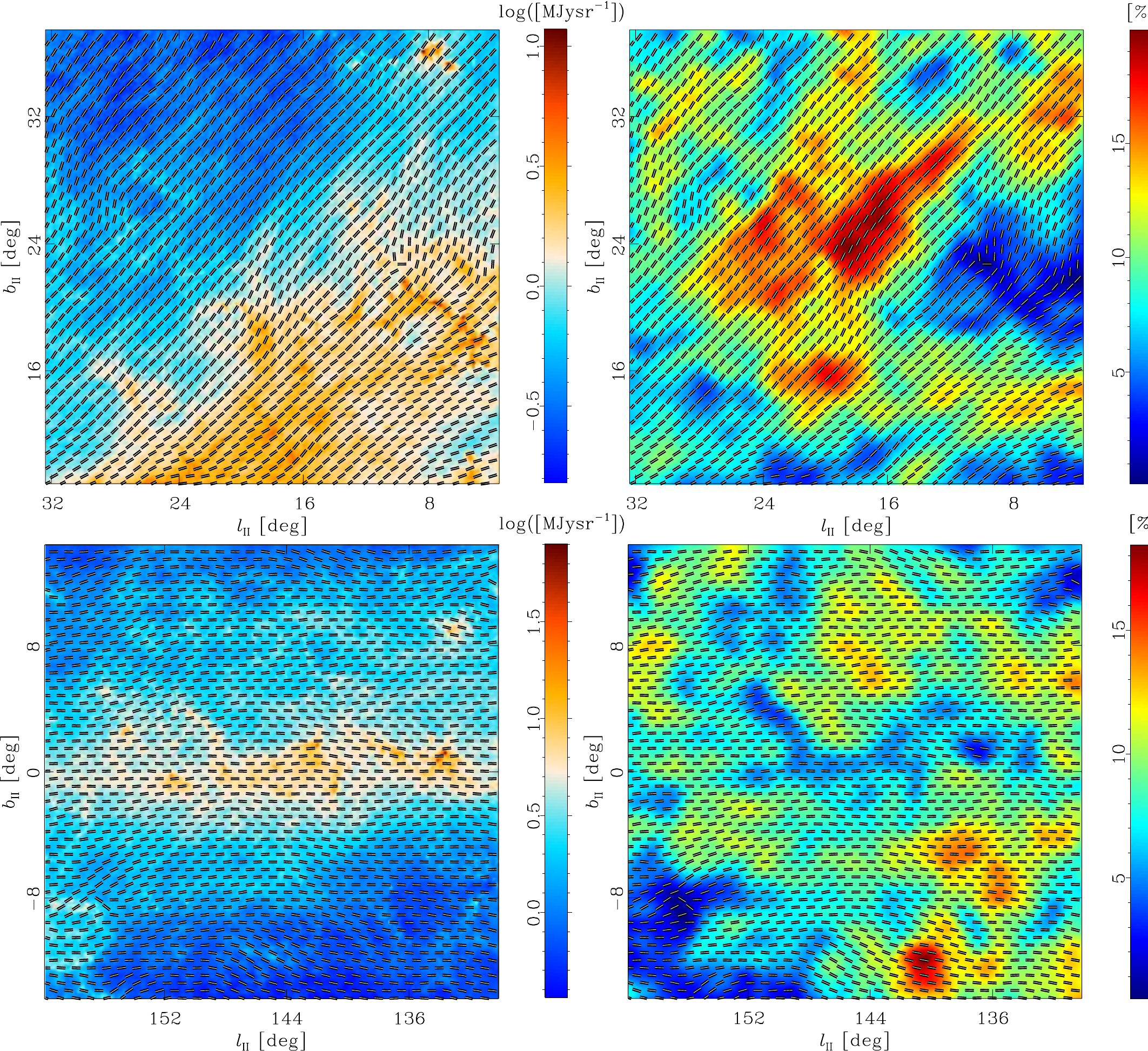}
\caption{\label{fig:highpolarmaps}
Maps of the intensity (\textit{left column}) and polarization fraction
(\textit{right column}) at 353{\GHz} for two of the most polarized regions: {\AquilaRift}
\textit{(upper)}, and the {\Fan} \textit{(lower)}.  The intensity map is
shown at the full {\Planck} resolution, while the polarization information
is shown at a resolution of $1\degr$. The normalized lines
show the orientation of the \appBfield.
The length of the polarization vectors is normalized and does not reflect polarization fraction.
}
\end{center}
\end{figure*}

\begin{figure*}[ht]
\begin{center}
\includegraphics[width=0.95\textwidth]{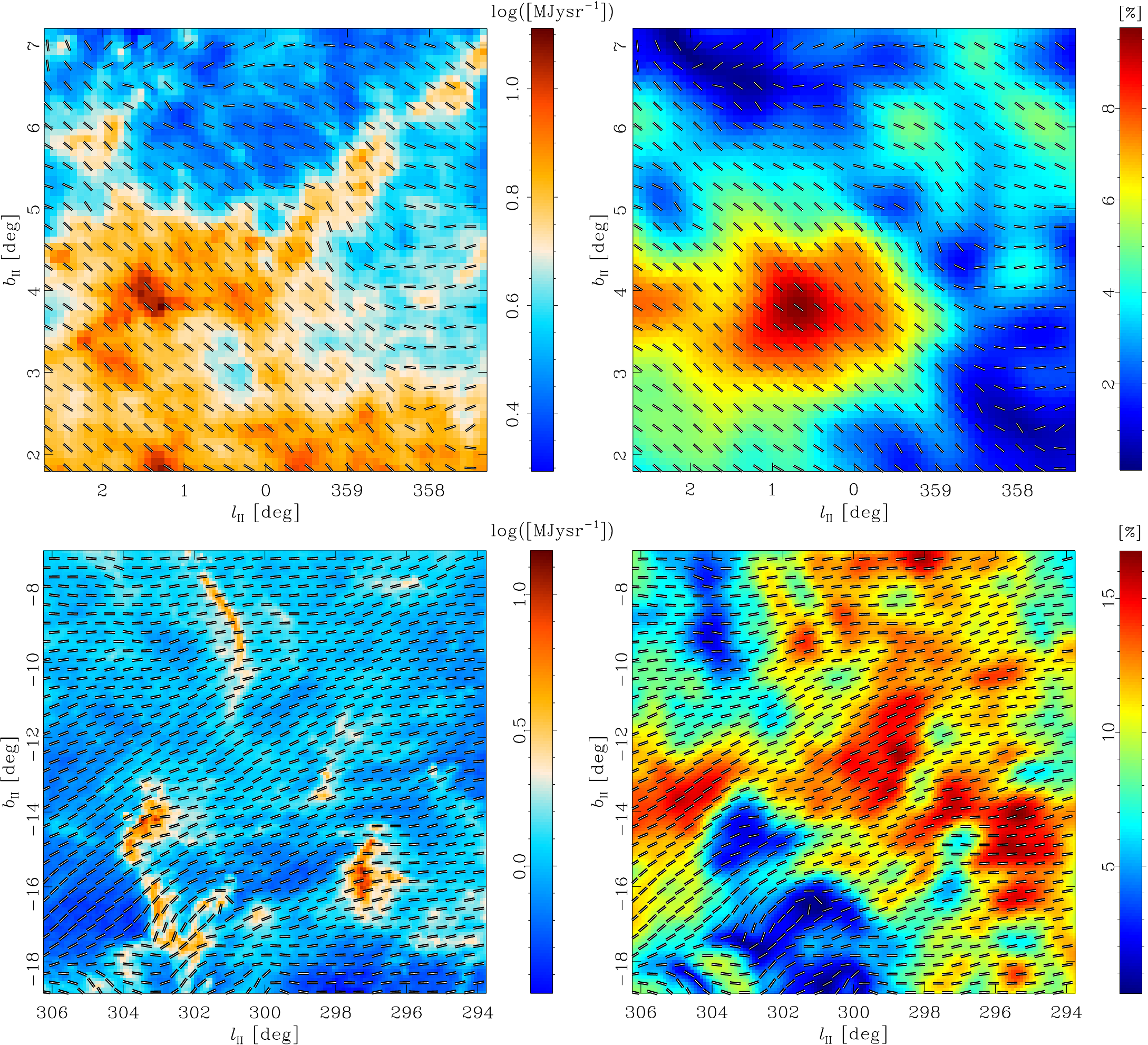}
\caption{\label{fig:zoomedmaps1}
Same as Fig.\,\ref{fig:highpolarmaps} for the Pipe Nebula
\textit{(upper)} and Musca \textit{(lower)} regions.
The polarization data is shown here at a resolution of $30\arcmin$.
}
\end{center}
\end{figure*}

\begin{figure*}[ht]
\begin{center}
\includegraphics[width=0.95\textwidth]{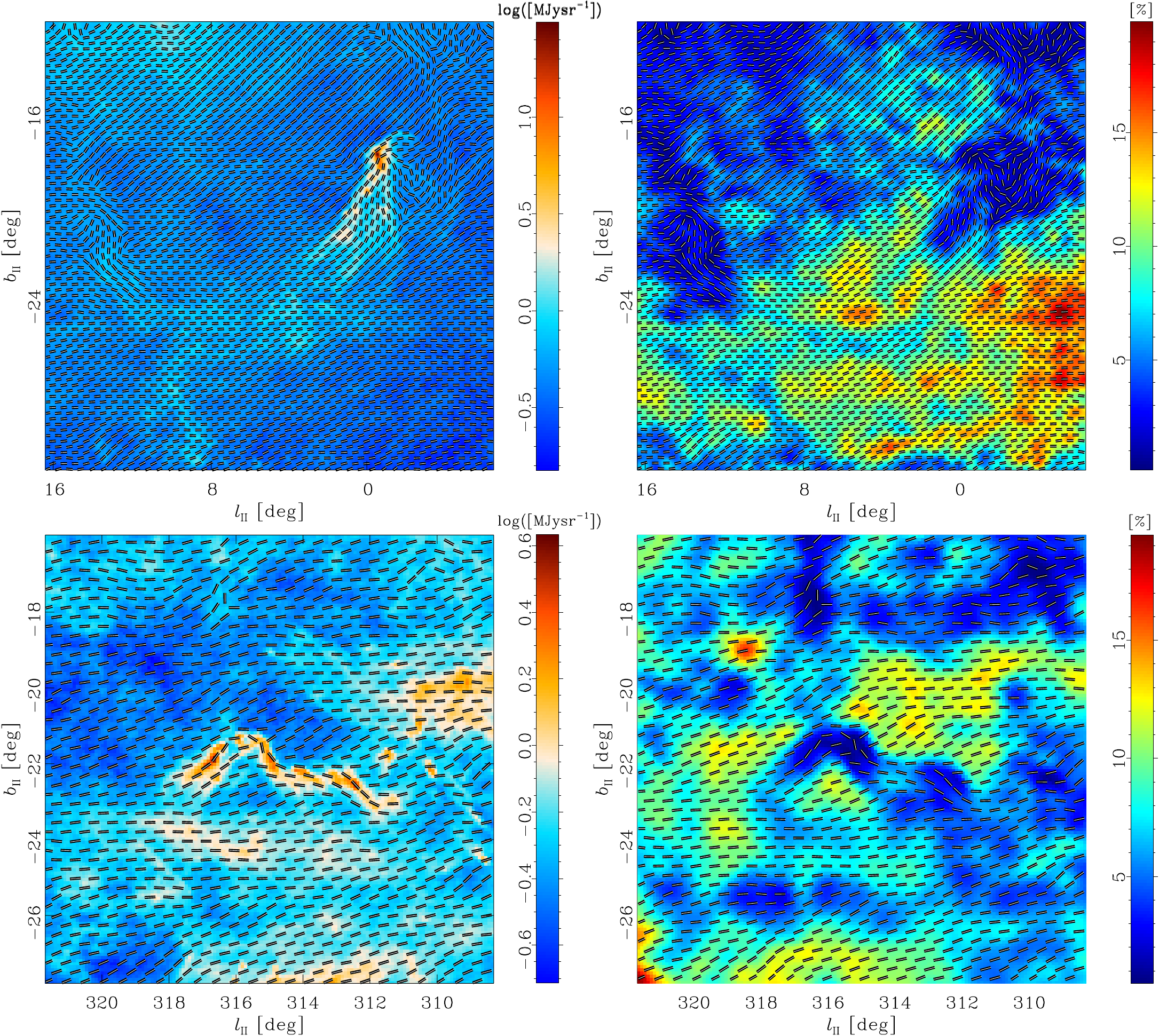}
\caption{\label{fig:zoomedmaps2}
Same as Fig.\,\ref{fig:zoomedmaps1} for regions RCrA and RCrA-Tail (\emph{upper}) and Cham-fil
(\emph{lower}).
The polarization data is shown here at a resolution of $30\arcmin$.
}
\end{center}
\end{figure*}

As seen from Fig.\,\ref{fig:polar_p_and_sigp}, the measured
polarization fraction shows significant variations on the sky. One of
the aims of this paper is to characterize those variations and to try
to understand their origin. These characteristics are compared to
those of polarized emission maps computed in simulations of
anisotropic MHD turbulence in a companion paper \citep{planck2014-XX}.

Figure\,\ref{fig:polar_p_and_sigp} shows that the polarization
fraction of the thermal dust emission can reach up to about $20\,\%$
in several large-scale regions of the sky. This is particularly the
case in the second Galactic quadrant (a region known as ``the Fan")
($\glon \simeq 140\degr$, $\glat \simeq 0\degr$), the Perseus area
($\glon \simeq 144\degr$, $\glat \simeq -30\degr$), the Loop~I area
($\glon \simeq 16\degr$, $\glat \simeq +24\degr$) and a region we call
{\RI} ($\glon \simeq 336\degr$, $\glat \simeq -20\degr$). The
large-scale distribution of these regions is consistent with
predictions from the Galactic magnetic field model used in the Planck
Sky Model \citep{delabrouille2012}. This model, based on a simple
description of the spiral magnetic field structure of the Milky Way,
was optimized to match the {\wmap} and {\archeops} data
\citep[e.g.,][]{Fauvet2011,Fauvet2012b}. It predicts a depolarization
factor that has a minimum in the Galactic plane towards the anticentre
at a position corresponding roughly to that of the Fan region and
shows two strong minima at mid-latitude toward the inner Galaxy $\glon
\simeq 0\degr$ and $|\glat| \simeq 45\degr$ which match fairly well
with the highly polarized regions detected with {\Planck} around the
{\AquilaRift} and {\Pavo}, above and below the galactic plane,
respectively.

Figure\,\ref{fig:phistograms} shows the histogram of the polarized
fraction $\polfrac$ over the sky fraction shown in
Fig.\,\ref{fig:polar_p_and_sigp}, the whole Galactic plane
($|\glat|<5\degr$) and the inner Galactic plane ($|\glat|<5\degr$,
$|\glon|<90\degr$) at a resolution of $1\degr$.  In the plane, the
most likely value of $\polfrac$ is a few percent while the rest of the
mid-latitude sky has a wider distribution, with a peak of the
histogram near $4\,\%$. The maximum $\polfrac$ values can reach about
$20\,\%$.  A more accurate determination of the maximum $\polfrac$
value $\pmax$, taking into account the effects of data resolution and
noise, is given in Sect.\,\ref{sec:maxpolfrac} and leads to a
similarly high value.  We note that this maximum value is much higher
than values reported previously from ground-based observations in the
submillimetre. This is mainly because such low brightness regions are
too faint to be observed from the ground, and because higher column
density and brighter regions, which can be observed from the ground,
have a tendency to be less polarized than faint regions (see
Sect.\,\ref{sec:polfrac_vs_NH}).  We also note that the high
polarization fractions observed here are more consistent with the
value inferred from the {\archeops} measurements at 353{\GHz}, which
was derived to be as high as 10--20\,\% \citep{Benoit2004} along the
outer Galactic plane, a region which includes the Fan region.

Figures\,\ref{fig:highpolarmaps}, \ref{fig:zoomedmaps1}, and
\ref{fig:zoomedmaps2} show maps around some of the regions outlined in
Fig.\,\ref{fig:overlays} and listed in
Table\,\ref{tab:pol_regions}. Figure\,\ref{fig:highpolarmaps} shows
the {\AquilaRift} and {\Fan} regions, which show high polarization
fraction. These highly polarized regions are generally located in
rather low intensity parts of the sky (e.g., {\RI}, {\Libra}, {\Pavo}
or {\Ara}), or on the edge of bright regions (e.g., the
{\AquilaRift}). They are also located in regions of the sky where the
polarization direction is rather constant over very large areas. For
instance, in the {\Fan} region, the magnetic field is oriented almost
completely parallel to the Galactic plane (i.e., polarization is
orthogonal to the plane) with high accuracy over a region spanning
more than $30\degr$, where the polarization fraction consistently has
$\polfrac>8\,\%$ and reaches $\polfrac\simeq 15\,\%$ in some
areas. Similarly, the highly polarized {\AquilaRift} region has a
$\Bfield$-field sky projection aligned with the elongated structure of
the ridge and the nearby {\LoopI} over most of the extent of the
source, and the polarization fraction there reaches up to
$20\,\%$. The highly polarized region is in fact located on the
gradient of the dust emission of the {\AquilaRift}, and mid-way
between the {\AquilaRift} itself and the radio emission of
{\LoopI}. In the {\Perseus} region, the large polarization also
appears in fairly low brightness regions, where the orientation of the
field is coherent over regions of the sky with typical sizes of a few
degrees.  Some of these structures have been detected in polarized
light at other wavelengths. For instance, the {\Fan}, Perseus, and
{\LoopI} regions seem to have counterparts detected in polarized
thermal dust and synchrotron emission, as well as in Faraday RM
surveys of polarized emission at radio frequencies, such as the Global
Magneto-Ionic Medium Survey \citep[GMIMS;][]{Wolleben2010a} and the
{\wmap} foreground emission \citep[][and references
therein]{Gold2011,Berkhuijsen1971c,Ruiz-Granados2010a,JanssonFarrar2012a}.
In particular, from the RM data of GMIMS a significant portion (about
5\,\%) of the sky has been identified to be dominated by the magnetic
field around a local {\HI} bubble (at a distance of 100 pc) whose
edges seem to coincide with the {\LoopI} region described above
\citep{Wolleben2010b}.  In general, such regions are identified with
nearby Galactic structures (e.g., supernova remnants and bubbles),
which can even distort the underlying more regular large-scale pattern
of the Galactic magnetic field.  Finally, other regions, such as
{\RI}, have almost no known counterpart structure in other
wavelengths. The area around {\Ara} and {\RIII} has been identified
only as a region with warmer dust in \cite{planck2011-7.0}. Here too,
the polarization fraction is typically $\polfrac>10\,\%$ (see also
Sect.\,\ref{sec:discussion}).

As seen in Figs.\,\ref{fig:polar_p_and_sigp} and \ref {fig:phistograms},
the inner Galactic plane shows much lower polarization fractions than the
highly polarized regions described above. This is partly due to the larger
depolarization factor caused by the overall structure of the MW magnetic field. It
is also likely due to the fact that the ISM in the MW contains a
collection of dense clouds, which have a general tendency to exhibit
lower polarization fractions (see Sect.\,\ref{sec:polfrac_vs_NH}).

Note that the polarization map exhibits narrow features where
polarization drops (see for instance the one crossing the Polaris Flare
region in Fig.\,\ref{fig:overlays}). These are sometimes regions with
higher gas column density $\NH$, but not
always. They can also be regions where the orientation of the field
changes more abruptly (see Sect.\,\ref{sec:polang} for a full discussion).

\subsection{Polarization angle}
\label{sec:polang}

Figure\,\ref{fig:polar_psi_and_sigpsi} shows the large-scale distribution
of the polarization direction. In the figure, the direction shown by
the normalized lines is that of the observed polarization direction
($\polang$) rotated by $90\degr$. The figure therefore shows the
orientation of the {\appBfield} ($\Bperp$).  In the simplified case
that the direction of $\Bfield$ remains homogenous along the LOS,
$\Bperp$ measures the projection of $\Bfield$ onto the plane of the
sky, i.e., perpendicular to the LOS.  However, in the more realistic
case of a disordered $\Bfield$ structure and inhomogeneous dust
distribution along the LOS, it is important to remember that $\Bperp$
is a LOS-averaged quantity, weighted by dust emission.

Figure\,\ref{fig:polar_psi_and_sigpsi} shows that, towards the
Galactic plane, $\Bperp$ is mostly oriented along the plane,
corresponding to a polarization angle close to $0^\circ$. This is
especially the case towards the outer MW regions.  There are a few
exceptions, in particular toward the tangent points (Cygnus X, $\glon
\simeq 81\degr$, $\glat \simeq 0\degr$; Carina, $\glon \simeq
277\degr$, $\glat \simeq -9\degr$), where the polarization signal is
actually the smallest in the plane due to the magnetic field pointing
along the LOS in those regions. This was already noticed in
\cite{Benoit2004}. We also note that the homogeneity of the field
orientation being parallel to the plane extends away from the plane
and up to $|\glat|\simeq10^\circ$ in many regions (in particular the
{\Fan}).  At intermediate latitudes, the field orientation follows a
few of the well known filamentary intensity structures of the local
ISM. In particular, this is the case for the {\AquilaRift} and
{\LoopI}, where the structure of $\Bperp$ follows the intensity flare
and loop elongation. As addressed earlier, this orientation of
$\Bperp$ in those regions was already noted in the synchrotron
polarized maps of {\wmap} \citep{Gold2011}.  Other regions, however,
show a variety of relative orientations between the field projection
and intensity structures, which can also be orthogonal in some
instances.

\subsection{Polarization angle dispersion function}
\label{sec:polstruct}

\begin{figure*}[!h!t]
\begin{center}
\includegraphics[width=0.95\textwidth]{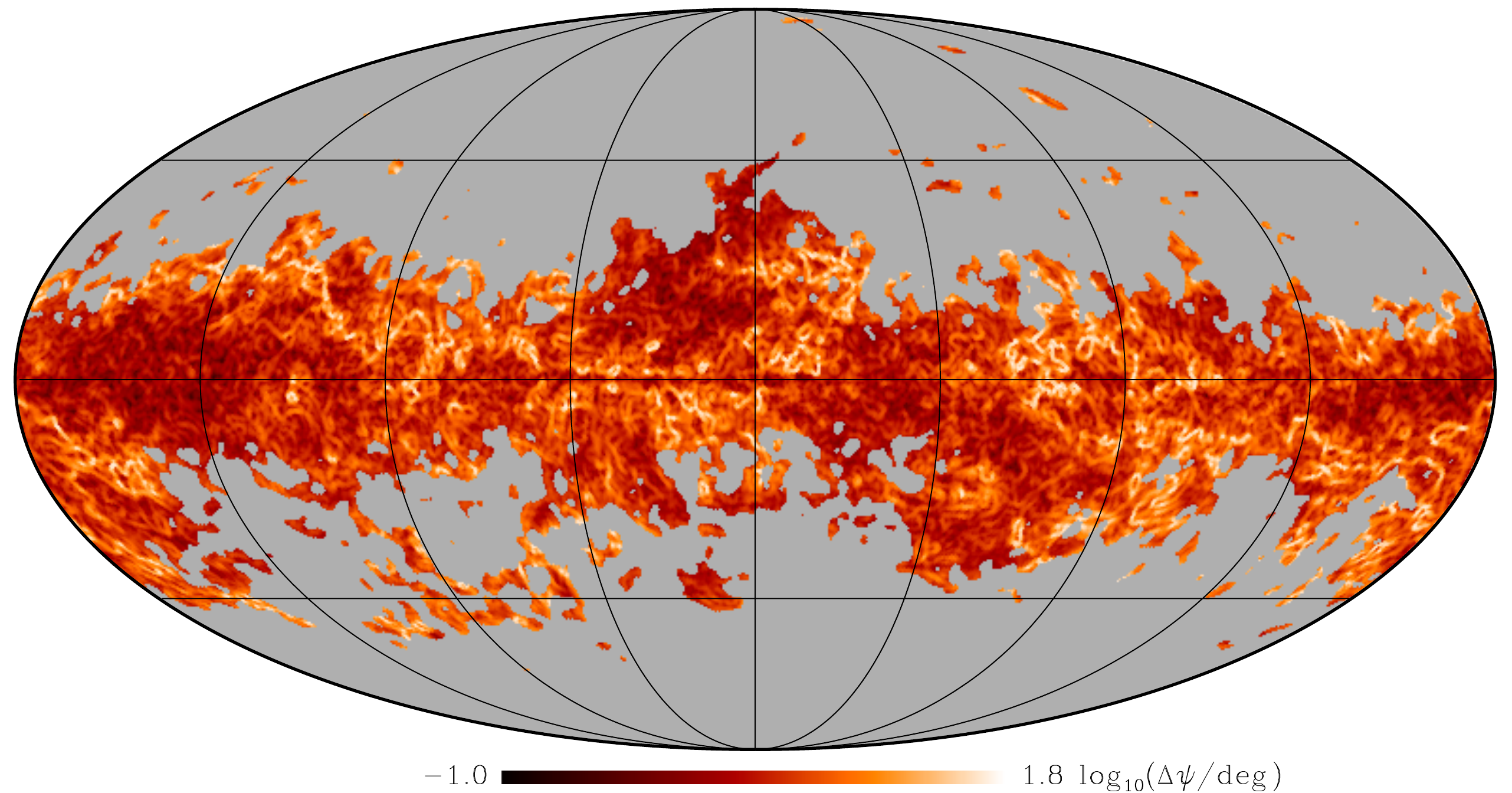}
\caption{
Map of the polarization {\DeltaAngName} $\DeltaAng$ at 353{\GHz} with
$1^\circ$ resolution and for
lag $\lag=30^{\prime}$. The map is shown in $\log_{10}$ scale over the
range $0.1\degr<\DeltaAng<70\degr$. Only sky regions where the {\SNR}
on $\DeltaAng$ is larger than 3 are shown (see text).
\label{fig:dphi_map}}
\end{center}
\end{figure*}

\begin{figure*}[!h!t]
\begin{center}
\includegraphics[width=0.95\textwidth]{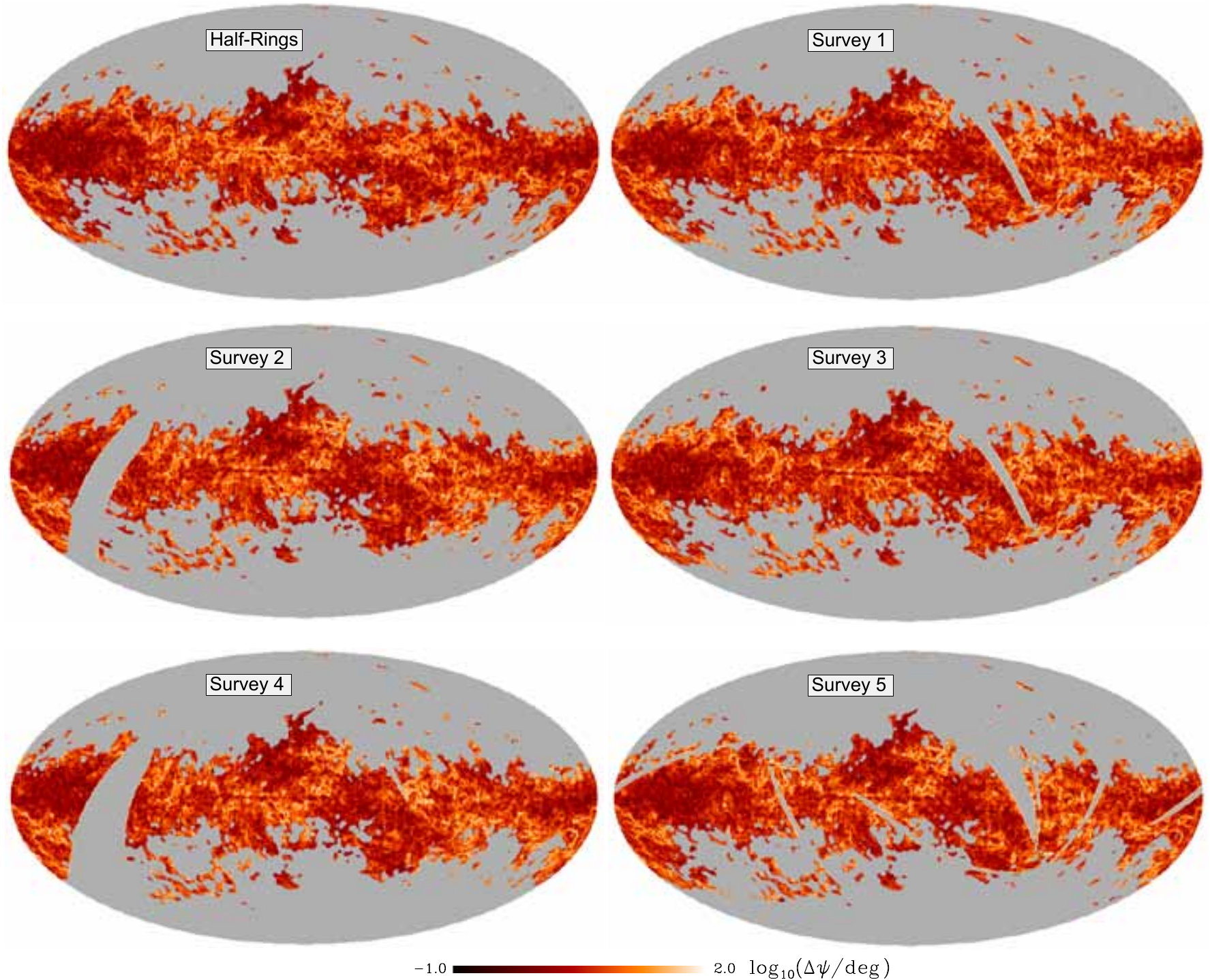}
\caption{
Maps of the polarization {\DeltaAngName} $\DeltaAng$ computed from half-ring correlations
($\DeltaAng_H$) and for individual {\Planck} surveys. The maps are
shown with a common $\log_{10}$ scale.
\label{fig:dphi_half_survey_maps}}
\end{center}
\end{figure*}

\begin{figure}[!h!t]
\begin{center}
\includegraphics[width=9cm]{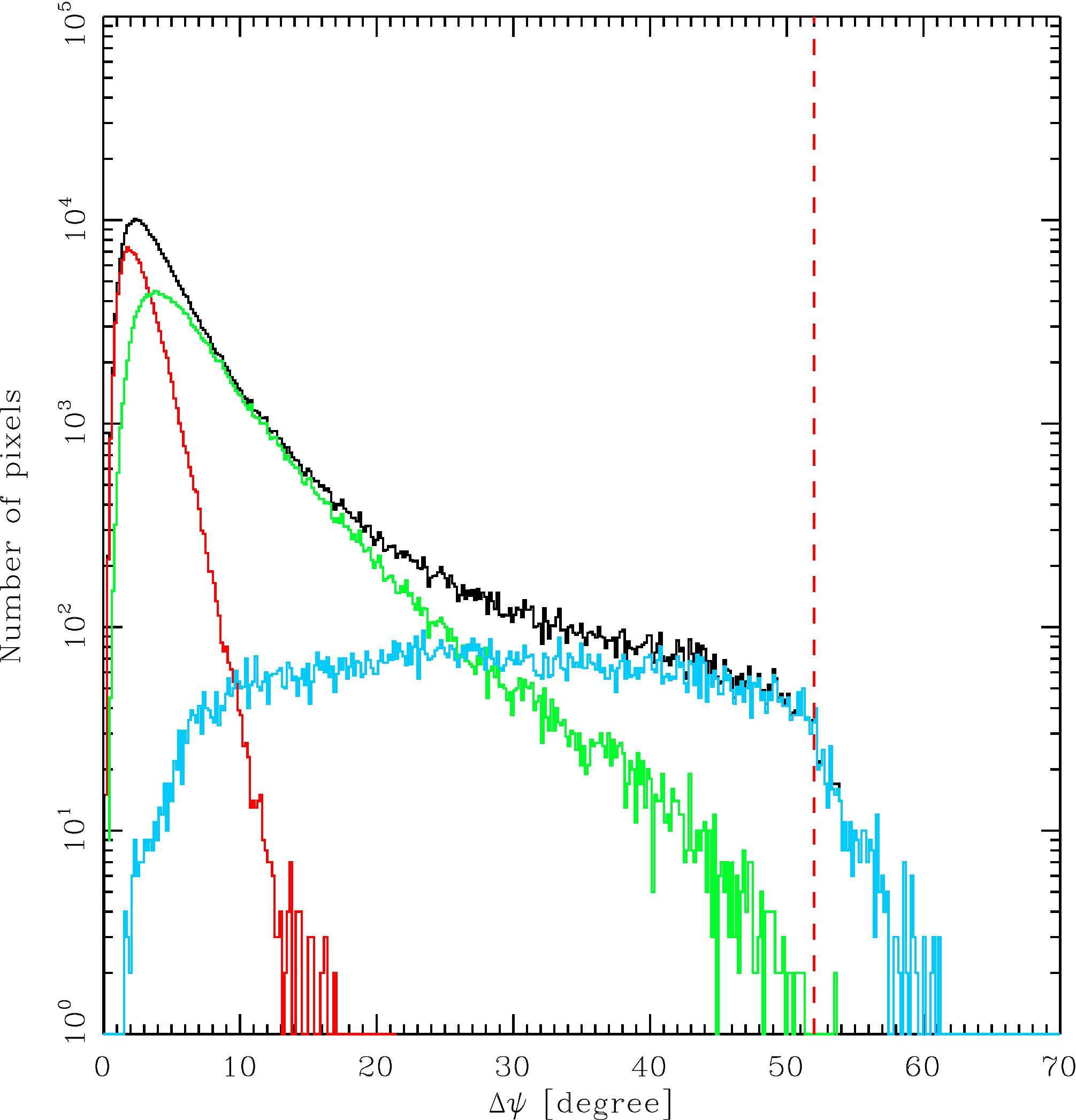}
\caption{
Histogram of $\DeltaAng$ at 353{\GHz} at $1^\circ$ resolution and a lag
$\lag=30\arcmin$. The black curve shows the full distribution over the
sky area covered in Fig.\,\ref{fig:dphi_map}. The red, green, and blue
curves show the histograms for regions covered in Fig.\,\ref{fig:dphi_map} with $\polfrac>5\,\%$,
$1\,\%<\polfrac<5\,\%$, and $\polfrac<1\,\%$, respectively. The vertical
dashed line shows $\DeltaAng=\randomdpsivalue\degr$, which is the limit for pure
random noise on $\DeltaAng$.
\label{fig:dphi_histo}}
\end{center}
\end{figure}

\begin{figure*}[ht]
\begin{center}
\includegraphics[width=0.95\textwidth]{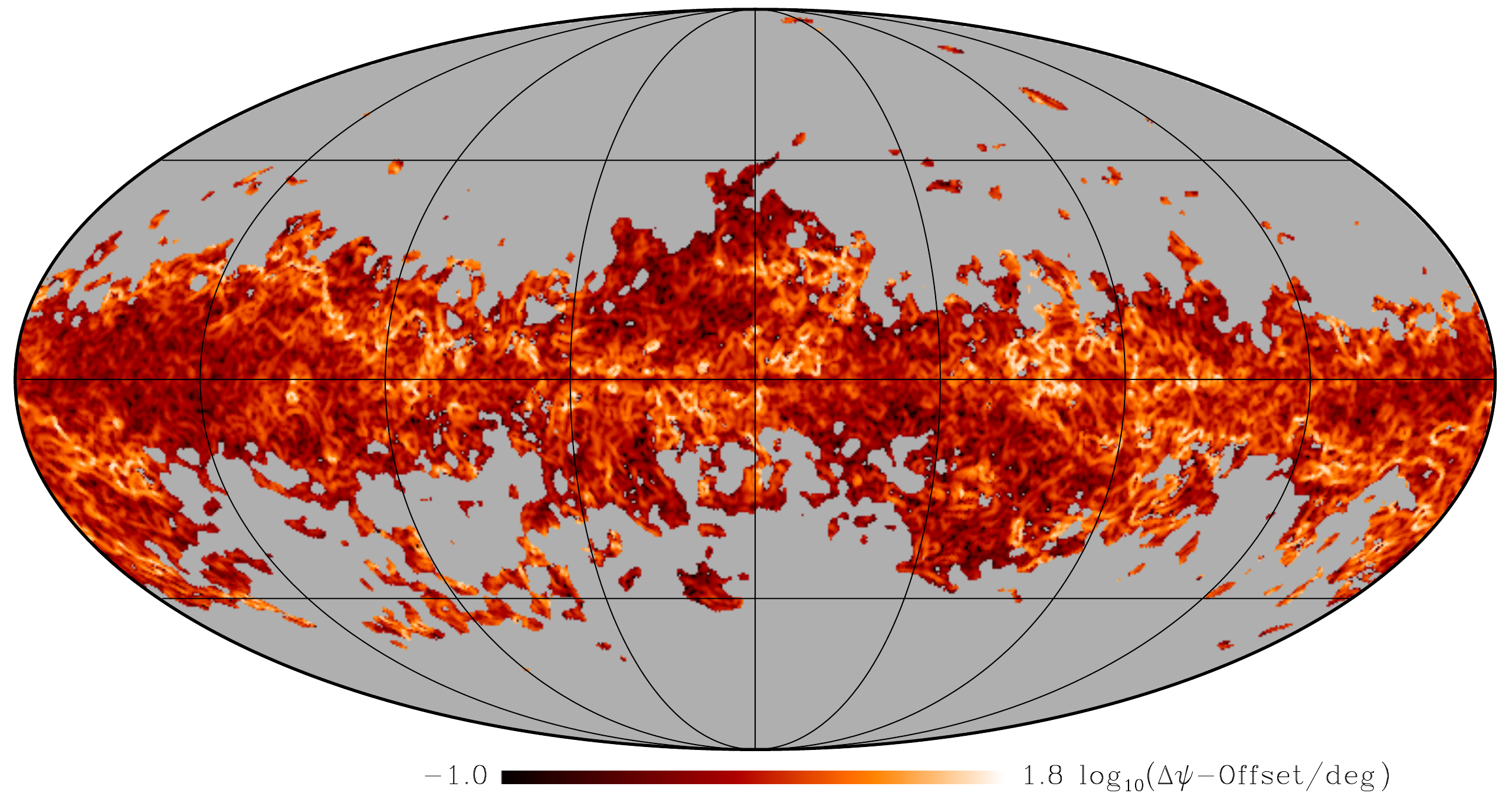}
\caption{
Same as Fig.\,\ref{fig:dphi_map} but with the noise-induced offset
subtracted, as derived from a test with $\DeltaAng=0\degr$ (see
Appendix\,\ref{sec:test_deltapsi_bias} for details).
\label{fig:dphi_moffset_map}}
\end{center}
\end{figure*}

In order to quantify the regularity of the $\Bfield$ field revealed by the
polarization measurements, we use the polarization {\DeltaAngName}
\citep[see][]{Serkowski1958,Kobulnicky1994,Hildebrand2009} given by
\begin{equation}
\DeltaAng^2 (\vect{x},\lag)=\frac{1}{N} \sum_{i=1}^{N}{ \DeltaAng_{xi}^2},
\label{equ:delta_phi1}
\end{equation}
where $\DeltaAng_{xi}=\polang(\vect{x})-\polang(\vect{x}+\vect{\lag}_i)$ is the angle difference
between $\polang(\vect{x})$, the polarization angle at a given sky
position $\vect{x}$ (the central pixel), and $\polang(\vect{x}+\vect{\lag}_i)$ the
polarization angle at a sky position displaced from the centre
by the displacement vector $\vect{\lag}_i$. The average in Eq.\,\ref{equ:delta_phi1}
is taken over an annulus of radius $\lag=|\vect{\lag}|$ (the
``lag'') and width $\Delta\lag$
around the central pixel and containing $N$ pixels. In practice, $\DeltaAng_{xi}$ is
computed from the Stokes parameters as
\begin{equation}
\DeltaAng_{xi}= \frac{1}{2} \arctan
  \left(\StokesQ_i\StokesU_x - \StokesQ_x  \StokesU_i
     ,\StokesQ_i \StokesQ_x + \StokesU_i
      \StokesU_x \right),
\label{equ:delta_phi2}
\end{equation}
where indices $x$ and $i$ stand for the central and displaced values, 
respectively.

The polarization {\DeltaAngName} measures the inhomogeneity of the
polarization direction orientation, irrespective of absolute
direction.  It provides important information on the magnetic field
distribution and orientation \citep[see,
e.g.,][]{Falceta-Goncalves2008,Poidevin2013}.  Regions where the sky
projection of the magnetic field is well ordered will have $\DeltaAng
\simeq 0^\circ$, while regions with a twisted or disordered $\Bfield$
field can in principle have up to $\DeltaAng=90^\circ$.  In addition,
since the {\Planck} convention for $\StokesQ$ and $\StokesU$ is
defined with respect to the Galactic coordinate system, even a
homogeneous field would produce $\DeltaAng \ne 0^\circ$, due to the
coordinate singularity at the poles. In order to avoid this, we have
rotated $\StokesQ$ and $\StokesU$ locally to a different coordinate
system so that the current point is at the equator of the new system,
before applying Eq.\,\ref{equ:delta_phi2}. When the signal is
dominated by noise, $\DeltaAng$ converges to $\DeltaAng=\pi/\sqrt{12}$
($\approx\randomdpsivalue\degr$), which can be identified as a bump in
the histograms of $\DeltaAng$.  The {\DeltaAngName} $\DeltaAng$ is
generally observed to increase with $\lag$, as the coherence is
gradually lost when going further away from a given point of the sky.
It is expected to increase linearly with lag in the presence of a
ordered magnetic field and to steepen at small lags due to either the
turbulent component of the magnetic field or the angular resolution of
the data used \citep[see, e.g.,][]{Hildebrand2009}.  The dependence of
$\DeltaAng$ on lag $\lag$ can be better probed from the analysis of
individual regions at higher resolution, either in emission or in
absorption towards stellar clusters
\citep{magalhaes&pmd_05,Falceta-Goncalves2008, Franco2010}.

Like other quadratic functions, $\DeltaAng$ is biased positively
when noise is present. As described in \cite{Hildebrand2009},
$\DeltaAng$ can be debiased using
\begin{equation}
\DeltaAng_{\db}^2 (\lag)=\DeltaAng^2 (\lag) - \sigma_{\DeltaAng}^2 \hspace{0.1cm},
\label{equ:delta_phi_debias}
\end{equation}
where $\sigma_{\DeltaAng}^2$ is the variance on $\DeltaAng$.
In the {\classical} approach, $\sigma_{\DeltaAng}^2$ can be expressed as a
function of $\sigpolang$ through partial derivatives as
\begin{equation}
\sigma_{\DeltaAng}^2 =\frac{1}{N^2\DeltaAng^2}
  \left(
        \left(\sum_{i=1}^{N}{\DeltaAng_{xi}}\right)^2\sigma_{\polang}^2+
        \sum_{i=1}^{N}{
\DeltaAng_{xi}^2
     \sigpolang(\lag_i)^2} \right).
\label{equ:sigma_delta_phi}
\end{equation}
However, this approximation is valid only close to the solution and
leads to a poor estimate of the bias at low {\SNR}.
Nonetheless, it is clear from Eqs.\,\ref{equ:sigma_delta_phi}
and \ref{equ:methodTwothetaRMS} that regions with low
polarization having higher values of $\sigma_p/\polfrac$ will have
higher $\sigpolang$ and therefore more biased $\DeltaAng$.

In order to assess the importance of the bias, we
use the two independent half-ring maps of the {\Planck} data to
compute an unbiased estimate of $\DeltaAng$ as
\begin{equation}
\DeltaAng_H^2 (\lag)=\frac{1}{N} \sum_{i=1}^{N}{\prod_{h=1}^{2} \DeltaAng^H_{xi}},
\label{equ:delta_surveys_phi1}
\end{equation}
where $\DeltaAng^H_{xi}$ is the angle difference for a half-ring map $H$, i.e.,
$\DeltaAng^H_{xi}=\polang^H(\vect{x})-\polang^H(\vect{x}+\vect{\lag}_i)$.
In practice, $\DeltaAng^H_{xi}$ is computed as
\begin{equation}
\DeltaAng^H_{xi}=\frac{1}{2} \arctan
  \left( \StokesQ^H_{i} \StokesU^H_{x} - \StokesQ^H_{x}  \StokesU^H_{i}
      ,\StokesQ^H_{i} \StokesQ^H_{x} + \StokesU^H_{i}
      \StokesU^H_{x} \right).
\label{equ:delta_surveys_phi2}
\end{equation}
Although $\DeltaAng_H$ is unbiased, it suffers from higher noise,
since only half of the {\Planck} data are used. Note also that, unlike
$\DeltaAng^2$, $\DeltaAng_H^2$ can be negative.

We evaluate $\DeltaAng$ from the full {\Planck} survey (we call this
estimate simply $\DeltaAng$ by default) and $\DeltaAng_H$ at each
pixel of the map using Eqs.\,\ref{equ:delta_phi2} and
\ref{equ:delta_surveys_phi2}, respectively. We also perform a Monte
Carlo noise simulation on $\StokesI$, $\StokesQ$, and $\StokesU$ for
each pixel using the full covariance matrix using
Eq.\,\ref{eq:noise_choldc}), and assuming that different pixels have
independent noise and that the half-ring maps have independent noise.
This simulation is used to construct the PDF of $\DeltaAng^2$ and
$\DeltaAng_H^2$ using 1000 noise samples. We then compute the mean
posterior estimates ($\overline{\DeltaAng}^2$ and
$\overline{\DeltaAng_H}^2$) and variances ($\sigma^2_{\DeltaAng^2}$
and $\sigma^2_{\DeltaAng_H^2}$) of $\DeltaAng^2$ and $\DeltaAng_H^2$
by integrating over the PDF.

Figure\,\ref{fig:dphi_map} shows the sky distribution of $\DeltaAng$
computed from the full survey at 353{\GHz} at $1^\circ$ resolution for
a lag of $\lag=30\arcmin$ and $\Delta\lag=15\arcmin$.
Figure\,\ref{fig:dphi_half_survey_maps} shows the same maps obtained
from the half-ring survey correlation ($\DeltaAng_H$), as well as for
individual {\Planck} surveys.  The mask used in the these figures was
obtained from the uncertainty on $\DeltaAng$, $\sigma_{\DeltaAng}$,
derived from the Monte Carlo analysis described above. The mask is
such that the {\SNR} on $\DeltaAng$ is larger than 3
($\DeltaAng/\sigma_{\DeltaAng}>3$) and retains 52\,\% of the sky at
the adopted analysis resolution of $1\degr$.  The differences between
individual panels of Fig.\,\ref{fig:dphi_half_survey_maps} are smaller
than the 33\,\% statistical uncertainty in the determination of
$\DeltaAng$ within the mask.  Figure\,\ref{fig:dphi_histo} shows the
histogram of $\DeltaAng$ within the above mask, as well as in subsets
of the data with various cuts in $\polfrac$. It shows that most sky
pixels with reliable $\DeltaAng$ have low $\DeltaAng$ values, and that
most of these pixels have large polarization fractions, above
$\polfrac=5\,\%$.

As can be seen in Figs.\,\ref{fig:dphi_map} and
\ref{fig:dphi_half_survey_maps}, a similar structure for $\DeltaAng$
appears in all estimates in the selection mask, clearly showing that
these structures are not caused by a single subsection of the data. We
note that, outside the mask, $\DeltaAng$ shows structures similar to
those observed in the mask. However, significant differences appear in
some regions, in particular between odd and even {\Planck} surveys. We
attribute those to an imperfect bandpass mismatch correction or to the
fact that no ADC correction has been applied here.  We have also
conducted tests in order to quantify the possible noise-induced bias
on $\DeltaAng$. Those are described in
Appendix\,\ref{sec:test_deltapsi_bias}.
Figure\,\ref{fig:dphi_moffset_map} shows the map of $\DeltaAng$ when
the resulting estimate of the bias has been subtracted. Comparison
with Fig.\,\ref{fig:dphi_map} shows that the effect of bias
essentially reduces low $\DeltaAng$ values, but does not explain the
patterns observed in the map.  We therefore conclude that the
structures seen in the map of the polarization {\DeltaAngName}
$\DeltaAng$ are real, rather than being induced by noise and/or bias.
In the rest of the analysis carried out here, we use the map of
$\DeltaAng$ derived from the full survey and only consider pixels
where the {\SNR} on $\DeltaAng$ as derived from our Monte Carlo
analysis is larger than 3. The resulting map is shown in
Fig.\,\ref{fig:dphi_map}. Further analysis of the angular distribution
function and the comparison with the polarization fraction are
presented in Sect.\,\ref{sec:polfrac_vs_deltaphi}.

\begin{figure}[!h!t]
\begin{center}
\includegraphics[width=9cm]{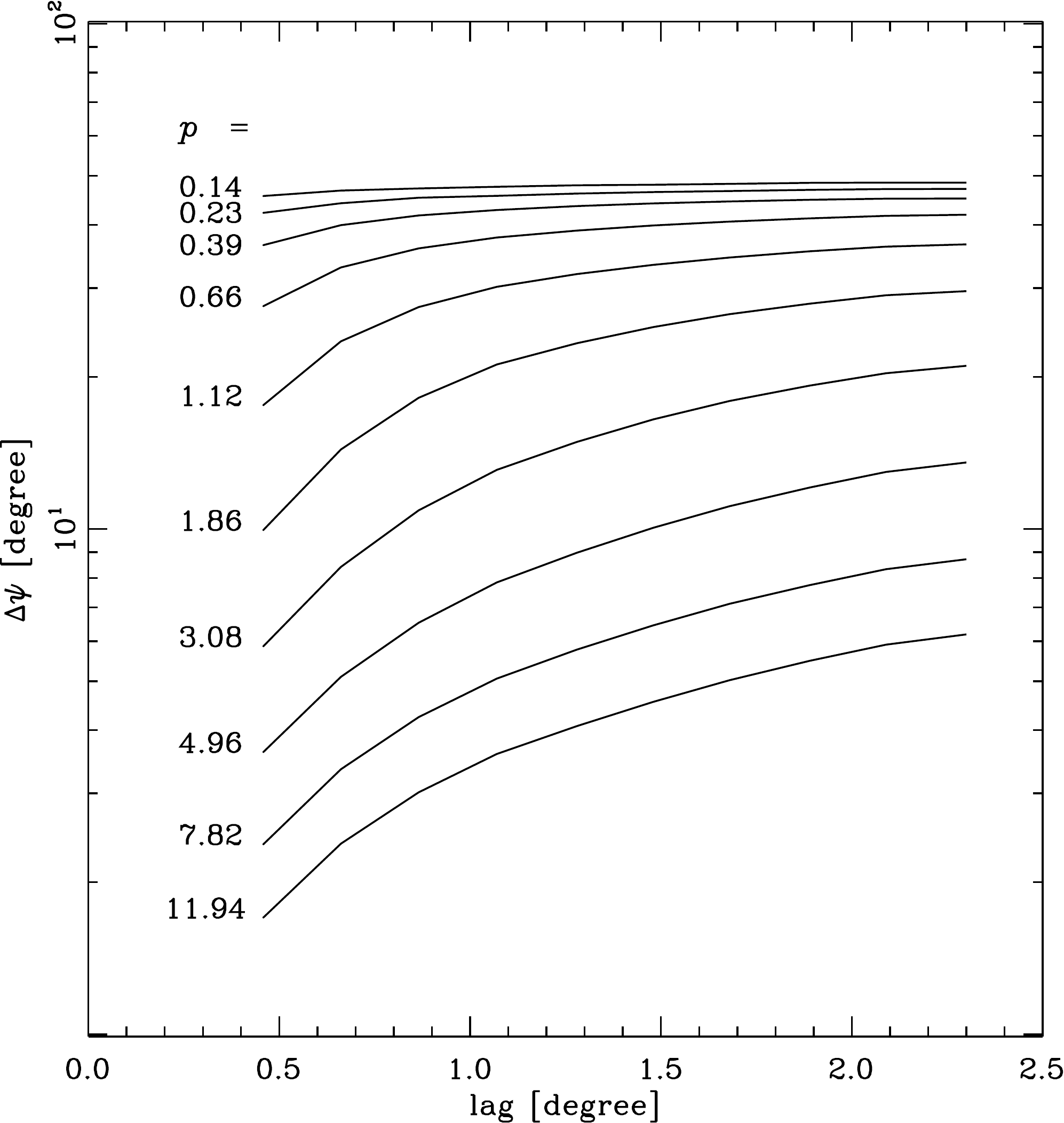}
\caption{
Evolution of the polarization {\DeltaAngName} ($\DeltaAng$) at 353{\GHz} as a function of lag $\lag$,
binned in intervals of the polarization fraction $\polfrac$.
The curves are labelled with the median polarization fraction in the
bin as a percentage.
\label{fig:delta_phi_vs_ell_vs_p}}
\end{center}
\end{figure}

\begin{figure*}[!h!t]
\begin{center}
\includegraphics[width=0.95\textwidth]{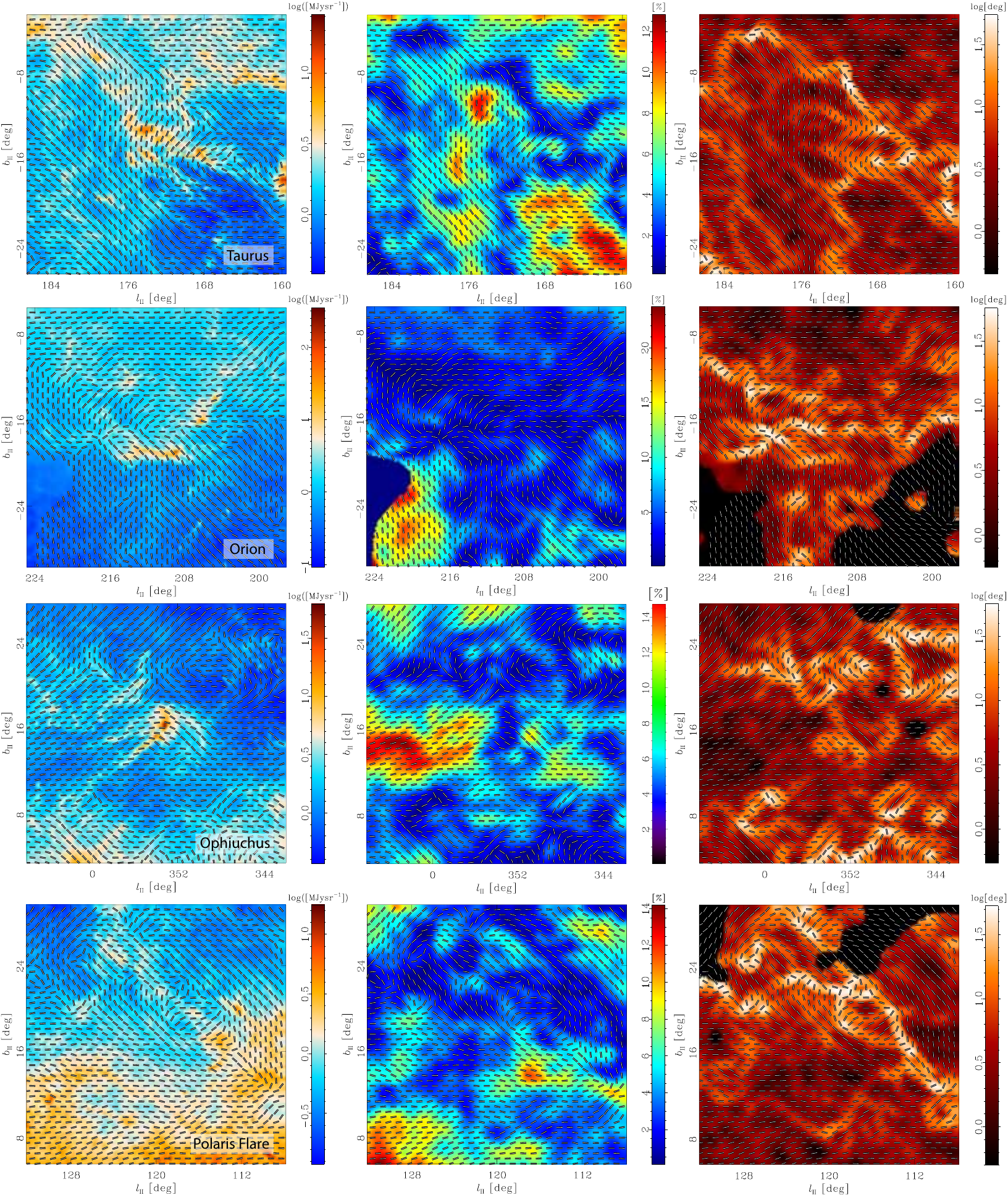}
\caption{\label{fig:zoomed_dphi_maps}
Maps of a few selected regions, illustrating the relation between
polarization fraction and polarization {\DeltaAngName}. \emph{Left}:
intensity at 353{\GHz}. \emph{Centre}: polarization fraction. \emph{Right}:
polarization {\DeltaAngName}, $\DeltaAng$, shown in
log scale. Regions are from top to bottom: Taurus, Orion, Ophiuchus, and Polaris.
}
\end{center}
\end{figure*}

Figure\,\ref{fig:delta_phi_vs_ell_vs_p} shows the values of the
observed $\DeltaAng$ averaged in bins of $\polfrac$ as a function of
the lag value. As expected, the {\DeltaAngName} increases steadily
with increasing lag. Lower values of $\DeltaAng$ systematically
correspond to higher $\polfrac$ values, as discussed in
Sect.\,\ref{sec:polfrac_vs_deltaphi}. Figure\,\ref{fig:zoomed_dphi_maps}
shows details of $\DeltaAng$ for a few selected regions.

\section{Discussion}
\label{sec:discussion}

In this section, we analyse the observed variations of the
polarization fraction and angle at 353{\GHz} and discuss the possible implications
in terms of dust physics and magnetic field structure.
 
\subsection{Maximum polarization fraction}
\label{sec:maxpolfrac}

The maximum dust polarization fraction ($\pmax$) is a parameter of
importance for the physics of dust and its alignment with respect to
the magnetic field, because it directly provides an upper limit to the
{\IntrinsicpName}, $\pzero$, for the optimal orientation of the
magnetic field, i.e., in the plane of the sky. It is also important
for the CMB component separation in polarization, as it sets the
maximum amplitude of dust polarization.  The observed $\polfrac$
values are, however, affected by averaging in the beam and along any
given LOS.  Variations of the $\Bfield$ direction within the beam or
along the LOS necessarily result in a decrease of the observed
$\polfrac$. Similarly, dilution by an additional unpolarized source of
emission, such as free-free or spinning dust emission, can only
decrease $\polfrac$. Therefore, derived values of $\pmax$ can only be
lower limits to the {\IntrinsicpName} $\pzero$.

\begin{figure}[!h!t]
\begin{center}
\includegraphics[width=9cm,angle=0]{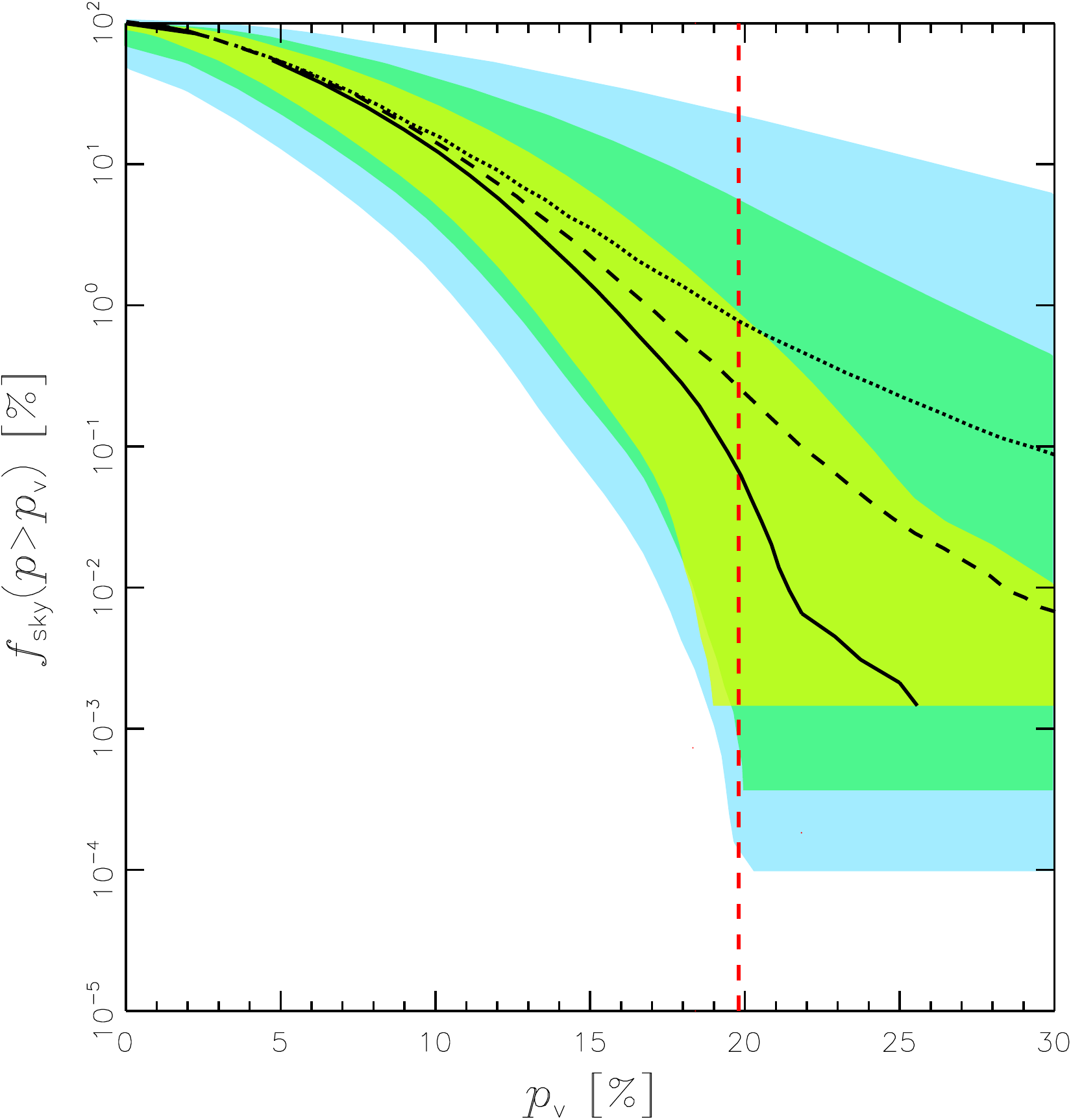}
\caption{
Fraction of the sky $f_{\rm sky}(\polfrac>\pv)$ above a given polarization fraction value
$\pv$, as a function of $\pv$ at the resolution of
$1\degr$ (\emph{solid line, yellow}), $30^\prime$ (\emph{dashed line, green}),
and $15^\prime$ (\emph{dotted line, blue}).
The range shown is the sky fraction corresponding to 
$\polfrac\pm\nsigused\sigma_{\polfrac}>\pv$.  
The \emph{vertical dashed line} shows the adopted common value
of $\pmax=\pmaxvalue\,\%$.
\label{fig:p_percentile}}
\end{center}
\end{figure}

\begin{table}[tmb]
\begingroup
\newdimen\tblskip \tblskip=5pt
\caption{
Statistics of the percentage polarization fraction $\polfrac$ at various
data resolutions, $\resolution$. The table gives the data resolution
(column 1) and the median and maximum values of $\polfrac$ (columns 2 and 3).
The last column (4) shows the maximum values for $\polfrac-\nsigused
\sigma_{\polfrac}$. The average value is computed in the last line and
used as the value for $\pmax$.
}
\label{tab:stat_polar_Bayesian}                            
\nointerlineskip
\vskip -3mm
\footnotesize
\setbox\tablebox=\vbox{
   \newdimen\digitwidth 
   \setbox0=\hbox{\rm 0} 
   \digitwidth=\wd0 
   \catcode`*=\active 
   \def*{\kern\digitwidth}
   \newdimen\signwidth 
   \setbox0=\hbox{+} 
   \signwidth=\wd0 
   \catcode`!=\active 
   \def!{\kern\signwidth}
%
\tabskip=0pt
\halign{
\hbox to 1.00in{#\leaderfil}\tabskip 2.2em&    
\tabskip=1em 
\hfil#\hfil&\hfil#\hfil&\hfil#\hfil\tabskip=0pt\cr
\noalign{\doubleline}
\omit$\resolution$\hfil&
med($\polfrac$) & max($\polfrac$) & max($\polfrac-\nsigused\sigma_{\polfrac}$) \cr
\noalign{\vskip 5pt\hrule\vskip 5pt}
       $15^\prime$  &        5.5  &       81.8  &       20.3 \cr
       $30^\prime$  &        5.3  &       48.7  &       20.0 \cr
       $1\pdeg0$  &        5.1  &       25.6  &       19.0 \cr
 Average  &                 &                  &   $\pmaxvalue \pm \pmaxuncert\,\%$\cr
\noalign{\vskip 5pt\hrule\vskip 3pt}}}
\endPlancktable                    
\endgroup
\end{table}

Here, we use the {\Planck} maps at 353{\GHz} to evaluate
$\pmax$. Since $\polfrac$ is a biased quantity and since noise depends
upon the data resolution, the observed maximum polarization fraction
depends upon resolution. It is therefore crucial to take uncertainties
into account. Figure\,\ref{fig:p_percentile} shows the sky fraction,
$\fsky(\polfrac>\pv)$, where the observed polarization fraction is
higher than a given value $\pv$ as a function of that $\pv$. The
various curves are for data resolutions of $1\degr$, 30\arcmin, and
15\arcmin. The coloured area shown correspond to
$\fsky(\polfrac\pm\nsigused\sigma_{\polfrac}>\pv)$ for the various
resolutions.

At low $\fsky$ values and high resolutions, high values of $\polfrac$
are observed. Inspection of the maps indicates that these are
point-like objects, either isolated pixels or actual point
sources. Since we are interested in diffuse emission only, these
isolated values are ignored in evaluating $\pmax$.
Table\,\ref{tab:stat_polar_Bayesian} lists the maximum and median
values of $\polfrac$ at different resolutions. It also shows the
maximum value of $\polfrac-\nsigused \sigma_{\polfrac}$ observed at
each resolution. We use the average of these values as a conservative
estimate of $\pmax$ and find $\pmax> \pmaxvalue\,\%$.  This indicates
that, in the most favourable conditions for dust alignment, the
{\IntrinsicpName} $\pzero$ is larger than $\pmaxvalue\,\%$.

\subsection{Polarization fraction vs column density}
\label{sec:polfrac_vs_NH}

\begin{figure}[!h!t]
\begin{center}
\includegraphics[width=9cm,angle=0]{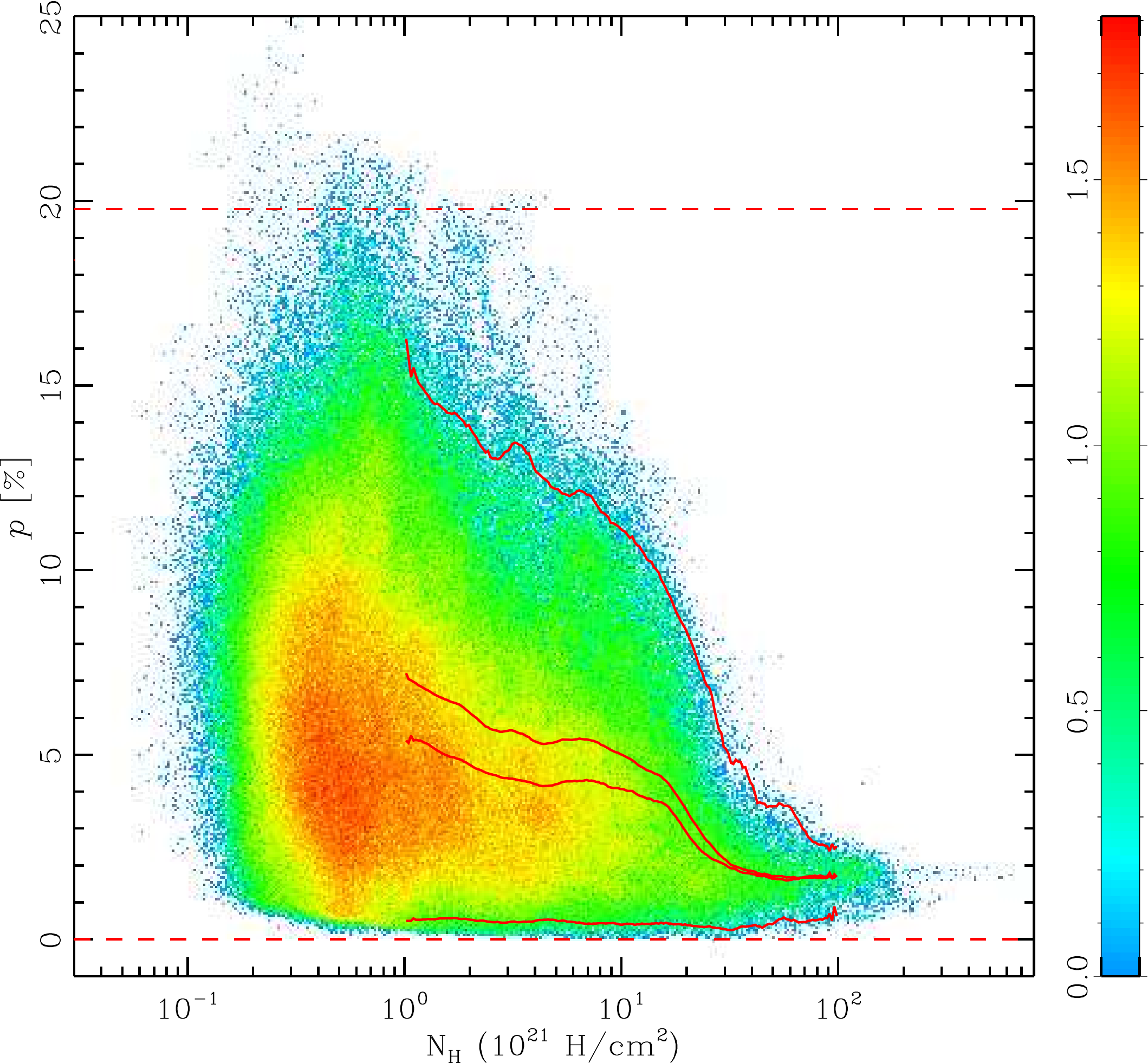}
\includegraphics[width=9cm,angle=0]{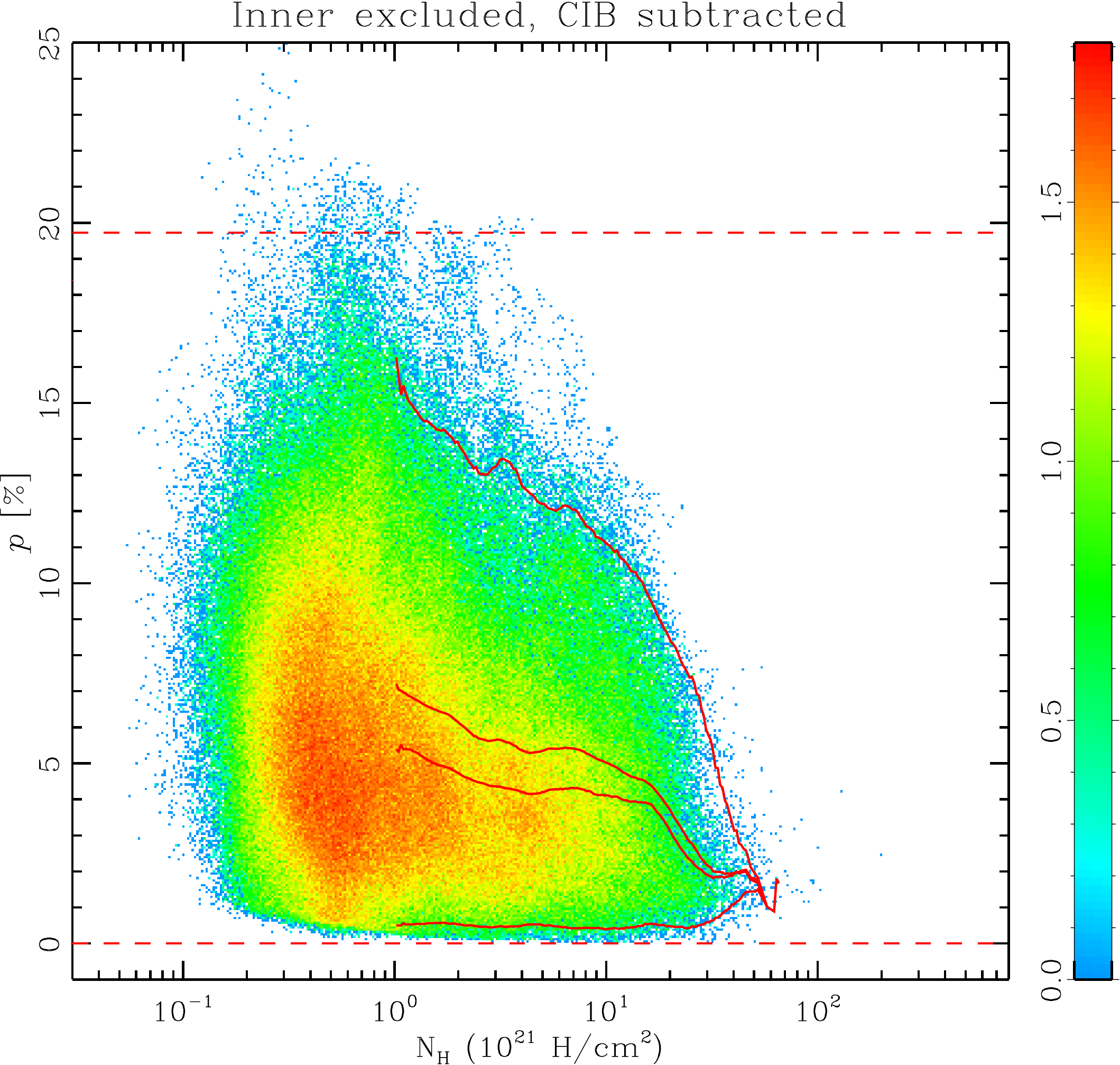}
\caption{
Distribution of the polarization fraction ($\polfrac$) as a function
of gas column density over the whole sky (\emph{upper panel}) and in
regions of the sky excluding the inner Galactic plane (excluding
$\glon<90\deg$ or $\glon>270\deg$, $|\glat|<2\degr$) (\emph{lower
panel}).  The values of $\polfrac$ were computed at $\plotresol$
resolution.  The gas column density is derived from the dust optical
depth at 353{\GHz} (see text). The colour scale shows the pixel
density in log$_{10}$ scale.  The curves show, from top to bottom, the
evolution of the upper $\nperc$ percentile, mean, median and lower
$\nperc$ percentile of $\polfrac$ for pixels with
$\NH>10^{21}\NHUNIT$.  Horizontal dashed lines show the location of
$\polfrac=0$ and $\pmax=\pmaxvalue\,\%$.
\label{fig:p_vs_nh}}
\end{center}
\end{figure}

We now analyse the variations of the polarization fraction $\polfrac$ with
dust column density.  We use the Bayesian mean posterior estimate of
$\polfrac$ described in Sect.\,\ref{sec:polarparam} and
shown in Fig.\,\ref{fig:polar_p_and_sigp}, computed at $1\degr$
resolution.
For the dust optical depth map, we use the map of $\dusttau$ derived
in \cite{planck2013-p06b}.
The maps were computed at $1\degr$ resolution.  Following
\cite{planck2013-p06b}, we adopt the
conversion factor from $\dusttau$ to $\Av$ or gas column density,
derived using the Galactic extinction from measurements of quasars,
\begin{equation}
\NH=(1.41\times10^{26}\,\NHUNIT) \ \dusttau,
\label{equ:tau353_NH}
\end{equation}
which leads to
\begin{equation}
\Av=4.15\times10^{4} \dusttau,
\end{equation}
when using $\Rv$=3.1.

Figure\,\ref{fig:p_vs_nh} shows the distribution of data for the
polarization fraction $\polfrac$ as a function of $\NH$, as derived
from dust optical depth, both for the sky fraction shown in
Fig.\,\ref{fig:polar_p_and_sigp} and for the same region but excluding
the inner Galactic plane (i.e., excluding $\glon<90\deg$ or
$\glon>270\deg$, $|\glat|<2\degr$).  As can be seen in the figure, the
plot shows both considerable scatter at a given $\NH$, and also
systematic trends with $\NH$.  The scatter in $\polfrac$ is likely due
to depolarization in the beam or along the LOS and/or to intrinsic
variations in $\polfrac$.  Possible origins of this scatter are
analysed in \cite{planck2014-XX}.  At low column densities
($2\times10^{20}\NHUNIT <\NH< 10^{21}\NHUNIT$, $0.1<\Av<0.5$\,mag),
$\polfrac$ values generally remain below $\polfrac\simeq15\,\%$.  The
maximum $\polfrac$ values are reached in this $\NH$ range. We observe
an ensemble average polarization of $7\,\%$ at $\NH=10^{21}\NHUNIT$
($\Av=0.5$\,mag).  The average values of $\polfrac$ at lower column
densities are not discussed in this paper, since a proper treatment
would require a careful analysis of the residual bias in the method
used to derive $\polfrac$.  This will be the subject of a future
paper.  At larger $\NH$ ($10^{21}\NHUNIT<\NH< 2\times10^{22}\NHUNIT$,
$0.5<\Av<10$\,mag), $\polfrac$ values are typically below $\polfrac
\simeq 10\,\%$ and show a steady decline, with $\langle \polfrac
\rangle$ decreasing down to $\simeq 4\,\%$.  We observe a sharp drop
in $\polfrac$ starting at $\NH\simeq 2\times10^{22}\NHUNIT$
($\Av\simeq10$\,mag). Above $\NH\simeq 4\times10^{22}\NHUNIT$
($\Av\simeq20$\,mag) values of $\polfrac$ are systematically below
$4\,\%$ with an average value of $\langle\polfrac\rangle \simeq$
1--2\,\%.

\begin{figure}[!h!t]
\begin{center}
\includegraphics[width=9cm,angle=0]{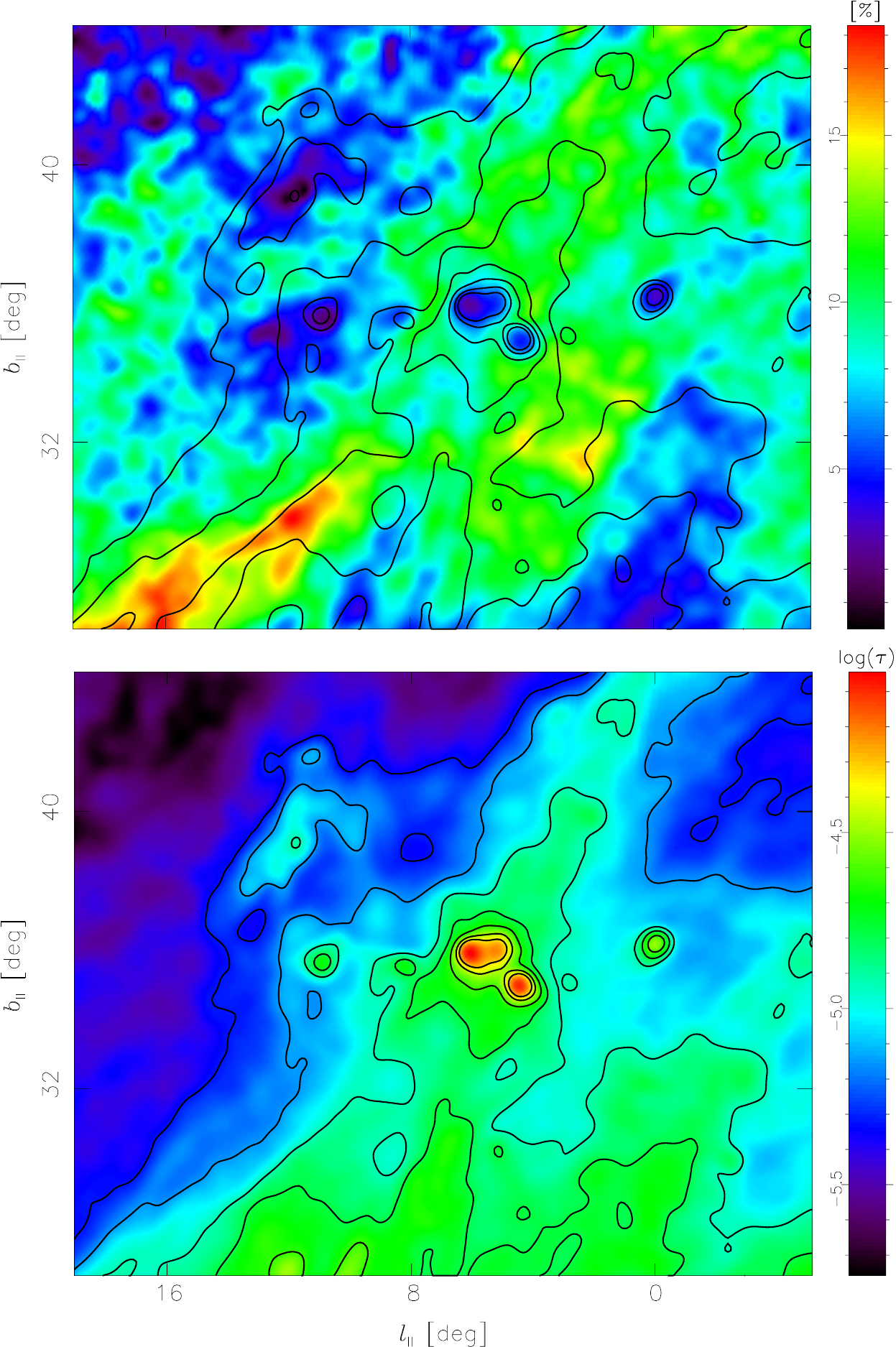}
\caption{
\emph{Top}: Map of the polarization fraction towards the dark molecular cloud
L134, overlaid with contours of the dust optical depth at 353{\GHz}. The
levels are $\dusttau=$1.4, 2.9, and $5.8\times10^{-5}$, corresponding to
$\Av=$ 0.6, 1.2, and 2.4\,mag.  \emph{Bottom}: Same for the dust
optical depth. The maps are shown at a common resolution
of $30\arcmin$.
\label{fig:p_vs_nh_L134}}
\end{center}
\end{figure}

Towards, nearby dense cores ($\nH>3 \times 10^4\nHUNIT$, size $\sim
0.1\,\mathrm{pc}$, $\NH>10^{22}\NHUNIT$) the polarization fraction is
observed to decrease systematically with $\NH$.  This effect
contributes to the sharp drop observed at $2\times10^{22}\NHUNIT$
(Fig.\,\ref{fig:p_vs_nh}, bottom panel).  Inspection of the {\Planck}
polarization map at 353{\GHz} shows many examples of such dips in
$\polfrac$ associated with nearby dense clouds.  A systematic
statistical study in the vicinity of {\Planck} cold clumps will be
presented in a forthcoming paper.  Figure\,\ref{fig:p_vs_nh_L134}
shows the example of the dark cloud L134
\citep{tucker1976,mattila1979} which is located at high Galactic
latitude in the otherwise highly polarized {\AquilaRift}. L134 is one
of the coldest Cold Clumps in the {\Planck} catalogue
\citep{planck2011-7.7b}. It is clearly seen that $\polfrac$ can be as
large as $10\,\%$ in the external regions and decreases to values as
low as $1\,\%$ at the column density peak.  This behaviour appears to
be common in the high-latitude sky and confirms previous studies.
Such a decrease of the polarization fraction towards large column
densities on cloud scales was reported previously in ground-based
measurements of polarization both in emission
\citep{WardThompson2000,MatthewsWilson2000} and extinction
\citep[e.g.,][]{GerakinesWhittet1995,Whittet2008}.  This is usually
interpreted as being due to a gradual loss of alignment of dust grains
in dense shielded regions.  In the likely hypothesis that dust
alignment processes involve UV and visible photons spinning up the
grains \citep{DraineWeingartner96,HoangLazarian2008}, polarization in
externally heated clouds is expected to drop off rapidly when outside
UV radiation cannot penetrate them.  The gradual decrease of
$\polfrac$ observed in Fig.\,\ref{fig:p_vs_nh} above $\Av=1$\,mag is
roughly consistent with such a scenario.

However, the decrease of the polarization fraction with increasing
column density could also be due to fluctuations in the orientation of
the magnetic field along a long LOS, causing depolarization.  In order
to shed light on this depolarization effect, the companion paper
\cite{planck2014-XX} compares the polarization properties of the
{\Planck} dust emission with maps of polarized emission computed in
simulations of MHD turbulence.  The simulations are anisotropic to
allow for an analysis of the influence of a large-scale magnetic field
combined with a turbulent field.  The polarized dust emission is
computed using a uniform dust {\IntrinsicpName} $\pzero= 20\,\%$. A
large scatter in the polarization fraction $\polfrac$ per bin of
column density and a decrease of the maximum (and mean) values of
$\polfrac$ with $\NH$ are found in the simulated maps, similar to
those observed.  Therefore, the variation of $\polfrac$ with $\NH$
resembles that inferred from MHD simulations, even though no loss of
dust alignment effiency due to radiative transfer is included.  This
indicates that the depolarization observed towards dense isolated
clumps such as L134 is not necessarily the result of a loss of dust
alignment, but could also be due to the tangling of magnetic field
lines.

As shown in Fig.\,\ref{fig:p_vs_nh}, which displays the dependence of
$\polfrac$ on $\NH$ over the whole sky and in regions excluding the
inner Galactic plane, most lines of sight with very low $\polfrac$
values are within the inner Galactic plane.  The large gas column
densities in the inner Galaxy ($\Av>20$\,mag) arise both in massive
star forming regions (i.e., dense gas with $\nH>3 \times 10^4\nHUNIT$
for regions around 0.3 to $1\,\mathrm{pc}$), but also along long lines
of sight (say $10\,\mathrm{kpc}$) sampling mostly low density gas in
the Molecular Ring.  We argue that the contribution from such star
forming regions in the inner Galaxy is small in the {\Planck} maps at
a resolution of $1\degr$, because such regions have angular sizes
smaller than $1\arcmin$ if they are located further than 2\,kpc from
the Sun.  The tail of high column densities in the inner Galaxy is
therefore mostly due to long lines of sight sampling low density gas.

\begin{figure}[!h!t]
\begin{center}
\includegraphics[width=9cm,angle=0]{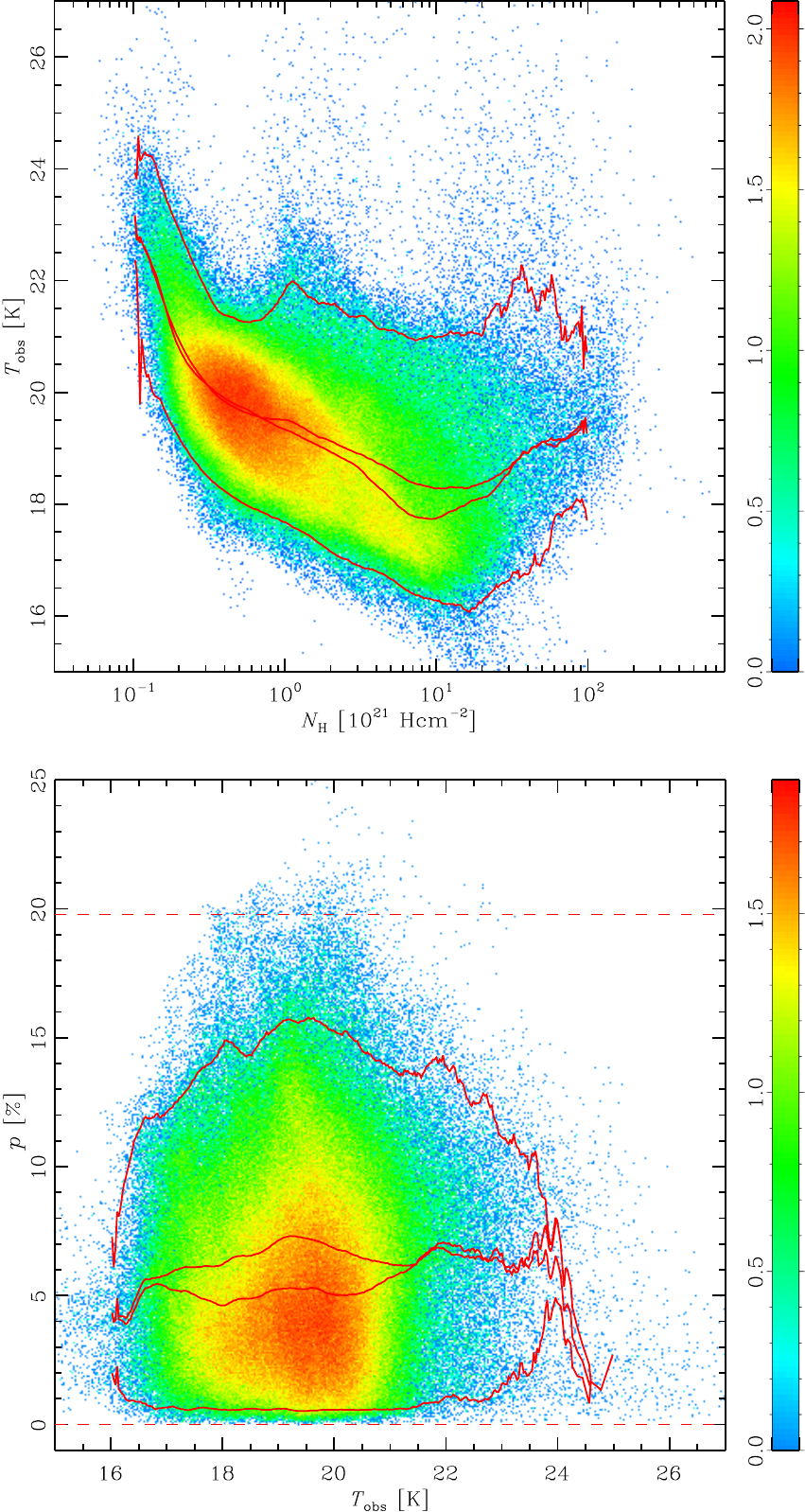}
\caption{
\emph{Upper}: Distribution of the apparent dust temperature ($\Td$)
and column density, as derived in \cite{planck2013-p06b}.
\emph{Lower}: Distribution of the polarization fraction ($\polfrac$)
as a function of $\Td$ in regions of the sky excluding the inner
Galactic plane (excluding $\glon<90\deg$ or $\glon>270\deg$,
$|\glat|<2\degr$).  Both plots are for pixels not masked in
Fig.\,\ref{fig:rawpolarmaps}.  The colour scale shows the pixel
density on a log$_{10}$ scale.  The curves show, from top to bottom,
the evolution of the upper $\nperc$ percentile, mean, median and lower
$\nperc$ percentile of $\polfrac$.  Horizontal dashed lines show the
location of $\polfrac=0$ and $\pmax=\pmaxvalue\,\%$.
\label{fig:delta_phi_vs_T}}
\end{center}
\end{figure}

For lines of sight towards the inner Galactic plane, the question is
whether they are probing a dense cold medium, shielded from the
ambient UV field, or if they result from the accumulation of low
density material distributed over large distances. The apparent dust
temperature can in principle be used to discriminate between these two
situations.  Figure\,\ref{fig:delta_phi_vs_T} shows the distribution
of the apparent dust temperature ($\Td$), as derived from the dust SED
fitting in \cite{planck2013-p06b} using a modified grey-body fit, as a
function of column density.  As discussed in \cite{planck2013-p06b}
the apparent dust temperature generally decreases with increasing
column density, up to $\NH\simeq 10^{22}\NHUNIT$.  The figure shows
that, at higher column densities, $\Td$ increases again with $\NH$.
The bulk of the large column densities above about $3 \times
10^{22}\NHUNIT$ therefore probe material in which dust is warmer than
in the cold shielded cores, because it resides either in the low
density medium, weakly shielded from the UV field of the inner Galaxy,
or close to star-forming regions. In this case, the observed decrease
of $\polfrac$ is unlikely to be due to radiative transfer effects
alone. This is taken as additional evidence that the structure of the
magnetic field could be the main interpretation for the apparent
decrease of $\polfrac$ with column density.

\subsection{Polarization fraction vs angle dispersion function}
\label{sec:polfrac_vs_deltaphi}

Figure\,\ref{fig:dphi_map} shows the distribution of
$\DeltaAng$ computed as described in Sect.\,\ref{sec:polstruct} from
the full survey at 353{\GHz} for $1^\circ$ resolution and with a $\lag=30\arcmin$
lag used in the analysis.

The map of $\DeltaAng$ exhibits a wide range of values.
A striking feature of the map is the existence of confined regions 
of high $\DeltaAng$ values, often reaching $50^\circ$ to $70^\circ$, 
which are organized in an intricate network of {\FilamentaryStructures}, 
some of which span more than $30^\circ$ in length. 
Figure\,\ref{fig:zoomed_dphi_maps} shows maps of selected regions around
some of these high $\DeltaAng$ regions.  Inspection of the
polarization maps shows that these {\Spaghettis} generally lie at the boundary 
between regions with uniform, but different, magnetic field
orientations on the sky.

Maps computed at larger lags look similar to those shown in
Fig.\,\ref{fig:dphi_map}, although with wider {\Spaghettis}, due to the
larger scale of the analysis. Maps computed at smaller lags show
{\Spaghettis} at the same locations as in Fig.\,\ref{fig:dphi_map}, which
indicates that the structures are in general unresolved. We also
derived maps of $\DeltaAng$ at higher resolution.
However, the noise and bias on $\DeltaAng$ increase quickly at higher resolution,
which makes it impossible to follow the structure of the {\Spaghettis} down to
the full {\Planck} resolution of $5^{\prime}$ in most regions of the sky.

\begin{figure}[!h!t]
\begin{center}
\includegraphics[width=9cm,angle=0]{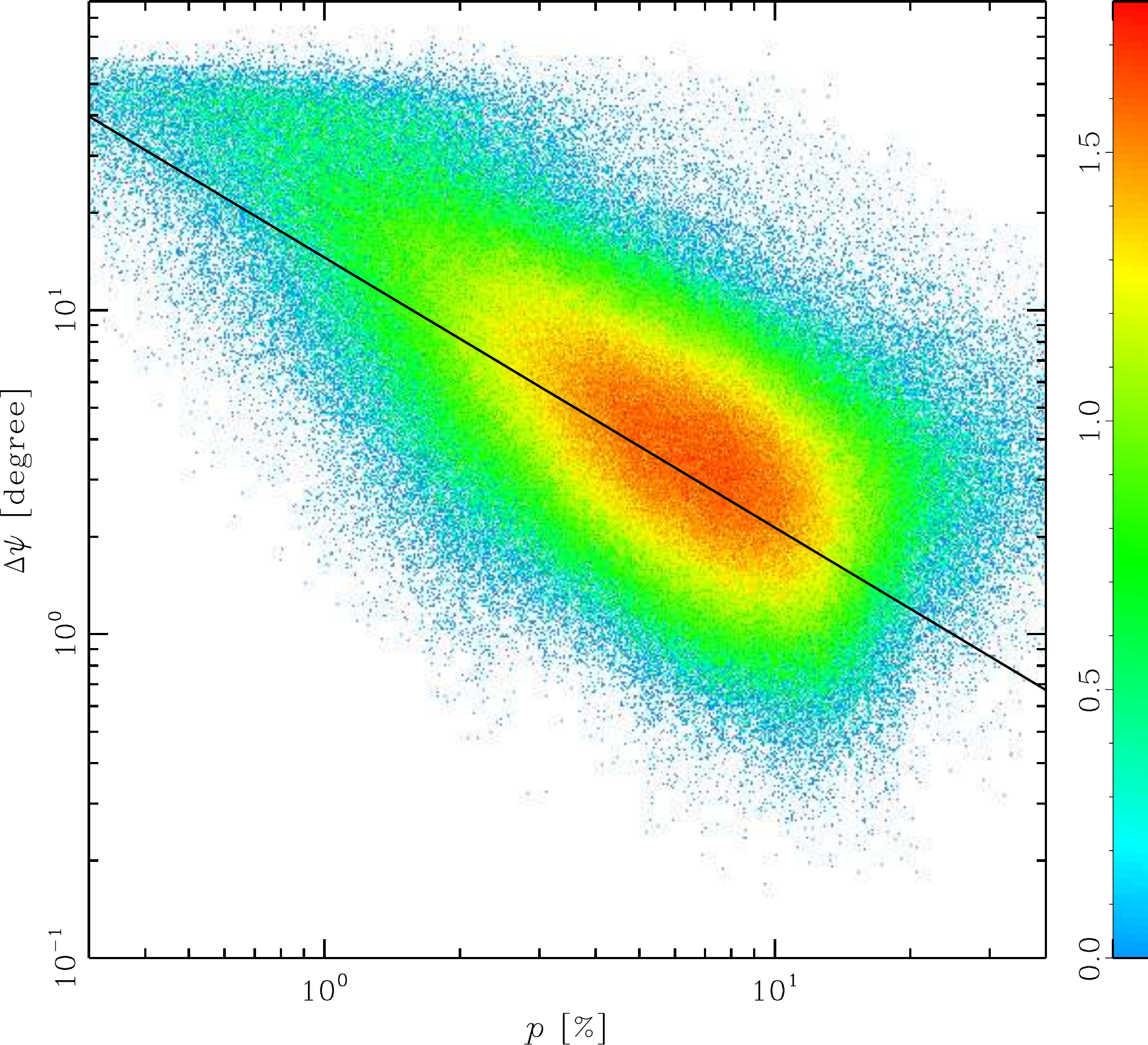}
\caption{
Scatter plot of the polarization {\DeltaAngName} $\DeltaAng$ 
as a function of polarization fraction $\polfrac$ at 353{\GHz}. 
The colour scale shows the pixel density on a log$_{10}$ scale.
The line indicates the best fit (see text). 
\label{fig:delta_phi_vs_p}}
\end{center}
\end{figure}

Comparison with the observed polarization fraction map of
Fig.\,\ref{fig:polar_p_and_sigp} on large scales clearly shows that,
overall, the {\Spaghettis} of high $\DeltaAng$ correspond to low
values of $\polfrac$. A similar trend was observed previously in the
OMC-2/3 molecular clouds regions by \cite{Poidevin2010}, using 14"
resolution polarimetry data at 353{\GHz}. The {\Planck} large-scale
maps show that this is a general trend, as confirmed by the plot in
Fig.\,\ref{fig:delta_phi_vs_p}, which shows that $\polfrac$ and
$\DeltaAng$ are approximately linearly anti-correlated in a log-log
plot. Low $\polfrac$ regions often correspond to regions where the
observed polarization direction $\polang$ changes.  This result is in
line with the findings of the previous section and further supports
the fact that variations in the magnetic field orientation play an
important role in lowering the observed polarization fraction.  The
best-fit correlation shown in Fig.\,\ref{fig:delta_phi_vs_p} is given
by
\begin{equation} {\log}_{10}(\DeltaAng)=\alphacorrel \times
{\log}_{10}(\polfrac)+\betacorrel,
\label{equ:delta_phi_vs_p_correlation}
\end{equation} with $\alphacorrel=\alphacorrelvalue$ and $\betacorrel=\betacorrelvalue$.

The above results are compared with those inferred from MHD
simulations in \cite{planck2014-XX}.  The simulations clearly show an
anti-correlation between $\DeltaAng$ and $\polfrac$, with a slope
similar to that observed in the data. It is worth noting that in the
noiseless simulations, the observed trend cannot be produced by
the bias on $\DeltaAng$ resulting from higher uncertainties in
polarization angles in regions of low signal and/or polarization
fraction. It clearly results from averaging effects of the
polarization angle within the beam and along the LOS. 
In brief, field line tangling weakens $\polfrac$, especially when the
large-scale field tends to be aligned with the LOS.

The regions of large $\DeltaAng$ bear some resemblance to the
so-called ``depolarization canals'' \cite[e.g.,][]{haverkorn&kd_00},
or more generally the regions of high polarization gradient
\citep{Gaensler2011}, detected in maps of radio polarized emission
from the warm ionized medium (WIM).  However, the two types of
features have different origins.  As explained earlier, the
{\Spaghettis} of large $\DeltaAng$ are generally associated with
discontinuities (at the resolution of the observations) in the
magnetic field orientation within dust-emitting regions.  In contrast,
the radio depolarization canals arise from Faraday rotation effects:
they are thought to be due to either differential Faraday rotation
(and hence depolarization) within synchrotron emission regions or
discontinuities in foreground Faraday rotation screens
\cite[e.g.,][]{fletcher&s_07}.  The first class of depolarization
canals do not correspond to true physical structures, and their
observed positions in the sky vary with radio frequency.  The second
class of depolarization canals, and more generally the regions of high
radio polarization gradient, are somewhat similar to our {\Spaghettis}
of large $\DeltaAng$, insofar as both can be traced back to physical
discontinuities.  But because the physical quantities that undergo a
discontinuity (free-electron density and LOS field component for the
former {\it versus} sky-projected field orientation for the latter) as
well as the places where the discontinuities occur (foreground Faraday
rotation screens for the former {\it versus} dust-emitting regions for
the latter) are unrelated, one does not expect any one-to-one
correspondence.

\subsection{Dust vs synchrotron polarization}
\label{sec:SYNCHROTRONcomp}
\begin{table}[tmb]
\begingroup
\newdimen\tblskip \tblskip=5pt
\caption{
Slope, intercepts and Pearson correlation coefficients
of the correlation between dust and sychrotron polarization fraction,
computed over Galactic quadrants in the MW plane and off
the plane.
}
\label{tab:dust_vs_synch} 
\nointerlineskip
\vskip -3mm
\footnotesize
\setbox\tablebox=\vbox{
   \newdimen\digitwidth 
   \setbox0=\hbox{\rm 0} 
   \digitwidth=\wd0 
   \catcode`*=\active 
   \def*{\kern\digitwidth}
   \newdimen\signwidth 
   \setbox0=\hbox{+} 
   \signwidth=\wd0 
   \catcode`!=\active 
   \def!{\kern\signwidth}
%
\tabskip=0pt
\halign{
\hbox to 0.40in{#\leaderfil}\tabskip 1.0em&    
\tabskip=0.6em 
\hfil#\hfil&
\hfil#\hfil&
\hfil#\hfil&
\hfil#\hfil&
\hfil#\hfil&
\hfil#\hfil\tabskip=0pt\cr
\noalign{\doubleline}
\omit  Quadrant \hfil&\multispan3\hfil$|\glat|<5\degr$\hfil&\multispan3\hfil$|\glat|>5\degr$\hfil\cr
\multispan7\hrulefill\cr  
\omit  \hfil&
\omit Slope\hfil&
\omit Intercept\hfil&
\omit Pearson\hfil&
\omit Slope\hfil&
\omit Intercept\hfil&
\omit Pearson\hfil\cr
Q1 & 0.310& $-0.551$&  0.341& 0.280& $-0.548$&  0.288\cr
Q2 & 0.355& $-0.379$&  0.470& 0.144& $-0.687$&  0.155\cr
Q3 & 0.229& $-0.646$&  0.300& 0.101& $-0.679$&  0.091\cr
Q4 & 0.135& $-0.835$&  0.170& 0.053& $-0.818$&  0.058\cr
All & 0.346& $-0.462$&  0.469&  0.137& $-0.704$&  0.144\cr
\noalign{\vskip 5pt\hrule\vskip 3pt}}}
\endPlancktable                    
\endgroup
\end{table}                        

\begin{figure*}[!h]
\begin{center}
\includegraphics[width=18cm,angle=0]{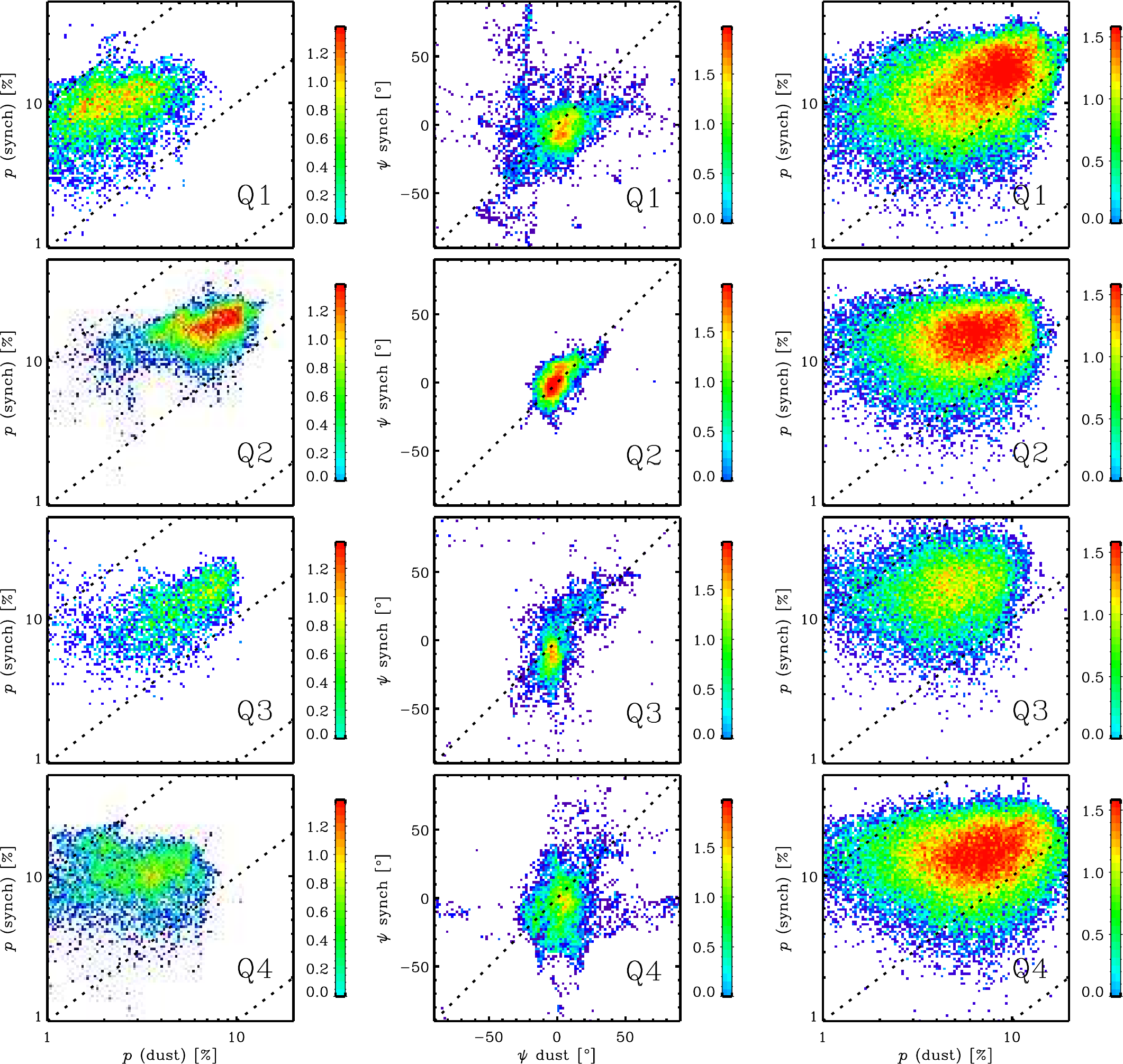}
\caption{
Comparison of dust and synchrotron polarization fraction
and polarization angle for
$|\glat|<5\degr$ (\emph{left} panels)  and off of the plane for $|\glat|>5\degr$
(\emph{right} panels), separated in the four Galactic quadrants
(\emph{top} to \emph{bottom}). The colour scale shows the pixel density on a log$_{10}$ scale.
\label{fig:p_pa_dust_vs_synch_quad}}
\end{center}
\end{figure*}

\begin{figure*}[!h]
\begin{center}
\includegraphics[width=18cm,angle=0]{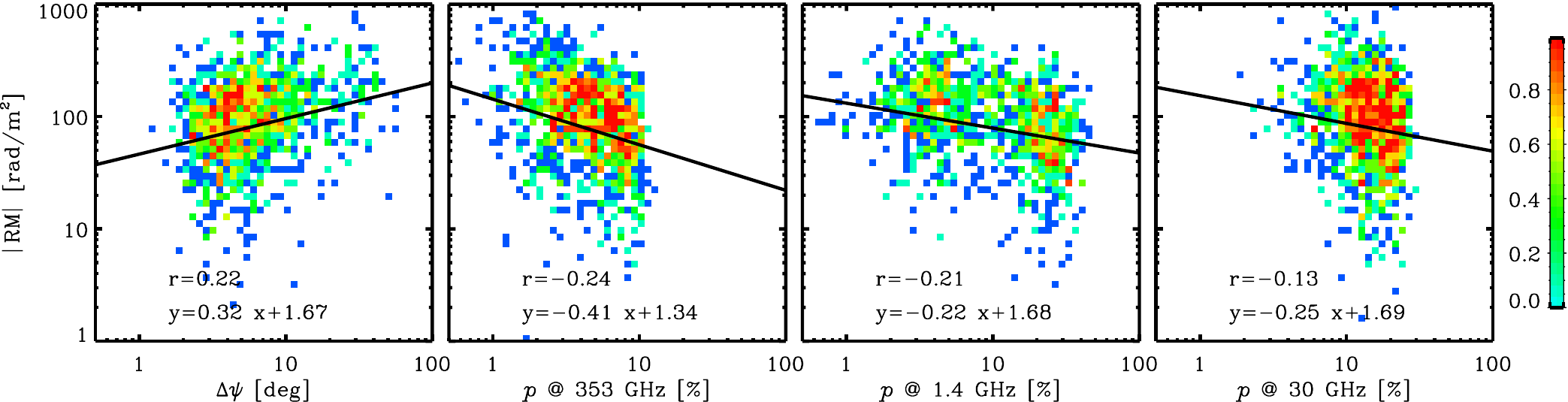}
\caption{
Faraday RMs in the Galactic plane ($|\glat| < 5\degr$) compared to
(\emph{left} to \emph{right}): dust angle dispersion; dust
polarization fraction at 353{\GHz}; synchrotron polarization fraction
at 1.4{\GHz}; and synchrotron polarization fraction at 30{\GHz}.  The
overplotted lines show the result of a simple linear fit between the
two datasets, with the Pearson's $r$ giving the measure of the degree
of correlation. The colour scale shows the pixel density on a log$_{10}$ scale.
\label{fig:dust_vs_RM}}
\end{center}
\end{figure*}

In this section we compare the dust with radio polarization data.  Our
aim is to test how much the complementary observables trace the same
magnetic fields and how their polarization properties are affected by
the irregular component of the field.  These comparisons tell us not
only about the fields but also about the relative distributions of
dust grains and relativistic electrons.

The synchrotron and dust emission are both linearly polarized
perpendicular to the local sky-projected magnetic field.  The
emissivities, however, have different dependencies on the magnetic
field strength: the dust emission does not depend on the field
strength, whereas the synchrotron emissivity is given by ${\cal
E}_{\rm syn} \propto n_{\rm e} \ B_\perp^{(\gamma + 1)/2}$, where
$n_{\rm e}$ is the density of relativistic electrons and $\gamma$ is
the power-law index of the relativistic-electron energy spectrum
(typically $\gamma \simeq 3$, so that ${\cal E}_{\rm syn} \propto
n_{\rm e} \ B_\perp^2$).  One therefore has to be careful in
interpreting the comparison between these observables.  Some
differences arise when the magnetic field in dust clouds differs from
that in the diffuse synchrotron-emitting medium.  Other differences
arise because of the emissivity dependence on the field strength that
weights the emission differently along the LOS.  Any single direction
may have a combination of these effects.  We would therefore expect to
see globally similar polarization structures where the particles
sample the same average field, but not identical structures.  A
correlation analysis between dust and synchrotron polarization is also
reported by \citet{planck2014-XXII}. Their cross-correlation between
the $\StokesQ$ and $\StokesU$ maps at {\wmap} and {\lfi} frequencies
with the corresponding maps at $353\,$GHz shows that some of the
polarized synchrotron emission at intermediate Galactic latitudes is
correlated with dust polarization in both intensity and angle.  We
might further expect to see statistical correlations even where the
irregular component perturbs the large-scale magnetic field, but the
degree of the correlation is complicated to predict.

The data sets are described in Sect.\,\ref{sec:SYNCHROTRONdata}.
Figure\,\ref{fig:p_pa_dust_vs_synch_quad} compares both the
polarization fraction and the polarization angle of the dust emission
at 353{\GHz} (the Bayesian estimates where the {\SNR} on $p$ is
greater than 3) with polarized synchrotron emission at 30{\GHz}.  The
comparison between the polarization angles is straightforward, as
synchrotron is dominant and there is no Faraday rotation at these
frequencies.  The comparison between the polarization fractions is
more complex, however, because in the microwave and radio data there
are additional total intensity components, such as free-free and
anomalous microwave emission.  To avoid contamination from anomalous
microwave emission, we use the 408\,MHz map of \citet{haslam:1982} for
synchrotron total intensity.  We correct for free-free emission as
described in Sect.\,\ref{sec:SYNCHROTRONdata}.  This correction is
approximate, but the synchrotron component dominates at low
frequencies.  We extrapolate the 408~MHz intensity to 30{\GHz} in
order to construct the polarization fraction, assuming a spectral
index of $-2.88$ \citep[see, e.g.,][]{Jaffe2011}.  Note that a change
in this constant index will simply shift the synchrotron polarization
fraction systematically up or down and not affect any observed
correlation.  Spatial variations in the index that are not accounted
for, however, remain a limitation of this simple method.

The left two columns of Fig.\,\ref{fig:p_pa_dust_vs_synch_quad} show
the Galactic plane ($|\glat|<5\degr$), while the right two show the
results for the rest of the sky.  The correlations are quantified by
linear fits and Pearson's correlations coefficients listed in
Table\,\ref{tab:dust_vs_synch}.  In all but the fourth quadrant, there
is clear correlation in the polarization fraction in the plane, where
the polarized intensity is strong.  The polarization angles remain
near zero, i.e., the magnetic field remains largely parallel to the
plane.  This confirms that at the largest scales probed through the
full disk, the synchrotron and dust sample roughly the same average
magnetic fields.  With a few notable exceptions, however, there is
little correlation away from the plane, where isolated local
structures and the irregular field component become more important.

The so-called Fan region in the second quadrant is one where the
comparison is most interesting, showing a relatively strong
correlation ($r=0.47$) in polarization fraction in the plane, as does
the third quadrant to a lesser degree.  Out of the Galactic plane, the
correlation in $p$ disappears.  But we still see correlation in the
polarization angles off the plane, where they remain concentrated
around zero, i.e., perpendicular to the plane, indicating that the
magnetic field is parallel to the plane even at latitudes above
$5\degr$.  The second interesting region for the comparison is the
first quadrant, where the sky is dominated by the radio loop I, i.e.,
the North Polar Spur (NPS).  Here the high-latitude polarization angles show
correlation where the two observables clearly trace the same magnetic
fields.

We also compare the dust polarization angle dispersion with the
polarized synchrotron emission at 1.4{\GHz} where it is subject to
significant Faraday rotation effects.  In Fig.\,1 of
\citet{burigana:2006}, the polarization fraction shows strong
depolarization of the synchrotron emission within $30\degr$ of the
plane, with the exception of the Fan region in the second quadrant.
Much of the depolarization is so-called ``beam depolarization''.  Even
a coherent diffuse background source viewed through the roughly
$1\degr$ beam results in emission co-added along slightly different
lines of sight that pass through different turbulent cells, get
Faraday-rotated differently, and cancel out.  One might then expect
that the resulting synchrotron polarization fraction would
anti-correlate with the dust polarization angle dispersion.  Lines of
sight toward highly turbulent regions should have low synchrotron
polarization due to Faraday effects and high dust polarization angular
dispersion.  Such correlations are not generally apparent, however in
some regions, such as the second quadrant dominated by the Fan region,
we see this effect, implying that the dust and synchrotron in the Fan
are tracing some of the same turbulent magnetic fields.

Finally, it is instructive to compare the dust polarization fraction
with Faraday RMs of extragalactic sources using the catalogues of
\citet{brown03}, \citet{brown07}, \citet{taylor09}, and
\citet{vaneck11}.  We compare the RM of each source with the polarized
emission in the corresponding map pixel as shown in
Fig.\,\ref{fig:dust_vs_RM}.  Remember that RMs are proportional to the
LOS field component (which is positive/negative if the field points
towards/away from the observer) times the free-electron density and
integrated along the LOS, whereas the dust polarization fraction is an
increasing function of the inclination angle of the magnetic field to
the LOS.  Therefore, if the large-scale field is reasonably coherent
along the LOS path through the Galaxy, then a field orientation
globally close to the LOS will tend to make extragalactic-source RMs
large (in absolute values) and the dust polarization fraction small,
whereas a field orientation globally close to the plane of the sky
will do the opposite.  As a result, one might expect a rough
anti-correlation between extragalactic-source RMs and dust
polarization fraction.  However, only a very loose anti-correlation
may be expected at best, since: (1) Faraday rotation and dust emission
take place in different environments, with possibly different magnetic
field directions; (2) RMs depend not only on the field inclination to
the LOS, but also on the total field strength and on the free-electron
column density; and (3) the LOS field component could undergo
reversals, which would decrease RMs without correspondingly increasing
the dust polarization fraction.  Similarly, one might expect a rough
positive correlation between extragalactic-source RMs (again in
absolute values) and dust polarization angle dispersion, because if
the large-scale field is globally oriented closer to the LOS, the dust
polarization angle will be more sensitive to the fluctuating field.
Figure\,\ref{fig:dust_vs_RM} confirms these expected trends in the
Galactic plane, which traces the large-scale field.  Away from the
plane (not shown) where more local structures dominate, we find no
correlations.

The {\Planck} polarization data at 353{\GHz} provide a new tracer of
magnetic fields and an important complement to radio observations due
to the independent source particle distribution.  This first look at
the comparison of these observables confirms the expected large-scale
correspondence as well as interesting correlations in the Fan and NPS
regions.  We find only weak correlations over much of the sky where
the effects of local structures and the irrregular field component
dominate.  This fact is not surprising but nonetheless has important
implications.  Though it is premature to draw physical conclusions
from these comparisons, they highlight the importance of, as well as the
challenges inherent in combining these data to form a coherent picture
of the Galactic magnetic fields.

\section{Conclusions}
\label{sec:conclusions}

We have presented the {\Planck} large-scale polarization maps at 353{\GHz},
where polarization is dominated by polarized thermal dust emission
from elongated grains aligned with the magnetic field. These data
allow us for the first time to study dust polarization over large
angular scales and opens the field for many detailed studies to come.

The dust polarization fraction $\polfrac$ is observed to range
from zero to more than $15\,\%$. We derive a lower limit to the maximum polarization
fraction of $\pmax=\pmaxvalue\,\%$.  These highest polarization fractions are
observed in a handful of individual regions, usually located in
intermediate to low column density parts of the sky.

The large-scale spatial distribution of $\polfrac$ shows a modulation
that is consistent with predictions of the general magnetic field
structure of the MW, as constrained previously from synchrotron and
RM data. At smaller scales, the variations of $\polfrac$
appear to be related to variations in both the total dust column density
and the magnetic field structure.

There is a clear tendency for $\polfrac$ to decrease with total column
density.  The variations associated with column density show
that $\polfrac$ starts to drop below around $\Av= 1$\,mag and show a
sharp drop above $\Av= 10$\,mag.  This is qualitatively consistent with
the prediction of models where dust alignment results from the
interaction with light. However, tangling of the $\Bfield$ field
geometry along the LOS is also very effective at suppressing the net
polarization fraction and possibly plays the major role in most cases,
as discussed in \cite{planck2014-XX}.

The {\Planck} polarization data at 353{\GHz} also allow precise
measurements of the polarization direction $\polang$ over most of the
sky. Rotated by $90\degr$, this direction is assumed to represent the
$\Bfield$-field orientation projected on the sky.  The polarization angles in
the Galactic plane are observed to be consistent with the expectation 
that $\Bfield$ lies mostly along the plane, as strongly suggested by previous
synchrotron measurements. This is particularly true in the inner MW
and in the highly polarized {\Fan} region.

In order to characterize the field rotation, we use the polarization
{\DeltaAngName} $\DeltaAng$, which measures the dispersion in angle
rotation at a given spatial scale. $\DeltaAng$ increases with lag
distance, as previous observations have shown at smaller scales in
specific regions. Away from the Galactic plane, the sky distribution
of $\DeltaAng$ reveals a spectacular complex of unresolved {\FilamentaryStructures}
of large $\DeltaAng$ values. This is the first time such
structures have been observed for dust polarization. These {\FilamentaryStructures}
anti-correlate with $\polfrac$, in the sense that regions
with maximal rotation generally correspond to the lowest polarization
fractions. We demonstrate that, over a large fraction of the sky, this
is not due to the noise-induced bias on $\DeltaAng$ and is therefore a
real effect.

We interpret the anti-correlation between $\DeltaAng$ and $\polfrac$ 
as depolarization due to the magnetic field structure. 
This is likely produced by field rotation both in the plane of the sky 
below the resolution of the data and along the LOS.
The {\FilamentaryStructures} often appear to be separating adjacent regions 
that have different but quite homogeneous field orientations. 
When they can be followed down to the {\Planck} resolution, 
their widths are smaller than the beam. 
Some of them span large angular distances at high latitudes,
which suggests that they are local. 
The {\FilamentaryStructures} bear a resemblance to the depolarization canals 
that are observed at radio frequencies and attributed to Faraday rotation effects, 
but we argue that they have a different origin here.
The regions of high $\DeltaAng$ are also observed in MHD simulations 
\citep[see details in][]{planck2014-XX}, with the same type of anti-correlation 
with $\polfrac$ as that observed in the {\Planck} data.
This similarity provides further support for the above interpretation.

We compared the dust polarization fraction and angle with polarized
synchrotron data. There are clear indications that the two tracers
globally see the same magnetic field orientation, particularly
interesting to see in the Fan region and the North Polar Spur, but that the
detailed distributions of dust and high-energy electrons must be
different in order to explain the observed maps. We infer a loose
statistical correlation between extragalactic-source RMs and both the
dust polarization fraction $\polfrac$ and the {\DeltaAngName}
$\DeltaAng$.  However, inspection of the maps shows that there is no
systematic spatial correspondence between depolarization {\FilamentaryStructures} in
dust and synchrotron emission at small angular scales.

\begin{acknowledgements}
The development of \Planck\ has been supported by: ESA; CNES and
CNRS/INSU-IN2P3-INP (France); ASI, CNR, and INAF (Italy); NASA and DoE
(USA); STFC and UKSA (UK); CSIC, MICINN, JA, and RES (Spain); Tekes,
AoF, and CSC (Finland); DLR and MPG (Germany); CSA (Canada); DTU Space
(Denmark); SER/SSO (Switzerland); RCN (Norway); SFI (Ireland);
FCT/MCTES (Portugal); and PRACE (EU). A description of the Planck
Collaboration and a list of its members, including the technical or
scientific activities in which they have been involved, can be found
at \url{http://www.sciops.esa.int/index.php?project=planck&page=Planck_Collaboration}.
We acknowledge the use of the Legacy Archive for Microwave Background
Data Analysis (LAMBDA), part of the High Energy Astrophysics Science
Archive Center (HEASARC). HEASARC/LAMBDA is a service of the
Astrophysics Science Division at the NASA Goddard Space Flight Center.
Some of the results in this paper have been derived using the
{\Healpix} package.  
\end{acknowledgements}

\bibliographystyle{aa}

\bibliography{PIP75,Planck_bib,Planck_bib_4polar_25apr.bib}

\appendix

\section{Noise estimates for  {\Planck} smoothed maps}
\label{sec:noise}

Here, we show how to smooth polarization maps 
and derive the covariance matrix associated to the smoothed maps.

\subsection{Analytical expressions for smoothing maps of the Stokes parameters and noise covariance matrices} \label{sec:SmoothCovar}

Smoothing intensity maps is straightforward, but this is not the case
for polarization maps. In principle, as the polarization frame rotates
from one centre pixel to a neighbouring one that will be included in
the smoothing, the $(\StokesQ,\StokesU)$ doublet must be also rotated
at the same time \citep[\eg][]{Keegstra1997}.  The issue is similar
for evaluating the effects of smoothing on the $3\times3$ noise
covariance matrix, though with mathematically distinct results.
In this Appendix, we present an exact analytical solution to the local
smoothing of maps of the Stokes $\StokesI$, $\StokesQ$, and
$\StokesU$, as well as the effects of smoothing on their
corresponding noise covariance matrix.

\begin{figure}
\center
\includegraphics[width=8cm,angle=0]{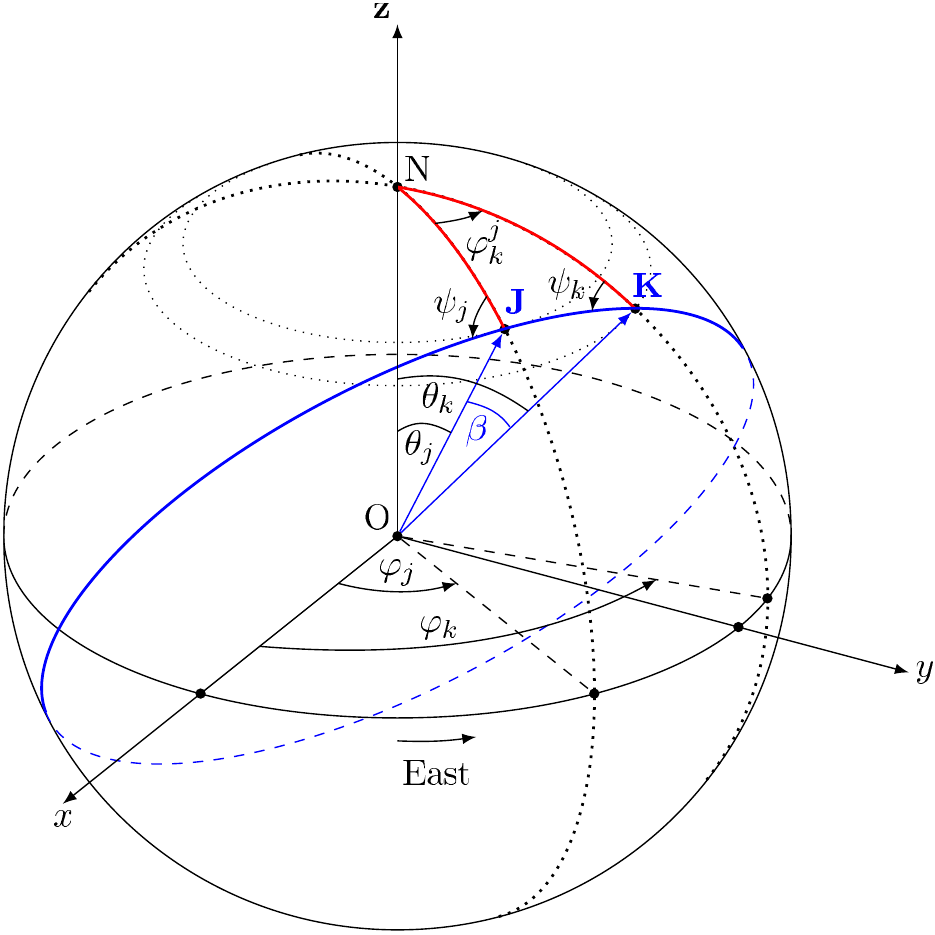}
\caption{Definition of points and angles on the sphere involved in the
geometry of the smoothing of polarization maps (adapted from
\citealp{Keegstra1997}). $\PointJ$ is the position of the centre of the smoothing beam, and
$\PointK$ a neighbouring pixel, with spherical coordinates
($\phiJ$,\,$\thetaJ$) and ($\phiK$,\,$\thetaK$),
respectively.  The great circle passing through $\PointJ$ and
$\PointK$ is shown in blue.  The position angles $\psiJ$ and
$\psiK$ are in the \healpix\ convention, increasing from north
through west on the celestial sphere as seen by the observer at O.
}
\label{Fig-K97}
\end{figure}

\subsubsection{Smoothing of Stokes parameters}

Fig.~\ref{Fig-K97} presents the geometry of the problem. Let us
consider a \healpix\ pixel $j$ at point $\PointJ$ on the celestial
sphere, with spherical coordinates ($\phiJ$,$\,\thetaJ$).  To perform
smoothing around this position with a Gaussian beam with standard
deviation $\sigFWHM = {\rm FWHM}/2.35$ centred at the position of this
pixel we select all \healpix\ pixels that fall within 5 times the FWHM
of the smoothing beam (this footprint is sufficient for all practical
purposes).  Let $k$ be one such pixel, centred at the point $\PointK$
with coordinates ($\phiK$,$\,\thetaK$), at angular distance $\betaSK$
from $\PointJ$ defined by
\begin{equation}
\cos{\betaSK} = \cos{\thetaJ}\cos{\thetaK}+\sin{\thetaJ}\sin{\thetaK}\cos{(\phiK-\phiJ)}
 \label{Eq-cosbeta}\,. 
\end{equation}
The normalized Gaussian weight is then
\begin{equation}
\w  =  \frac{\exp^{-\left(\betaSK/\sigFWHM\right)^2/2}}{\sum_i \exp^{-\left(\betaJI/\sigFWHM\right)^2/2}} \,, \label{Eq-w}
\end{equation}
and $\sum_k \w = 1$.
Before averaging in the Gaussian beam, we need to rotate the polarization reference frame in
$\PointK$ so as align it with that in $\PointJ$.  For that the reference
frame is first rotated by $\psiK$ into the great circle running through
$\PointK$ and $\PointJ$, then translated to $\PointJ$, and finally rotated
through $-\psiJ$. The net rotation angle of the reference frame from point $\PointK$ to
point $\PointJ$ is then
\begin{equation}
\psiSK= \psiK - \psiJ\,. 
\label{Eq-psiSK}
\end{equation}
Due to the cylindrical symmetry around axis $z$, evaluating $\psiSK$
does not depend on the longitudes $\phiJ$ and
$\phiK$ taken separately, but only on their difference
\begin{equation}
\phiSK = \phiK-\phiJ\,.
\end{equation}
Using spherical trigonometry in Fig.~\ref{Fig-K97} with the \healpix\
convention for angles $\psiJ$
and $\psiK$, we find:
\begin{eqnarray}
\sin{\psiJ}  & = & \sin{\thetaK}\sin{\phiSK} \,/ \sin{\betaSK} \label{Eq-sinpsiJ} \,; \\
\sin{\psiK}  & = & \sin{\thetaJ}\sin{\phiSK} \,/ \sin{\betaSK} \label{Eq-sinpsiK} \,; \\
\cos{\psiJ}  & = & - \left(\cos{\thetaK}\sin{\thetaJ} - \cos{\thetaJ}\sin{\thetaK}\cos{\phiSK} \right) \,/ \sin{\betaSK} \label{Eq-cospsiJ} \,;\\
\cos{\psiK}  & = & ~\,\,\left(\cos{\thetaJ}\sin{\thetaK} - \cos{\thetaK}\sin{\thetaJ}\cos{\phiSK} \right) \,/ \sin{\betaSK}  \label{Eq-cospsiK} \,.
\end{eqnarray}
To derive $\psiK$ and $\psiJ$ we use the two-parameter $\arctan$ function that resolves the $\pi$ ambiguity in angles:  
\begin{equation}
\psiSK = \arctan{(\sin{\psiK}, \cos{\psiK})}-\arctan{(\sin{\psiJ}, \cos{\psiJ}}) \label{Eq-psijkarctan}\,.
\end{equation}
Because of the $\tan$ implicitly used, $\sin{\betaSK}$ (a positive
quantity) is eliminated in the evaluation of $\psiJ$, $\psiK$, and
$\psiSK$.

We can now proceed to the rotation. It is equivalent to rotate the
polarization frame at point $\PointK$ by the angle $\psiSK$, or to
rotate the data triplet ($I_k$,\,$Q_k$,\,$U_k$) at point $\PointK$ by
an angle $-2\psiSK$ around the axis $I$. The latter is done with the
rotation matrix \cite[\eg][]{Tegmark2001}
\begin{equation} 
\MatRpsi =  \left[ \begin{array}{ccc}
1 & 0 & 0 \\
0 & \cos{2\psiSK} & \sin{2\psiSK} \\
0 & -\sin{2\psiSK} & \cos{2\psiSK} \end{array} \right]\,.
\label{RotMatrix}
\end{equation} 
Finally, the smoothed $\StokesI$, $\StokesQ$, and $\StokesU$ maps are calculated by:
\begin{eqnarray}
\left( \begin{array}{c}
I \\ Q \\ U
 \end{array} \right)_\star
 & = & \sum_k \w \
\MatRpsi \ 
\left( \begin{array}{c}
I \\ Q \\ U
 \end{array} \right)_k \label{eq:smoothIQU}\,.
\end{eqnarray}

\subsubsection{
Computing the noise covariance matrix for smoothed polarization maps
} 
\label{sec:var_smoothing}

We want to compute the noise covariance matrix $\MatS$ at the position
of a \healpix\ pixel $j$ for the smoothed polarization maps, given the
noise covariance matrix $\MatC$ at the higher resolution of the
original data.  We will assume that the noise in different pixels is
uncorrelated.
From the given covariance matrix $\MatC_k$ at any pixel $k$ we can
produce random realizations of Gaussian noise through the Cholesky
decomposition of the covariance matrix:
\begin{eqnarray}
\MatC_k & = & \MatL_k \times \MatL_k^\Trans \, \label{eq:chol_def},\\ 
\VectN_k & = & \MatL_k \times \VectG_k \,,
\end{eqnarray}
where in the decomposition $\MatL_k^\Trans$ is the transpose of the
matrix $\MatL_k$ and $\VectG_k = (G_I, G_Q, G_U)_k$ is a vector of
normal Gaussian variables for $I$, $Q$, and $U$.

Applying Eq.~\ref{eq:smoothIQU} to the Gaussian noise realization, we
obtain
\begin{eqnarray}
\VectN_\star 
 & =  & \sum_k \w \,\MatRpsi \ 
\left( \begin{array}{c}
N_I \\ N_Q \\ N_U
 \end{array} \right)_k = \sum_k \w \, \MatRpsi \,\VectN_k \,.
\end{eqnarray}
The \ifdefined\CHECK \reply{ \fi
covariance matrix of the smoothed maps
\ifdefined\CHECK } \fi
at the position $\PointJ$ is given by
\begin{eqnarray}
\MatS & =  &\left< \VectN_\star \,\VectN_\star^\Trans \right> \nonumber \\
& =  &\left < \sum_k \w \,\MatRpsi \,\,\MatL_k \, \VectG_k \sum_i \wi  \,\VectG^\Trans_i \, \MatL^\Trans_i \, \MatRpsiji^\Trans \right >  \nonumber\\
& = & \sum_{k,i}  \w \,\MatRpsi \,\MatL_k \, \left < \VectG_k \,  \VectG^\Trans_i \right > \, \wi  \,\MatL^\Trans_i \, \MatRpsiji^\Trans  \,.
\end{eqnarray}
\normalsize

If the noise in distinct pixels is independent, as assumed, then
$\langle \VectG_k\,\VectG_i \rangle = \delta_{ki}$, the Kronecker
symbol, and so
\begin{eqnarray}
\MatS & =  
& \sum_k \w^2 \,\MatRpsi \, \MatC_k \, \MatRpsi^\Trans\,,
 \end{eqnarray}
which can be computed easily with Eq.~\ref{RotMatrix}.

Developing each term of the matrix, we can see more explicitly how the
smoothing mixes the different elements\footnote{For example, {$\sII$
is the first element of matrix $\MatS$ which is being evaluated at the
pixel centered on $\rm J$.}}  of the noise covariance matrix:
\begin{eqnarray}
\sII & = & \sum_k \w^2 \, \CII \label{Eq-sII} \\
\sIQ & = & \sum_k \w^2 \, \left(\co\,\CIQ + \si\,\CIU\right) \label{Eq-IQ}\\
\sIU & = & \sum_k \w^2 \, \left(-\si\,\CIQ + \co\,\CIU\right)\\
\sQQ & = & \sum_k \w^2 \left(\coco\,\CQQ + 2\,\co\si\,\CQU + \sisi\,\CUU \right)\\
\sQU & = & \sum_k \w^2 \,\left(\left(\coco-\sisi\right)\CQU + \co\si\left(\CUU-\CQQ\right)\right)\label{Eq-QU}\\
\sUU & = & \sum_k \w^2 \left(\sisi\,\CQQ - 2\,\co\si\,\CQU + \coco\,\CUU \right) \label{Eq-UU} \,,
\end{eqnarray}
\normalsize
where we note that $\co = \cos{2\psiSK}$ and $\si=\sin{2\psiSK}$
depend on $j$ and $k$.  The mixing of the different elements of the
covariance matrix during the smoothing is due not to the smoothing
itself, but to the rotation of the polarization frame within the
smoothing beam.


\subsubsection{Smoothing of the noise covariance matrix with a Monte Carlo approach}

For the purpose of this paper, we obtained smoothed covariance matrices using a
Monte Carlo approach.

We first generate correlated noise maps ($n_{\rm l}$, $n_{\rm Q}$, $n_{\rm U}$) on
$\StokesI$, $\StokesQ$, and $\StokesU$ at the resolution of the data using
\begin{equation}
\left(\begin{array}{l}
n_{\rm l} \\
n_{\rm Q} \\
n_{\rm U}
\end{array}\right)
=
\left(\begin{array}{lll}
L_{11} & 0       & 0 \\
L_{12} & L_{22} & 0 \\
L_{13} & L_{23} & L_{33}
\end{array}\right)
\times
\left(\begin{array}{l}
G_{\rm l} \\
G_{\rm Q} \\
G_{\rm U}
\end{array}\right)\,.
\label{eq:noise_choldc}
\end{equation}
where $G_{\rm l}$, $G_{\rm Q}$, and $G_{\rm U}$ are Gaussian normalized random
vectors and $L$ is the Cholesky decomposition of the covariance matrix $\MatC$ defined in Eq. \ref{eq:chol_def}

The above noise $\StokesI$, $\StokesQ$, an $\StokesU$ maps are then
smoothed to the requested resolution using the {\tt{smoothing}}
procedure of the {\Healpix} package. The noise maps are further
resampled using the {\tt{udgrade}} procedure of the {\Healpix}
package, so that pixelization respects the Shannon theorem for the
desired resolution.  The smoothed covariance matrices for each sky pixel
are then derived from the statistics of the smoothed noise maps. The
Monte Carlo simulations have been performed using 1000 realizations.\\

Both the analytical and the Monte Carlo approaches have been estimated
on the {\Planck} data and shown to give equivalent results.

\section{Debiasing methods}
\label{sec:debiasing}

Since $\polfrac$ is a quadratic function of the observed Stokes
parameters  (see Eq. \,\ref{equ:pem}) it is affected by a positive bias 
in the presence of noise. The bias becomes dominant at low {\SNR}.
Below, we describe a few of the techniques that have been
investigated in order to correct for this bias.

\subsection{Conventional method (method 1)}
\label{sec:methodOne}

This method is the {\classical} determination \citep[see][for a
summary]{planck2014-XXIII98} often used on extinction polarization data.
It uses the internal variances provided with the {\Planck} data, which
includes the white noise estimate on the intensity ($\sigII$) as well
as on the $\StokesQ$ and $\StokesU$ Stokes parameters ($\sigQQ$ and $\sigUU$) and the
off-diagonal terms of the noise covariance matrix
($\sigIQ$,$\sigIU$,$\sigQU$).

The debiased $\polfrac^2$ values are computed using
\begin{equation}
\polfrac^2_{\db}=\polfrac^2_{\rm obs}-\sigma_p^2\,,
\end{equation}
where $\sigma_p^2$ is the variance of $\polfrac$ computed from the
observed Stokes parameters and the associated variances as follows:
\begin{eqnarray}
\sigma_p^2 &=&\frac{1}{\polfrac^2 \StokesI_{obs}^4} \times
(\StokesQ_{\obs}^2\sigQQ+\StokesU_{\obs}^2\sigUU+\frac{\sigII}{\StokesI_{\obs}^2}
\times (\StokesQ_{\obs}^2+\StokesU_{\obs}^2)^2      \nonumber \\
   && +2 \StokesQ_{\obs}U_{\obs}\sigQU \nonumber \\
   && -2\StokesQ_{\obs}\frac{(\StokesQ_{\obs}^2+\StokesU_{\obs}^2)}{\StokesI_{\obs}}\sigIQ-2\StokesU_{\obs}\frac{(\StokesQ_{\obs}^2+\StokesU_{\obs}^2)}{\StokesI_{\obs}}\sigIU\,.
\label{equ:pclassic}
\end{eqnarray}

The uncertainty on $\polang$ is given by
\begin{equation}
\sigma_{\polang}= 28.65^{\circ}\times
\sqrt{\dfrac{\StokesQ^{2}\sigUU+\StokesU^{2}\sigQQ-2\StokesQ\StokesU\sigQU}{\StokesQ^{2}\sigQQ+\StokesU^{2}\sigUU+2\StokesQ\StokesU\sigQU}}
\times\sigma_{\polint}\,,
\label{equ:methodTwothetaRMS_polint}
\end{equation}
where $\sigma_{\polint}$ is the uncertainty on the polarized intensity:
\begin{equation}
\sigma_{\polint} = \dfrac{1}{\polint^2}(\StokesQ_{\obs}^2\sigQQ+\StokesU_{\obs}^2\sigUU+2\StokesQ_{\obs} \StokesU_{\obs} \sigQU)\,.
\label{equ:methodTwopRMS}
\end{equation}

In the case where $I$ is supposed to be perfectly known, $\sigII = \sigIQ = \sigIU = 0$,
\begin{equation}
\sigma_{\polang}= 28.65^{\circ}\times
\sqrt{\dfrac{\StokesQ_{\obs}^{2}\sigUU+\StokesU_{\obs}^{2}\sigQQ-2\StokesQ_{\obs}\StokesU_{\obs}\sigQU}{\StokesQ_{\obs}^{2}\sigQQ+\StokesU_{\obs}^{2}\sigUU+2\StokesQ_{\obs}\StokesU_{\obs}\sigQU}}
\times\dfrac{\sigma_{p}}{\polfrac}\,.
\label{equ:methodTwothetaRMS}
\end{equation}

It is to be noted that, since it is based on derivatives around the
true value of the $\StokesI$, $\StokesQ$, and $\StokesU$ parameters, it is only valid in the high
{\SNR} regime. The {\classical} values of uncertainties derived above are compared to
the ones obtained using the Bayesian approach in Fig.\,\ref{fig:classical_vs_ml}.

\subsection{Time cross-product method (method 2)}
\label{sec:methodTwo}

This method consists in computing cross products between estimates of $\StokesQ$
and $\StokesU$ with independent noise properties. In the case of
{\Planck} HFI, each
sky pixel has been observed at least four times and the four independent
surveys can be used for this purpose. Another option is to use
half-ring maps which have been produced from the first and second halves of
each ring. These methods have the disadvantage of using only part of the
data, but the advantage of efficiently debiasing the data if the noise is effectively
independent, without assumptions about the $\StokesQ$ and $\StokesU$
uncertainties.

In that case, $\polfrac^2_{\db}$ can be computed as
\begin{equation}
\polfrac^2_{\db}=\frac{\sum_{i>j} \StokesQ_i \StokesQ_j + \StokesU_i \StokesU_j}{\sum_{i>j} \StokesI_i \StokesI_j}\,,
\label{equ:methodTwopRMS}
\end{equation}
where the sum is carried out either over independent survey maps or half-ring
maps.

The uncertainty of $\polfrac^2$ can in turn be evaluated from the dispersion
between pairs
\begin{equation}
\sigma^2 (\polfrac^2_{\db})=\frac{ \sigma^2(\StokesQ^2)+ \sigma^2(\StokesU^2) + (\StokesQ^2+\StokesU^2)/\StokesI^2\sigma^2(\StokesI^2)}{\StokesI^4}\,.
\label{equ:methodTwothetaRMS}
\end{equation}

\subsection{Bayesian methods (method 3)}
\label{sec:Bayesian}

\begin{figure*}[!h!t]
\begin{center}
\includegraphics[width=9cm,angle=0]{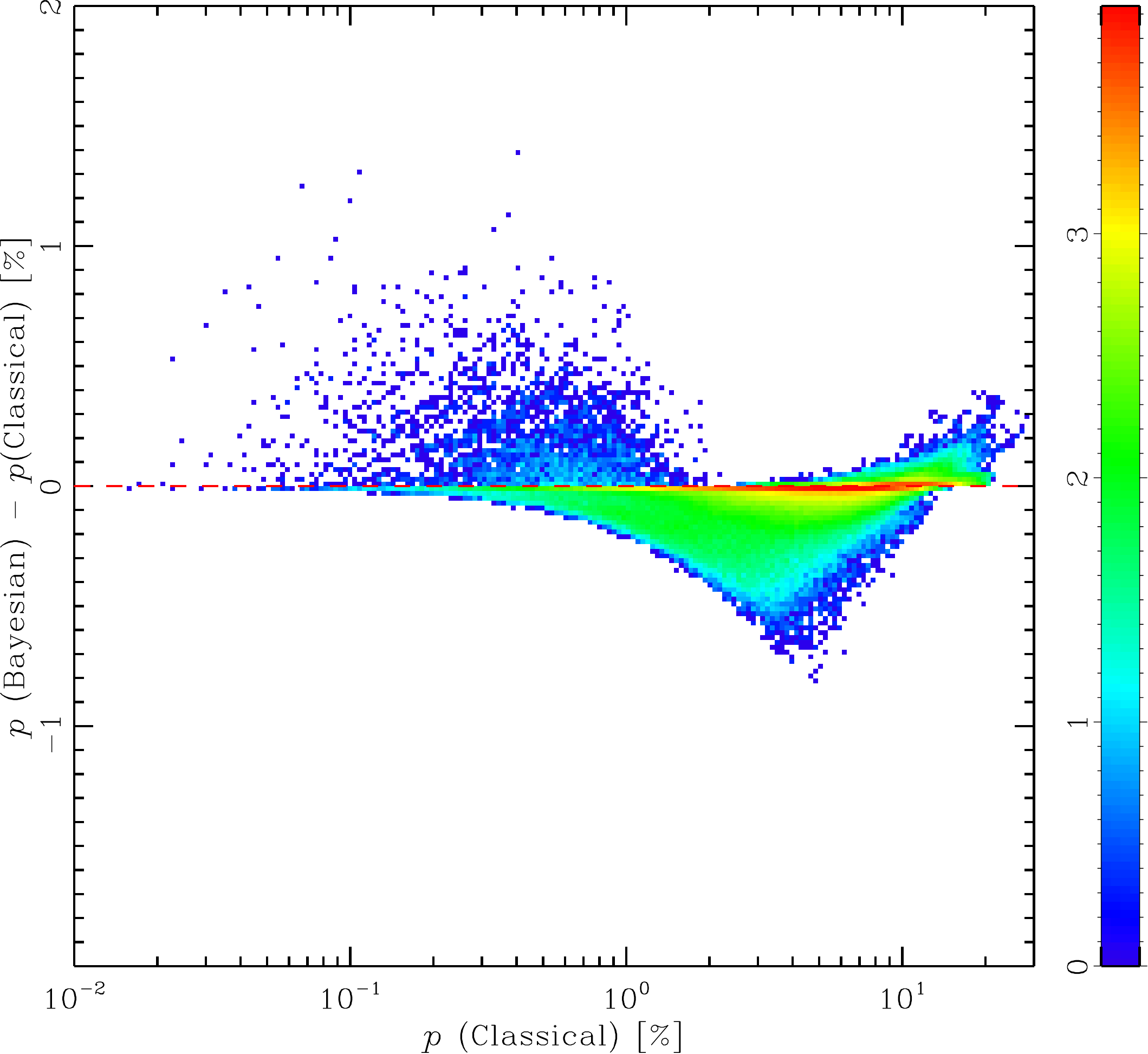}
\includegraphics[width=9cm,angle=0]{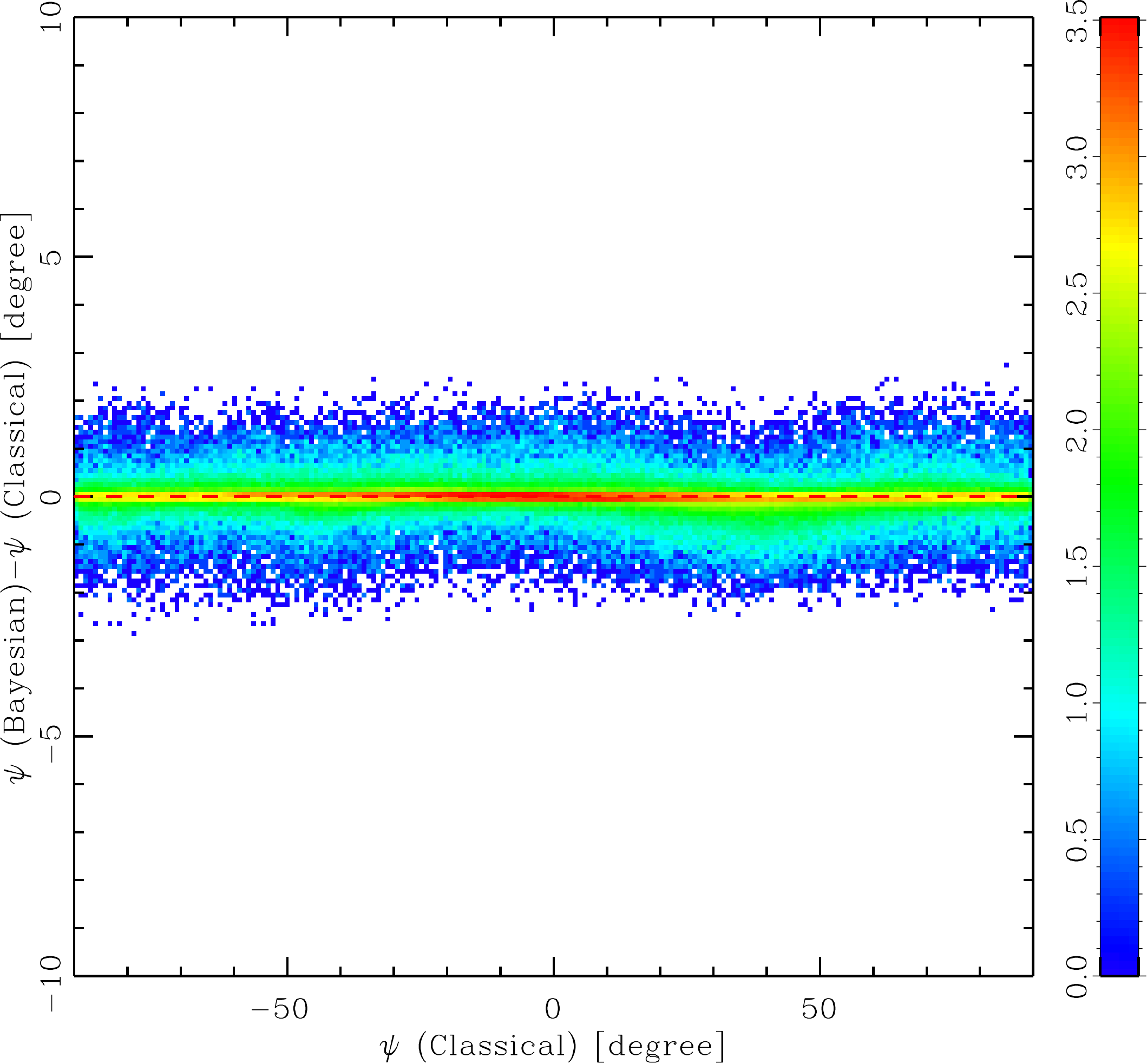}
\includegraphics[width=9cm,angle=0]{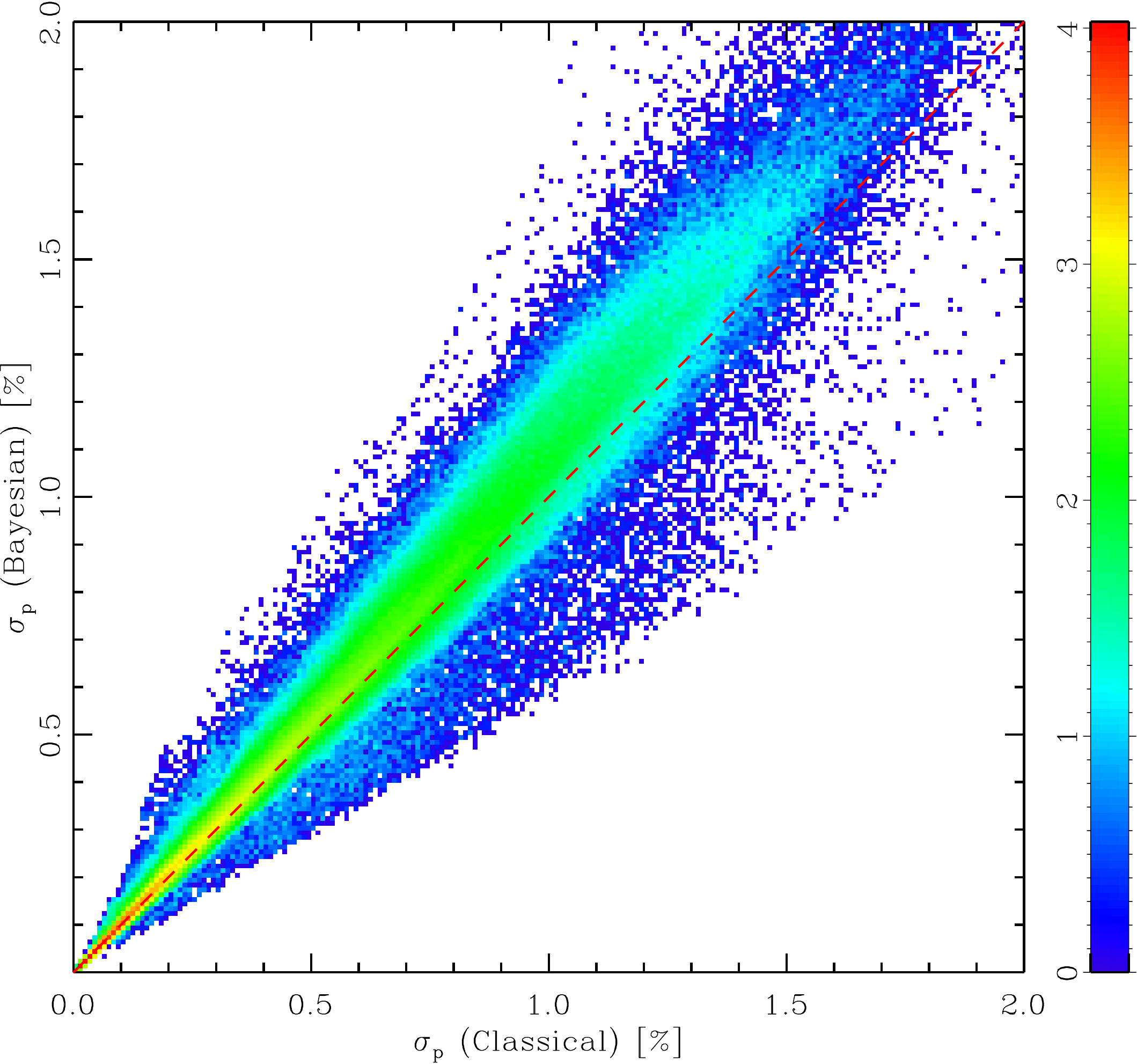}
\includegraphics[width=9cm,angle=0]{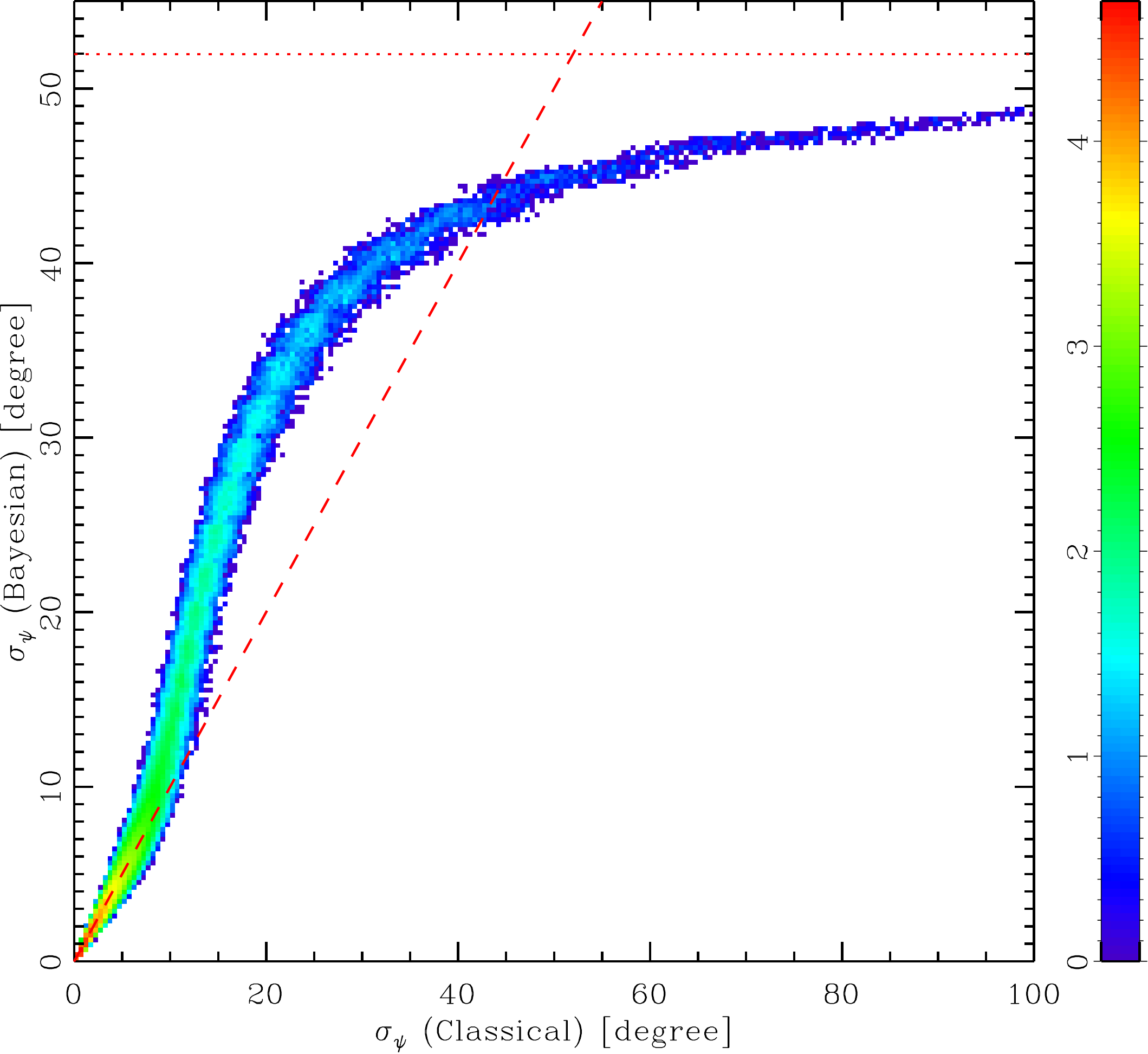}
\caption{
\emph{Upper panels}: difference between the {\classical} and the
Bayesian mean posterior estimates of $\polfrac$ and $\polang$ as a
function of the {\classical} estimate.  \emph{:Lower panels}: Bayesian
mean posterior estimates of $\sigpolfrac$ and $\sigpolang$ as a
function of the {\classical} estimate.  The dashed blue lines show
where the two methods give the same result.  Each plot shows the
density of points in log-scale for the {\Planck} data at $1\degr$
resolution.  The dotted line in the lower right plot shows the value
for pure noise. The colour scale shows the pixel density on a
log$_{10}$ scale.
\label{fig:classical_vs_ml}}
\end{center}
\end{figure*}

We use a method based on the one proposed by \cite{Quinn2012}
and extended to the more general case of an arbitrary covariance matrix
by \cite{planck2014-XXIII98}. Unlike the {\classical} method presented in
Sect.\,\ref{sec:methodOne}, this method is in principle accurate at any
signal-to-noise ratio.
Figure\,\ref{fig:classical_vs_ml} compares the Bayesian
predictions for $\polfrac$ and $\polang$ and their uncertainties
$\sigpolfrac$ and $\sigpolang$ with those obtained from the {\classical}
method (Eq.\,\ref{equ:pem}, \ref{equ:thetam}, 
\ref{equ:methodTwothetaRMS}, and \ref{equ:methodTwopRMS}) as predicted
from for the {\Planck} data at $1\deg$ resolution.  As can be seen in the
figure, the bias on $\polfrac$ is generally important even at this low
resolution. The {\classical} uncertainties are accurate only at low
uncertainties, as expected since \ref{equ:methodTwothetaRMS} and
\ref{equ:methodTwopRMS} are obtained from Taylor expanssion
around the true values of the parameters.  The difference
in the uncertainties is the greatest for $\sigpolang$ as the true
value can only reach $\randomdpsivalue\deg$ for purely random oreintations.

\section{Tests on $\DeltaAng$ bias}
\label{sec:test_deltapsi_bias}

\begin{figure}[!h!t]
\begin{center}
\includegraphics[width=9cm]{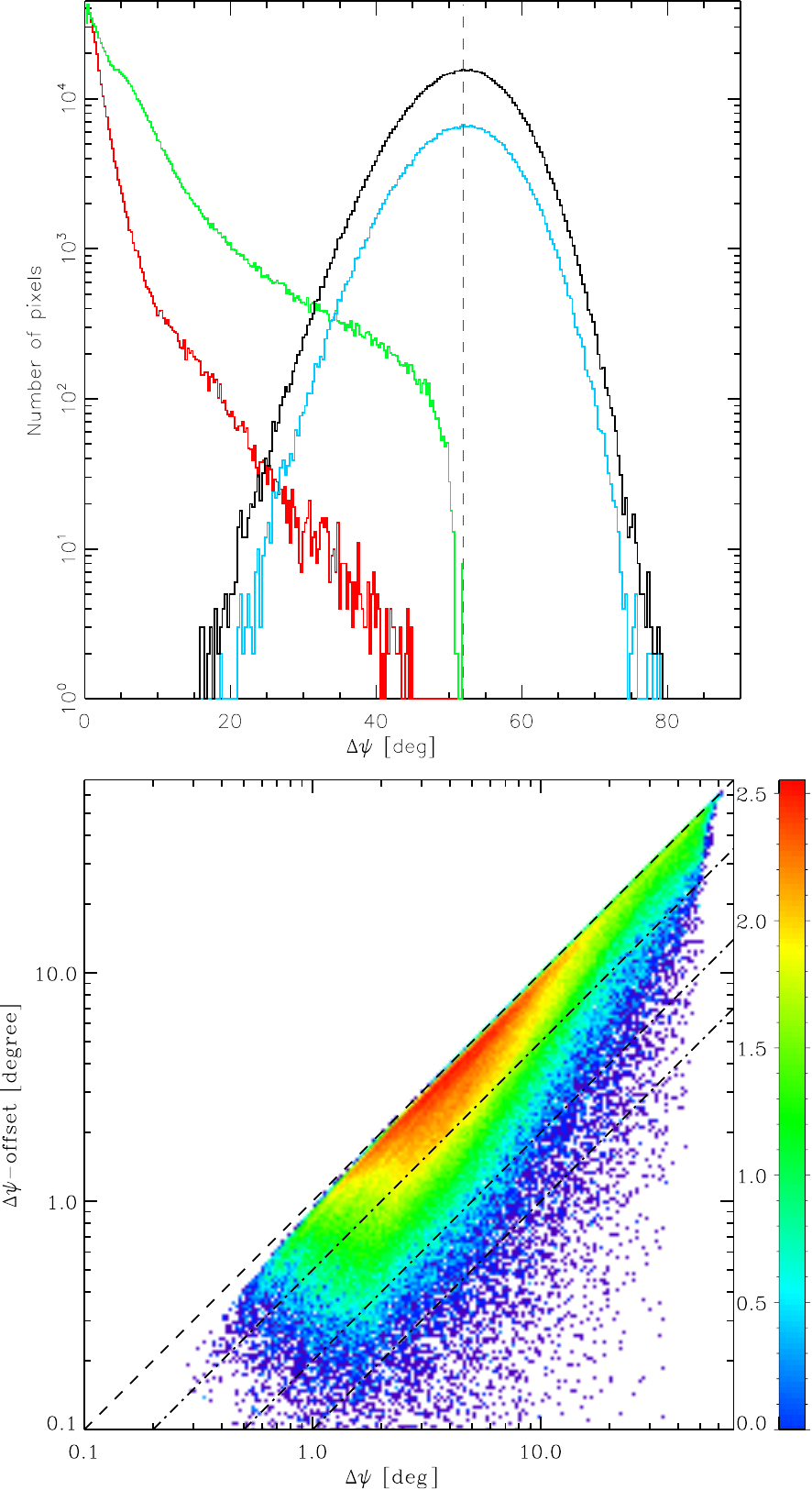}
\caption{
\emph{Upper}: histogram of $\DeltaAng$ obtained on simulated data assuming
either $\DeltaAng=0\degr$ (curves peaking at $\DeltaAng=0\degr$) or a
random value for $\DeltaAng$ (curves peaking at
$\DeltaAng=\randomdpsivalue\degr$) and noise simulated using the
actual {\Planck} noise covariance matrices.  The green and black curves
show the histograms and the blue and red curves show histograms
where $\DeltaAng/\sigpolang>3$.  The vertical dashed line shows
$\DeltaAng=\randomdpsivalue\degr$, which is the value for pure random
noise on $\StokesQ$ and $\StokesU$.  \emph{Lower}: distribution of $\DeltaAng$ minus the
offset derived from a simulation with $\DeltaAng=0\deg$, with respect
to $\DeltaAng$, in the region where $\DeltaAng/\sigpolang>3$.
Dashed lines show $\DeltaAng$=$n\times(\DeltaAng$-offset), with
$n$=1,2,5, and 10. The colour scale shows the pixel density on a
log$_{10}$ scale.
\label{fig:histogram_simu_dpsi_all}
}
\end{center}
\end{figure}

We have performed tests in which we used the {\Planck} noise
covariance matrices in order to check that the structures we observe
in the maps of the polarization {\DeltaAngName} $\DeltaAng$ are not
caused by systematic noise bias.  One of the tests (called
\testnoise) consisted of assigning each pixel a random  polarization angle
$\polang$. The second one, (called \testzero) consisted of
setting $\polang$ to a constant value over the whole sky map, which
leads to $\DeltaAng=0\degr$ (except near the poles).  In that case,
changing $\polang$ in the data was done while preserving the value of
$\polfrac$ and $\sigma_p$ computed as in Eq.\,\ref{equ:pclassic},
through the appropriate modification of $\StokesI$, $\StokesQ$, and
$\StokesU$.  The tests also use the noise covariance of the data, so
that the tests are performed with the same sky distribution of the
polarization {\SNR} as in the data.  This is critical for investigating
the spatial distribution of the noise-induced bias on $\DeltaAng$. In
both tests, we added correlated noise on $\StokesI$, $\StokesQ$, and
$\StokesU$ using the actual noise covariance matrix at each pixel, and
computed the map of $\DeltaAng$ using Eq.\,\ref{equ:delta_phi1} and
the same lag value as for the {\Planck} data.

Figure\,\ref{fig:histogram_simu_dpsi_all} shows the histograms of the
$\DeltaAng$ values obtained for these two tests, both for the whole
sky and in the mask used in the analysis of the real data.
As expected, the {\testnoise} test histograms peak at the value of
$\DeltaAng$ for Gaussian noise only
(no signal, $\DeltaAng=\randomdpsivalue\degr$). The corresponding map of
$\DeltaAng$ does not exhibit the filamentary structure of the actual
data shown in Fig.\,\ref{fig:dphi_map}. Similarly, the test histograms
of $\DeltaAng$ do not resemble that of the real data shown in
Fig.\,\ref{fig:dphi_histo}.
The {\testzero} test is important for assessing the amplitude of the
noise-induced bias, as Monte Carlo simulations show that assuming a true value of
$\DeltaAng_{0} = 0\degr$ maximizes the bias.
We therefore use this test as a determination of the upper limit for
the bias given the polarization fraction and noise properties of the
data.  Figure\,\ref{fig:histogram_simu_dpsi_all} shows that the
histograms of the recovered values peak at $\DeltaAng=0\degr$.  The
histogram is also narower in the high $\DeltaAng$ {\SNR} region than
over the full sky.  In the high $\DeltaAng$ {\SNR} mask, $60\,\%$ of the
data points have a noise-induced bias smaller than $1.6\degr$, and
$97\,\%$ have a bias smaller than $9.6\degr$.  The maps of the bias
computed for this test show a correlation with the map of
$\DeltaAng$. However, as shown in
Fig.\,\ref{fig:histogram_simu_dpsi_all} (lower panel), the effect of
the bias is small at high values of $\DeltaAng$ for most pixels and
can reach up to 50\,\% for a larger fraction of points at lower
$\DeltaAng$ (say below $\DeltaAng=10\degr$) values.  The map of
$\DeltaAng$ with the bias derived using test {\testzero} subtracted
does not significantly change the structure of the map shown in
Fig.\,\ref{fig:dphi_map} and in particular does not explain the
{\FilamentaryStructures} observed.  We note, however, that the
noise-induced bias can change the slope of the correlation between
$\DeltaAng$ and $\polfrac$.

\raggedright
\end{document}